\documentclass[fleqn,usenatbib]{mnras}

\usepackage[T1]{fontenc}
\usepackage{ae,aecompl}

\usepackage{graphicx}	
\usepackage{amsmath}	
\usepackage{amssymb}	
\usepackage{multicol}
\usepackage{ulem}

\title[Conservation of radial actions]{Conservation of radial actions in time-dependent spherical potentials}
\author[Burger, Pe\~{n}arrubia, Zavala]{
Jan D. Burger$^{1}$\thanks{E-mail: jdb5@hi.is (JB)},
Jorge Pe\~{n}arrubia$^{2}$ and Jes\'us Zavala$^{1}$
\\
$^{1}$Centre for Astrophysics and Cosmology, Science Institute, University of Iceland, Dunhagi 5, 107 Reykjav\'ik, Iceland\\
$^{2}$Institute for Astronomy, University of Edinburgh, Royal Observatory, Blackford Hill, Edinburgh EH9 3HJ, UK
}

\date{Accepted XXX. Received YYY; in original form ZZZ}

\pubyear{2021}

\begin{document}
\label{firstpage}
\pagerange{\pageref{firstpage}--\pageref{lastpage}}
\maketitle


\begin{abstract}
In slowly evolving spherical potentials, $\Phi(r,t)$, radial actions are typically assumed to remain constant. Here, we construct dynamical invariants that allow us to derive the evolution of radial actions in spherical central potentials with an arbitrary time dependence. We show that to linear order, radial actions oscillate around a constant value with an amplitude $\Delta J_r \propto \dot{\Phi}/\Phi\,P(E,L)$.
Using this result, we develop a diffusion theory that describes the evolution of the radial action distribution of ensembles of tracer particles orbiting in generic time-dependent spherical potentials. Tests against restricted $N$-body simulations in a varying Kepler potential indicate that our linear theory is accurate in regions of phase-space in which the diffusion coefficient $\Tilde{D}(J_r) < 0.01\,J_r^2$. 
For illustration, we apply our theory to two astrophysical processes. We show that the median mass accretion rate of a Milky Way (MW) dark matter (DM) halo leads to slow global time-variation of the gravitational potential, in which the evolution of radial actions is linear (i.e. either adiabatic or diffusive) for $\sim 84$ per cent of the DM halo at redshift $z=0$. This fraction grows considerably with lookback time, suggesting that diffusion may be relevant to the modelling of several Gyr-old tidal streams in action-angle space. 
As a second application, we show that dynamical tracers in a self-interacting DM (SIDM) dwarf halo (with $\sigma/m_\chi = 1\,{\rm cm^2g^{-1}}$) have invariant radial actions during the formation of a cored density profile. 

\end{abstract}

\begin{keywords}
galaxies: statistics -- diffusion -- galaxies: kinematics and dynamics -- dark matter
\end{keywords}



\section{Introduction}\label{sec_intro}

Cosmological $N$-body simulations 
of the gravitational growth and collapse of primordial cold dark matter (CDM) density perturbations 
have been very successful in reproducing the observed large scale structure of the Universe (e.g \citealt{Springel2005Nature}). 
Such CDM N-body simulations also make concordant predictions on smaller scales, including for the abundance of small haloes and the inner structure of haloes in general (see \citealt{2019Galax...7...81Z}). Moreover, N-body simulations are frequently used to model the dynamical evolution of virialized self-gravitating systems, such as globular clusters, individual DM haloes, or galaxies of different shapes and sizes. However, an interpretation of the results of such N-body simulations is not always straightforward. 
The behaviour of collisionless CDM haloes on small scales cannot be directly compared against observations owing to poorly understood baryonic effects.
Therefore, it is difficult to assess whether mismatches between simulations and observations pose serious challenges to the CDM paradigm or not (\citealt{Bullock2017}). Another issue is that on small scales, the results of N-body simulations, while concordant, often stand unaccompanied by a  theoretical explanation derived from fundamental principles. To understand the small scale predictions of CDM simulations and to disentangle baryonic effects from the long term gravitational evolution of virialized systems, it is thus desirable to derive from fundamental principles a suitable theoretical description of the latter. 

One approach to modeling the evolution of self-gravitating systems is Hamiltonian perturbation theory (see e.g. \citealt{1972MNRAS.157....1L}, \citealt{1984MNRAS.209..729T}, \citealt{2008gady.book.....B}). Hamiltonian perturbation theory is most useful if the evolution is due to a small perturbation over a smooth and stationary distribution of particles whose symmetry allows the Hamiltonian to be written as a function of a set of invariant actions. Such a Hamiltonian may then be expressed in terms of a power series of $\epsilon$:
\begin{align}
    H(\mathbf{J},\mathbf{\theta},t) = H_0(\mathbf{J})+\epsilon H_1(\mathbf{J},\mathbf{\theta},t) + ... + \epsilon^k H_k(\mathbf{J},\mathbf{\theta},t),\label{eq:ham}
\end{align}
where $\epsilon\ll 1$, $t$ is the time coordinate, $\mathbf{J}$ is the three-vector of action variables, and $\mathbf{\theta}$ is the three-vector of conjugate angles.
The equations of motion derived from this expanded Hamiltonian are solved iteratively starting at the lowest order. Formulating Hamiltonian perturbation theory in action-angle space is particularly attractive, given that actions are adiabatic invariants (\cite{2008gady.book.....B}). 
Recently, Hamiltonian perturbation theory has been formulated into a fully self-consistent kinetic theory in which the evolution of a self-gravitating system is governed by a set of equations akin to the \cite{1960PhFl....3...52B}-\cite{1960AnPhy..10..390L} equations of Plasma physics (\citealt{2010MNRAS.407..355H}). Hamiltonian perturbation theory in general, and this formalism in particular, have been quite successful in describing the secular evolution of self-gravitating systems. For instance, (\citealt{2015A&A...584A.129F}) and (\citealt{2019MNRAS.484.3198D}) have modeled the formation of spiral arms in isolated rotating disk galaxies. While it is a very powerful approach from a conceptual point of view, it can prove rather difficult to solve the differential equations arising in Hamiltonian perturbation theory. In some cases more physical insight may be gained using a different approach, particularly if the evolution of the dynamical system is governed by a globally evolving gravitational potential instead of a localized perturbation. 

A possible alternative to Hamiltonian perturbation theory is to treat self-gravitating objects as thermodynamical ensembles of particles. However, attempts at predicting the evolution of self-gravitating systems with the tools of statistical mechanics face a number of well-known difficulties.
For one, particles interacting with each other gravitationally have negative specific heat (\citealt{1961SvA.....4..859A},\citealt{1977MNRAS.181..405L},\citealt{1989ApJ...344..848P}; for a review see \citealt{1999PhyA..263..293L}). 
As a consequence, the evolution of systems of particles which are subject to long range forces cannot be described using canonical or grand canonical ensembles. Another property which is specific to large systems of particles which interact via the gravitational force and are not in dynamical equilibrium is non-ergodicity \citep{1999PhyA..263..293L}, which implies that the time-average and the ensemble average of dynamical quantities are not equivalent. Yet a further problem 
was highlighted by \citet{1990PhR...188..285P} who argued that the long range nature of the gravitational force forbids the division of the system into non-interacting macrocells, since the energy of the system is now 
non-extensive. 
As a result of this, gravitating systems cannot be described by standard thermodynamics, a conclusion which has been supported by \citet{2008PhRvE..78b1130L} and \citet{2014PhR...535....1L}. 

Despite these challenges, several attempts have been made to derive a valid statistical description of collisionless systems under gravity. \citet{1967MNRAS.136..101L} sought to construct the equilibrium distribution function of a particle ensemble subject to a strongly time-dependent gravitational force of a newly formed galaxy. The author finds that in general, the most probable coarse-grained distribution function of said particle ensemble is that of a Fermi-Dirac gas, save a normalization factor. In the non-degenerate limit, which is applicable for galaxies, this distribution function can be approximated as a Maxwell-Boltzmann distribution, i.e. the distribution function of an isothermal sphere. The Maxwell-Boltzmann distribution is also obtained by \citet{2000ApJ...531..739N} in a different approach using \citet{1957PhRv..106..620J} information theory. While this result is accurate in the center of the potential, it fails to reproduce numerical experiments of violent relaxation \citep{2005MNRAS.362..252A} in the outskirts. Moreover, it also implies that the system of particles has infinite total mass. 

More recently, \cite{2013MNRAS.430..121P} attempted to derive the equilibrium distribution function of a virialized DM halo. Following arguments by \citet{1957PhRv..106..620J}, the authors state that statistical mechanics can still be applied to gravitating systems, provided that additional physical constraints other than just the conservation of energy are taken into account. In their formalism, they maximize the entropy of the system to derive its equilibrium configuration, using the additional constraint that the ensemble average of the DM particle's radial actions is approximately conserved. The resulting distribution function matches the properties of simulated haloes over several orders of magnitude. However, the authors need to include a second population of particles in order to accurately predict the abundance of particles with very low angular momentum in DM haloes (see their Fig.~4). Without this second population their formalism fails to explain the inner density cusps that are ubiquitous in simulated DM haloes (\citealt{2020Natur.585...39W}).

N-body simulations consistently show that a cuspy density profile is formed immediately after gravitational collapse of DM haloes. Subsequently, haloes evolve towards a universal mass distribution, which is well described by the single two parameter Navarro-Frenk-White (NFW, \citealt{1996ApJ...462..563N,1997ApJ...490..493N}) profile for virtually all simulated haloes\footnote{A slightly improved fit can be obtained with the three parameter Einasto profile (\citealt{2010MNRAS.402...21N}), but the NFW profile still works remarkably well.}, through minor mergers and diffuse accretion. It thus appears that \citet{2013MNRAS.430..121P}'s formalism captures the late evolution of the halo, but cannot explain how particles with low radial actions and low angular momenta (the ``cusp'' particles) are created during -- or immediately after -- the initial, impulsive gravitational collapse. 
The mechanism(s) that drive the accumulation of low angular momentum material in the central density cusp are not yet understood.

While central density cusps are ubiquitous in N-body CDM simulations of structure formation, the observed kinematics of several dwarf galaxies (\citealt{Moore1994}, \citealt{deBlok2008}, \citealt{Kuzio2008}, \citealt{Walker2011}) favour DM haloes with constant-density cores. Several scenarios have been proposed to reconcile the success of the CDM paradigm at explaining the large scale structure of the Universe with the apparent failure on smaller scales. Two frequently discussed mechanisms are supernova (SN) feedback \citep[e.g.][]{Navarro1996,Pontzen:2011ty} and self-interacting DM (SIDM, \citealt{Spergel2000}, \citealt{Yoshida2000}, \citealt{Dave2001}, \citealt{Colin2002}, \citealt{Vogelsberger2012}, \citealt{Rocha2013}). For SN feedback to be a feasible mechanism of cusp-core transformation, star formation needs to be bursty and cyclical (\citealt{Pontzen:2011ty}). Moreover, supernovae need to be energetic enough to unbind the cusp (\citealt{Penarrubia:2012bb}) and feedback is more efficient if baryons are more concentrated towards the center of the galaxy (\citealt{2021arXiv210301231B}). SIDM is a feasible core formation mechanism if the momentum transfer cross section per unit mass $\sigma/m$ is of the right magnitude $\sim1$cm$^2$g$^{-1}$ (\citealt{Zavala2013}, \citealt{Kaplinghat2016}). \citet{2019MNRAS.485.1008B} demonstrated that the key difference between those two mechanisms is that cores are formed impulsively through SN feedback and adiabatically through SIDM. Hence, luminous tracers conserve radial actions in SIDM haloes, but not in CDM haloes with impulsive SN feedback. 

In this article, we aim to develop a statistical theory for the evolution of the radial action distribution of an ensemble of tracer particles orbiting in a generic time-dependent spherical potential. In contrast to Hamiltonian perturbation theory, our statistical theory does not focus directly on the evolution of the distribution function but rather on individual particles; and then derives the evolution of the distribution function by treating the particles as a microcanonical ensemble. With this approach, we aim to clearly characterize the difference between the adiabatic and the impulsive regime at the level of individual particles. We further seek to determine how distributions of radial actions behave in potentials whose rate of evolution lies between those two regimes. By characterizing the behaviour of tracers in those three regimes, we aim to qualitatively understand how cusps form in collapsing haloes, how radial action distributions evolve in a typical MW halo at the current time, and whether cusp-core transformation due to SIDM is truly an adiabatic process. At the current stage, our formalism does not take into account deviations from isotropy and instead focuses on globally evolving potentials. 
To develop our theory, we closely follow the work presented in \citet{2013MNRAS.433.2576P} and \citet{2015MNRAS.451.3537P}. \citet{2013MNRAS.433.2576P} generalized an argument of \citet{1982Obs...102...86L}, who found a coordinate transformation relating the equations of motion in Dirac's cosmology with a time-dependent gravitational constant $G$ to the standard equations of motion in a frame in which $G$ is constant. \citet{2013MNRAS.433.2576P} showed that using a similar coordinate transformation, the equations of motion of particles orbiting in any time-dependent central potential can be solved in a frame in which the potential is static, provided that one is able to solve an auxiliary differential equation for a scale factor $R$ that relates the spatial coordinates in the static frame to the original time-dependent frame. The evolution of a particle's energy in the time-dependent frame is then fully determined by $R(t)$ and the particle's phase space coordinates, while the energy in the static frame is a constant of motion (a.k.a. dynamical invariant). 
In a subsequent paper, \citet{2015MNRAS.451.3537P} showed that if the evolution of the potential is slow enough, the evolution of the energy distribution of a set of tracers is diffusive and can be calculated statistically by treating the tracers as a microcanonical ensemble. In fact, the evolution of the energy distribution of tracers is fully determined by the drift and diffusion coefficients $\Tilde{C}(E,t)$ and $\Tilde{D}(E,t)$, which are related to microcanonical averages of the difference between a particle's energy and its dynamical energy invariant. To perform this average, it is not necessary to assume a phase-mixed particle distribution and hence this approach to statistical physics does not rely on the assumption of ergodicity.


Our paper is structured as follows: In Section \ref{section_one}, we derive the first order Taylor expansion of the time-dependent radial action $J_r$ in the parameter $\dot{R}/R$. We show that to first order, the radial action oscillates with an amplitude $\Delta J_r$ around a dynamical action invariant $J_r'$ -- the radial action in the static frame. The oscillation amplitude depends linearly on the radial period and can be calculated in general time-dependent spherical potentials, provided $R(t)$ is known. We test our model on a time-dependent Kepler potential, where analytic expressions for both $R(t)$ and the radial period are known. 
In Section \ref{sec_form}, we derive the diffusion equation in radial action space and  define drift and diffusion coefficients $\Tilde{C}(J_r,t)$ and $\Tilde{D}(J_r,t)$. Furthermore, we show that we can classify the evolution of radial action distributions as linear or non-linear, depending on whether $\sqrt{\Tilde{D}} \ll J_r$ or $\sqrt{\Tilde{D}} \geq J_r$.
In Section \ref{sec_results}, we test our diffusion theory, using restricted $N$-body simulations of five different tracer particle ensembles in a time-dependent Kepler potential. 
In particular, we test whether the diffusion formalism yields accurate results, both in cases where $\sqrt{\Tilde{D}}\ll J_r$ (linear) and in cases where $\sqrt{\Tilde{D}}\sim J_r$ (non-linear) on average. Based on the results obtained in Section \ref{sec_results}, in Section \ref{sec_applications} we apply our theory to two different astrophysical processes. We discuss if radial actions can be considered conserved quantities in Milky-Way (MW) size CDM haloes. To this end, we apply our formalism using the median mass accretion history of MW size haloes reported in \citet{10.1111/j.1365-2966.2010.16774.x} to estimate the fraction of DM particles within the MW halo whose radial actions are expected to show a linear rather than a non-linear evolution. 
We discuss implications of our results for the analysis of tidal streams in the MW today and the formation of central density cusps in DM haloes shortly after gravitational collapse.

Furthermore, we also simulate core formation in a dwarf size SIDM halo.
For $\sigma/m_\chi = 1\,{\rm cm^2g^{-1}}$, we quantify how adiabatic core formation proceeds by determining the fraction of DM particles whose radial actions evolve linearly and applying the diffusion formalism developed in Section \ref{sec_form} to model the evolution of an initially Gaussian radial action distribution of tracer particles in the SIDM halo.
We draw our conclusions in Section \ref{sec_conclusions}. Appendix \ref{app1} outlines how the scale factor is calculated numerically, Appendix \ref{app2} discusses a modification to the diffusion formalism developed in Sections \ref{section_one} and \ref{sec_form}, and in Appendix \ref{app3} we discuss an approximation to the scale factor in potentials that are not scale-free. 

\section{Radial actions in a time-dependent potential}\label{section_one}
\subsection{Time-dependent radial action distributions}\label{sec:sub_revo}

The radial action is an integral of motion in spherical static potentials \citep{2008gady.book.....B}. It is defined as 
\begin{align}
    J_{r} = \frac{1}{\pi}\int_{r_{\rm{peri}}}^{r_{\rm{apo}}}dr\,\sqrt{2\left[E-\Phi(r)\right]-\frac{L^2}{r^2}},\label{action_integral}
\end{align}
where $E$ denotes the particle's energy, $L$ its angular momentum and $r_{\rm{peri,apo}}$ the peri- and apocentre of the particle's orbit.

\begin{figure*}
	\includegraphics[height=6.5cm,width=8.5cm,trim=1.0cm 0.5cm 0.75cm 0.3cm, clip=true]{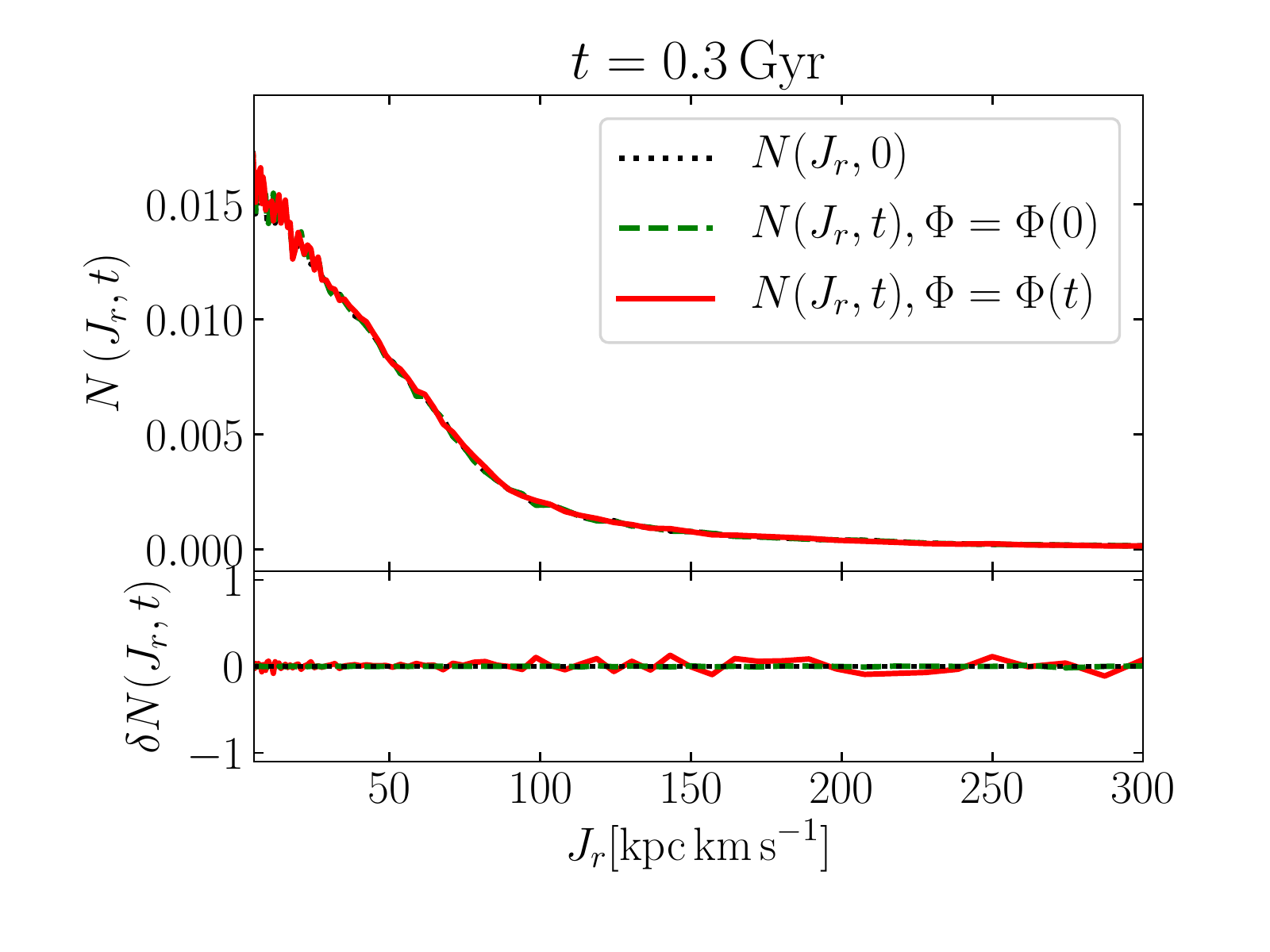}
	\includegraphics[height=6.5cm,width=8.5cm,trim=1.0cm 0.5cm 0.75cm 0.3cm, clip=true]{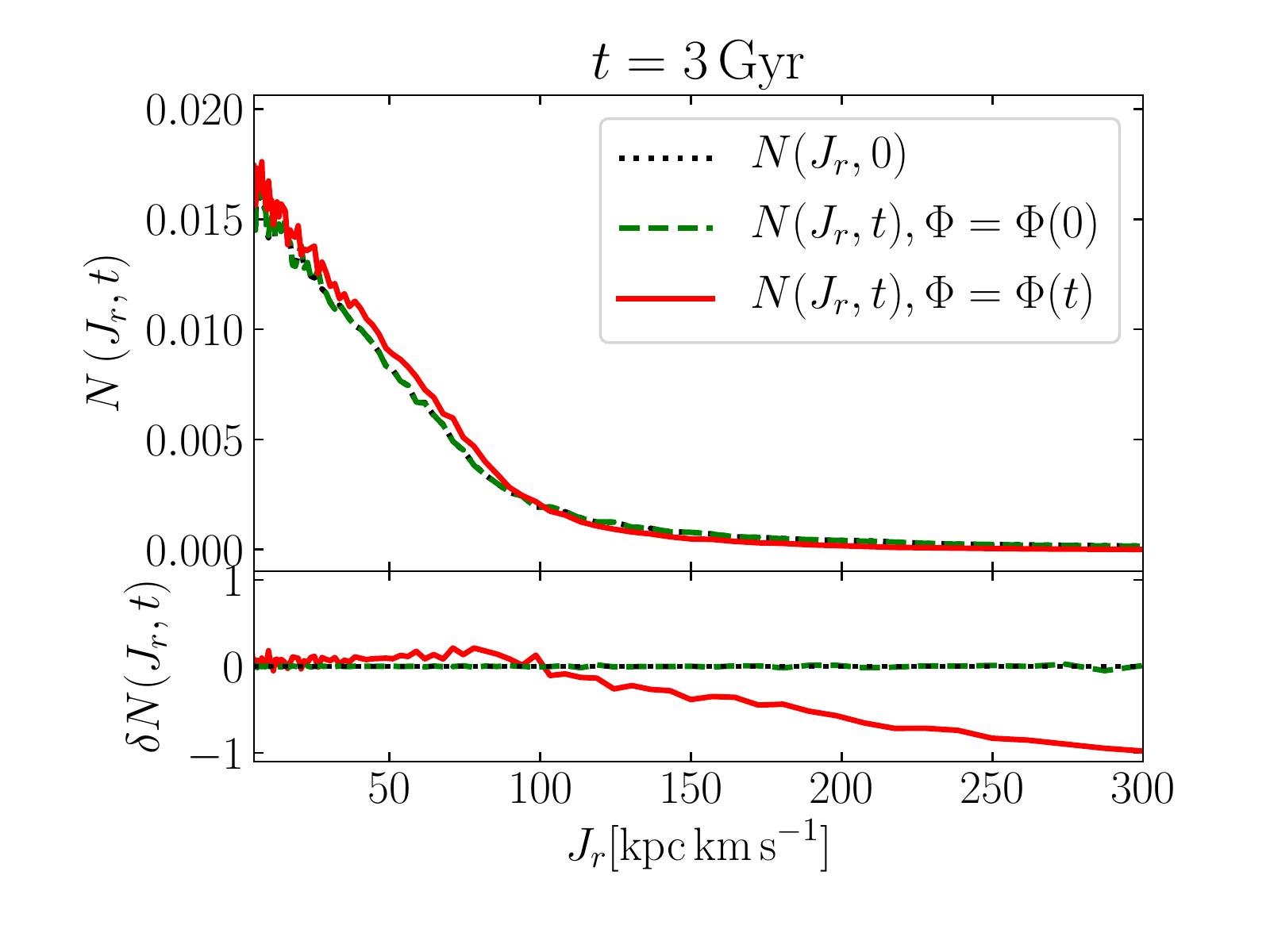}
	\caption{Evolution of the radial action distribution calculated from the orbits of $10^5$ tracer particles in a time-dependent Kepler potential (see Eq.~\ref{td_kepler_pot}). The black dotted lines show the initial distribution, the dashed green lines show the distribution evolved in a frame with a static potential, and the solid red lines show the distribution in the time-dependent frame. The simulation times at which the distributions are measured are indicated in the title of each panel. The top of either panel shows the distribution, whereas the bottom shows the fractional change 
	with respect to the initial distribution. When evolving the distribution from 0.3 to 3 ${\rm Gyr}$, the radial actions of tracers in the simulation with a time-dependent potential tend to systematically drift towards smaller values.
     }\label{act_dist_evolution}
\end{figure*}

If the potential varies with time while retaining its spherical symmetry, radial actions are approximately conserved insofar as the change is `slow' (\citealt{2008gady.book.....B}). 
The question of exactly how slow the change in potential has to be for actions to be adiabatic invariants, however, is non-trivial. If the evolution of the gravitational potential is too fast, radial actions can no longer be considered adiabatic invariants and this can cause an asymmetric drift of radial action distributions. In Fig. \ref{act_dist_evolution} we illustrate this process on a population of tracer particles orbiting in a time-dependent Kepler potential,  
\begin{align}
    \Phi(r,t) = -\frac{GM(t)}{r}.\label{ind_pot}
\end{align}
Using a Kepler potential facilitates the calculation of radial actions as Eq.~(\ref{action_integral}) has an analytic solution (e.g. \citealt{1984MNRAS.207..511G}):
\begin{align}
    J_r = \frac{GM}{\sqrt{-2E}}-L. \label{action_kepler}
\end{align}
We run a simple test simulation of $10^5$ tracer particles orbiting in a time-dependent Kepler potential with a linear time dependence of the mass
\begin{align}
    \Phi(r,t) = -\frac{GM_0\left(1+\epsilon t\right)}{r}.\label{td_kepler_pot}
\end{align}
The tracers' initial phase space coordinates are chosen at random. We require the initial orbital radius to be smaller than $300\,\rm{kpc}$ and that the angular momentum of each particle is smaller than $150\,\rm{kpc\,km\,s}^{-1}$ to avoid very extended and energetic orbits. For definiteness, we choose $M_0 = 10^8\,M_{\odot}$ and $\epsilon = 1/30\,\rm{Gyr}^{-1}$ in Eq.~(\ref{td_kepler_pot}). 
The mass of the Kepler potential grows by 10 per cent over $3 \,{\rm Gyr}$.
In the following, this is the adopted benchmark model whenever we use simulations in a time-dependent Kepler potential to test our theory.

Using the N-body code AREPO (\citealt{Springel:2009aa}) we follow the tracer orbits and compute the distribution of the tracers' radial actions at different times to determine its time evolution. 

In Fig. \ref{act_dist_evolution}, we show the evolved radial action distribution at two different times:  $t=300\,$Myr (left panel), and $t=3\,$Gyr (right panel). In the upper part of each panel, we show the final distribution $N(J_r,t)$ as a red solid line and the distribution at $t=0$ as a black dotted line. We also show the final radial action distribution 
in a simulation in which the host potential is kept constant as a green dashed line. In the bottom part of each panel, we present the fractional change in the distribution $\delta N(J_r,t)$, defined as
\begin{align}
    \delta N(J_r,t) = \frac{N(J_r,t)-N(J_r,0)}{N(J_r,0)}.
\end{align}
We find almost no evolution when using a static Kepler potential, as should be the case, since actions are integrals of motion. 
In the simulation with a time-dependent host potential, however, we see a striking evolution of the radial action distribution. 
As time goes by, we observe a progressive flattening of the distribution's tail at $J_r > 100\,\rm{kpc}\,\rm{km\,s}^{-1}$. At larger values, we find that $\delta N(J_r,t)$ slowly tends towards the limiting value of -1 (at $J_r \sim 300\,\rm{kpc}\,\rm{km\,s}^{-1}$), indicating that almost no particles with radial actions larger than that remain.
As a result of this, we find that $\delta N(J_r,t)$ is continuously larger than $0$ at radial actions smaller than $100\,\rm{kpc}\,\rm{km\,s}^{-1}$. Overall, this indicates a significant net drift of the initial distribution towards smaller radial actions. 
 
\subsection{First-order expansion of time-dependent radial actions}\label{sub_tay_one}
To attempt to understand the evolution observed in Fig. \ref{act_dist_evolution}, we start by deriving the time evolution of the radial actions of individual tracers in time-dependent spherical potentials.
Our calculation closely follows the one for energies presented in \citet{2013MNRAS.433.2576P}, which is based on a coordinate transformation found by \citet{1982Obs...102...86L}.

In a time-dependent potential, particles are subject to a time-dependent force
\begin{align}
    \ddot{\mathbf{r}} = \mathbf{F}(\mathbf{r},t).\label{force_eg}
\end{align}
If the force is conservative, \citet{2013MNRAS.433.2576P} and \citet{1982Obs...102...86L} show that for each phase space trajectory there exists a canonical transformation $\mathbf{r}\to R(t)\mathbf{r}'$ and a complementary transformation of the time coordinate $dt\to d\tau R^2(t)$ such that the time-dependence vanishes from the equation of motion, which can then be written as 
\begin{align}
    \frac{d^2\mathbf{r}'}{d\tau^2} = \mathbf{F'}(\mathbf{r'})\label{new_force_eq}
\end{align}
The scale factor $R(t)$ is then a solution to the differential equation 
\begin{align}
    \ddot{R}R^3\mathbf{r}'-R^3\mathbf{F}(R\mathbf{r}',t) = -\mathbf{F}'(\mathbf{r}')\label{scalefactor}.
\end{align}
Notice that here, the time-evolution of the scale factor is coupled to the phase space trajectory. \citet{2013MNRAS.433.2576P} shows that in the case of a slowly changing potential, an approximate analytic first-order solution to Eq. (\ref{scalefactor}) exists for scale-free potentials. In Appendix \ref{app3}, we demonstrate that a modification of this solution may be used for general spherical potentials.

Up to first order in $\dot{R}/R$, the energy of a particle which is subject to the force in Eq.~(\ref{force_eg}) is 
\begin{align}
    E \approx \frac{I}{R^2}+\frac{\dot{R}}{R}\left(\mathbf{r}\cdot\mathbf{v}\right),\label{linear_energy}
\end{align}
where $I$ is a dynamical invariant equal to the energy in the frame in which Eq.~(\ref{new_force_eq}) is the equation of motion, i.e., the "time-independent", or static, frame.

Eq.~(\ref{linear_energy}) expresses the energy in the time-dependent frame as a function of the invariant energy and a first order correction that depends on the orbit of each particle. In the following, we aim to derive a similar expression for the radial action. A possible way to do so, presented in \citet{2013MNRAS.433.2576P}, is to simply insert Eq. (\ref{linear_energy}) into Eq.~(\ref{action_kepler}) and identify the first order correction from there. However, since the radial action is not usually analytic in generic spherically symmetric potentials, it is desirable to derive a more general expression. 

We start by integrating Eq.~(\ref{scalefactor}) on both sides to define 
\begin{align}
    \Tilde{\Phi}(\mathbf{r}') = -\int d\mathbf{r}'\cdot \mathbf{F}'(\mathbf{r}')
    = \frac{1}{2}\ddot{R}R^3r'^2+R^2\Phi(R\mathbf{r'},t).\label{shifted_phi}
\end{align}
The radial action in the time-independent frame is 
\begin{align}
    J_{r'} = \frac{1}{\pi}\int_{r'_{\rm{peri}}}^{r'_{\rm{apo}}}dr'\,\sqrt{2I-2\Tilde{\Phi}(\mathbf{r}')-\frac{L^2}{r'^2}},\label{invariant_action}
\end{align}
where $J_{r'}$ is a dynamical invariant akin to $I$.
We now seek to relate Eq.~(\ref{invariant_action}) to the radial action that one would de-facto measure in a time-dependent potential. 
To measure $J_r$ in a time-dependent frame, one usually fixes the gravitational potential at the time one measures $J_r$. For this reason, we shall also refer to $J_r$ as the "instantaneous" action. Fixing $\Phi$ at the time we measure $J_r$, we can define effective apo- and pericentre radii in the time-dependent frame. Formally, this means that we consider quantities in the time-dependent frame at a fixed time $t_1$. At this time, the invariant energy is  
\begin{align}
    I = R^2(t_1)E(t_1)-R^2(t_1)\frac{\dot{R}(t_1)}{R(t_1)}\left(\mathbf{r}(t_1)\cdot \mathbf{v}(t_1)\right).\label{invariant}
\end{align}
We then obtain
\begin{align}
    J_{r'} =\frac{1}{\pi} \int_{r'_{\rm{peri}}}^{r'_{\rm{apo}}} dr'\,R(t_1)\sqrt{f(\mathbf{r}',t_1)}, 
\end{align}
where 
\begin{align}
 \nonumber   f(\mathbf{r}',t_1) &= 2E(t_1)-2\frac{\dot{R}(t_1)}{R(t_1)}\left(\mathbf{r}(t_1)\cdot \mathbf{v}(t_1)\right)\\
\nonumber &-\ddot{R}(t_1)R(t_1)r'^2-2\Phi(R(t_1)\mathbf{r}',t_1)\\
&-\frac{L^2}{R^2(t_1)r'^2}.
\end{align}
Now, we can perform the coordinate transformation $\mathbf{x} = R(t_1)\mathbf{r}'$, which is simply a shift of the radial coordinate. We write the effective apo- and pericentre radii in the time-dependent frame as    
\begin{align}
    R(t_1)r'_{\rm{apo,peri}} = r_{\rm{apo,peri}}(t_1) 
\end{align}
which is physically exact in the fully adiabatic limit. 
With these definitions, 
\begin{align}
    J_{r'} = \frac{1}{\pi}\int_{r_{\rm{peri}}(t_1)}^{r_{\rm{apo}}(t_1)}dx \sqrt{f(\mathbf{x},t_1)}
\end{align}
with 
\begin{align}
\nonumber    f(\mathbf{x},t_1) &=
    2E(t_1)-2\frac{\dot{R}(t_1)}{R(t_1)}\left(\mathbf{r}(t_1)\cdot \mathbf{v}(t_1)\right)\\
\nonumber    &-\frac{\ddot{R}(t_1)}{R(t_1)}x^2-2\Phi(\mathbf{x},t_1)-\frac{L^2}{x^2}.
\end{align}
In the limit $|\ddot{R}/R| \ll |\dot{R}/R|$ we can perform a Taylor expansion to first order in $\dot{R}/R$ to find
\begin{align}
\nonumber    J_{r'} &= \frac{1}{\pi}\int_{r_{\rm{peri}}(t_1)}^{r_{\rm{apo}}(t_1)}dx\,\sqrt{2E(t_1)-2\Phi(\mathbf{x},t_1)-\frac{L^2}{x^2}}\\
\nonumber &- \frac{\dot{R}(t_1}{R(t_1)}\left(\mathbf{r}(t_1)\cdot \mathbf{v}(t_1)\right)\\
\nonumber &\times \frac{1}{\pi}\int_{r_{\rm{peri}}(t_1)}^{r_{\rm{apo}}(t_1)}dx\,\frac{1}{\sqrt{2E(t_1)-2\Phi(\mathbf{x},t_1)-\frac{L^2}{x^2}}}\\
&+\mathcal{O}\left(\left[\frac{\dot{R}(t_1)}{R(t_1}\right]^2,\frac{\ddot{R}(t_1)}{R(t_1)}\right).\label{radial_action_expansion}
\end{align}
Dropping all the higher order terms, this leads to 
\begin{align}
    J_{r'} \approx J_{r,t_1} - \frac{\dot{R}(t_1)}{R(t_1)}\left(\mathbf{r}(t_1)\cdot \mathbf{v}(t_1)\right)\frac{P(E,L,t_1)}{2\pi}\label{almost_app}
\end{align}
In this equation, $J_{r,t_1}$ and $P(E,L,t_1)$ refer to instantaneous actions and radial periods measured at the fixed time $t_1$. Now, if the system is adiabatic, we can chose $t_1$ to be any time $t$ and simply write
\begin{align}
    J_r \approx J_{r'} + \frac{\dot{R}}{R}(\mathbf{r}\cdot\mathbf{v})\frac{P(E,L)}{2\pi}\label{linear_order}
\end{align}
up to linear order in perturbation theory. 
As long as the change in the potential is slow enough, the quantity $J_{r'}$ in Eq.~(\ref{linear_order}) is a dynamical invariant. In case of a faster change in the potential, higher order terms of the perturbative expansion have to be taken into account until at some point the evolution becomes non-perturbative.

\subsection{Numerical tests of the first-order expansion}\label{sec_3particles}
We test the performance of Eq.~(\ref{linear_order}) by following the orbits of three tracers in our benchmark Kepler potential. An example of the performance of Eq.~(\ref{linear_order}) in a more general potential is shown in Appendix \ref{app3}.
The three particles are initially at a distance of 5 ${\rm kpc}$ from the host's centre, but have different initial energies and angular momenta. 

\begin{figure*}
	\includegraphics[height=4.5cm,width=5.7cm,trim=0.1cm 0.7cm 1.5cm 0.1cm, clip=true]{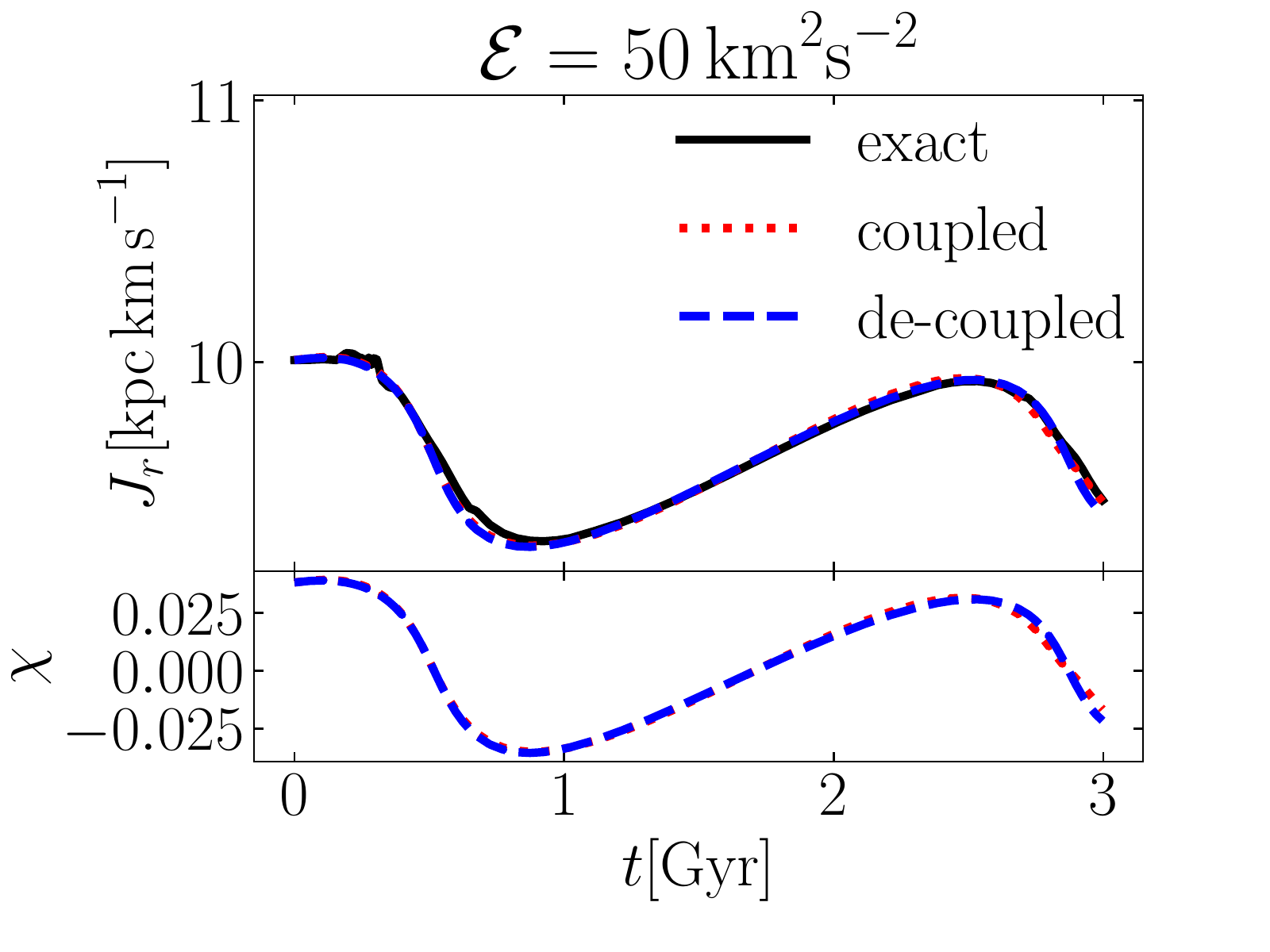}
	\includegraphics[height=4.5cm,width=5.7cm,trim=0.1cm 0.7cm 1.5cm 0.1cm, clip=true]{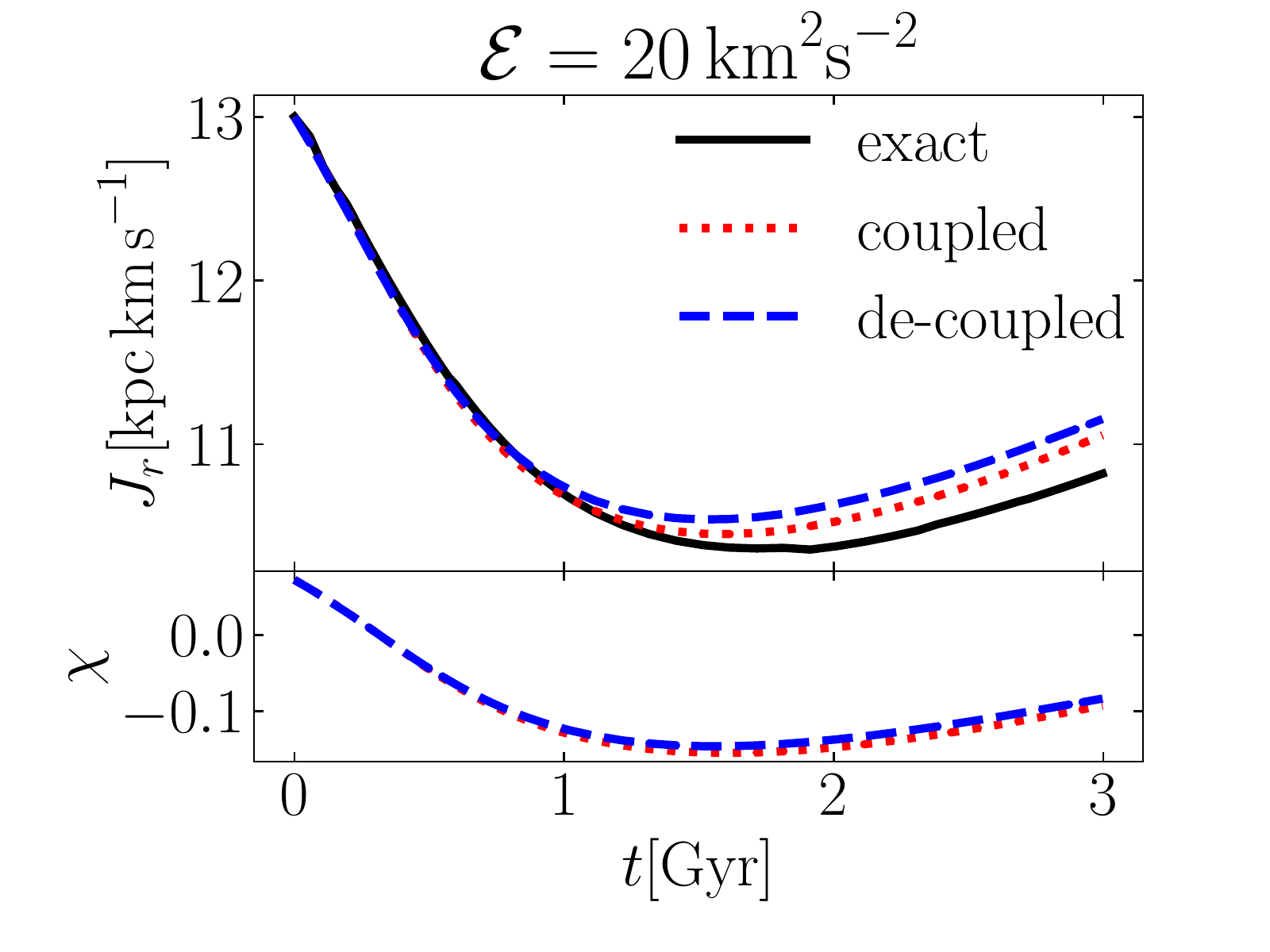}
	\includegraphics[height=4.5cm,width=5.7cm,trim=0.1cm 0.7cm 1.5cm 0.1cm, clip=true]{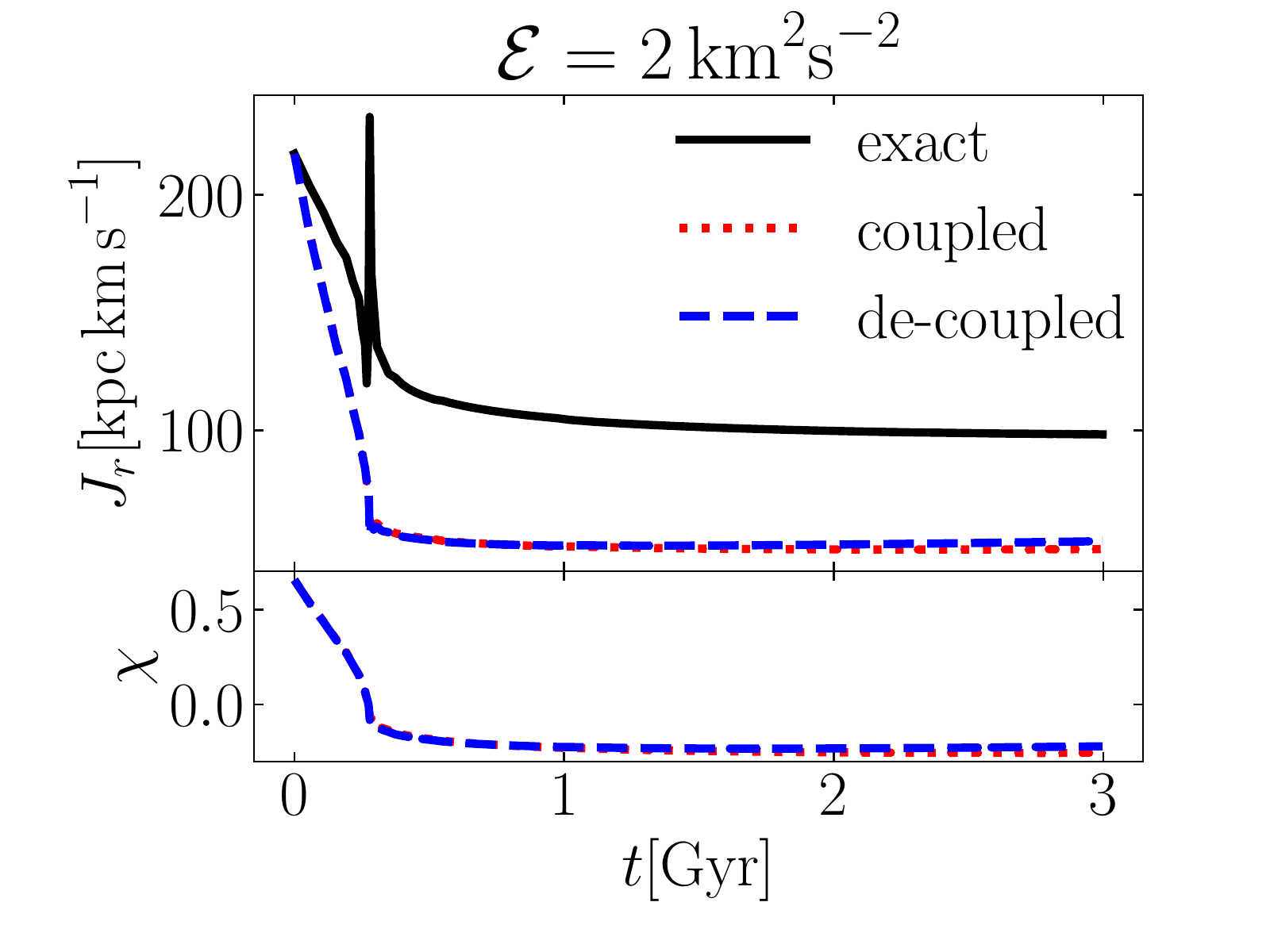}
	\caption{The upper panels show the time evolution of the radial action of three different particles orbiting in the benchmark Kepler potential.
	Initial energies are shown at the top of each plot. Black solid lines denote a direct measurement of the radial action from energies and angular momenta according to Eq.~(\ref{action_kepler}). Red dotted lines and blue dashed lines both correspond to the linear approximation given by Eq. (\ref{linear_order}), using two different estimates for $\dot{R}/R$. ``Coupled" and ``de-coupled" refer to an exact (dotted red) or approximate (dashed blue) way of solving Eq.~(\ref{scalefactor}). Angular momenta are chosen such that the initial radial actions are $J_r\sim 10\,\rm{kpc\,km\,s}^{-1}$ in the left and middle panel, and $J_r\sim 200\,\rm{kpc\,km\,s}^{-1}$ in the right panel. The lower panels show the ratio $\chi$ between the linear approximation and the exact value of the radial action as a function of time.}\label{3particles}
\end{figure*}

Since negative energies correspond to gravitationally bound particles, we define $\mathcal{E} = -E$ as our energy variable of reference. In a Kepler potential, the radial period and the azimuthal period coincide and can, in terms of the energy, be written as 
\begin{align}
    P(E,L) \equiv P(\mathcal{E}) = 2\pi\frac{GM}{\sqrt{2\mathcal{E}}^3}\label{Kepler_period}
\end{align}
Eq.~(\ref{linear_order}) implies that the amplitude with which $J_r$ oscillates around the dynamical invariant $J_{r'}$ is directly related to the orbital period. Given that the first order correction to $J_{r'}$ has to be small compared to the value of $J_{r'}$ itself for the linear approximation to be valid, we expect from Eq.~(\ref{Kepler_period} that the first order Taylor expansion will become progressively less accurate as $\mathcal{E}\to 0$. This region is known as the ``fringe" of a self-gravitating system. 

The initial energies and angular momenta of the three test particles are
\begin{align}
    (\mathcal{E}[\rm{km}^2\rm{s}^{-2}],L[\rm{kpc\,km\,s}^{-1}]) = \{(50,33),(20,55),(2,15)\}. 
\end{align}
The first two tracers initially have similar radial actions but different energies. The third tracer has a much larger initial radial action and represents a particle in the ``fringe".

The upper panels of Fig. \ref{3particles} show the time evolution of the radial actions of the three particles. Black lines in each figure denote an exact calculation of the radial action following Eq.~(\ref{action_kepler}). Red-dotted and blue-dashed lines both refer to the linear approximation for the radial action given in Eq.~(\ref{linear_order}), using expression (\ref{Kepler_period}) to calculate the radial period. The difference between the latter two cases lies in the way $\dot{R}/R$ is estimated. 
The scale factor used to obtain the red line is calculated directly from Eq.~(\ref{scalefactor}), using a KDK leapfrog algorithm (see Appendix \ref{app1}) to solve the differential equation. The scale factor's evolution is thus ``coupled" to the particle's phase space trajectory. The dashed blue lines refer to an approximate analytic solution to the scale factor that can be calculated for scale-free potentials (see \citet{2013MNRAS.433.2576P}). In this approximation the equation of motion is ``de-coupled" from Eq.~(\ref{scalefactor}).

In the three lower panels of Fig.~\ref{3particles} we show $\chi=\Delta/J_r$, 
where $\Delta\equiv  (\dot{R}/{R})\left(\mathbf{r}\cdot\mathbf{v}\right)P/(2\pi)$. Notice that $\chi$ quantifies the fractional size of the first order correction relative to the radial action, and can be positive or negative depending on the orbital phase.

Comparison of the left and middle panels of Fig. \ref{3particles} shows that the orbital period is shorter in the left panel, which is in line with our expectation from Eq.~(\ref{Kepler_period}). 
Furthermore, the results confirm our expectation that the amplitude of the first order correction increases with the energy at a fixed invariant action. We find that for the most bound particle (left panel), the first order Taylor expansion is an excellent fit to the measured radial action. We furthermore find that there is almost no difference between the red and blue lines, which indicates that the ``de-coupled" approximation to the scale factor provides a good approximation to its true value. The success of the Taylor expansion here is reflected in $\chi$ as well, which peaks at $\sim 2.5$ per cent when the magnitude of the radial velocity is maximal. As the period increases, the agreement between the measured radial action and the prediction from Eq.~(\ref{linear_order}) starts to deteriorate after roughly $1\,$Gyr of integration (notice that in the middle panel $|\chi|$ has risen to $\sim 10$ per cent). Nonetheless, the evolution of the radial action is still captured well by this approximation. We furthermore find that the numerical solution to Eq.(\ref{scalefactor}) now provides a slightly more accurate result than the approximate analytic solution for $\dot{R}/R$.

The right panel of Fig.~\ref{3particles} shows the evolution of the radial action of the third tracer. That this particle is in the ``fringe" of the potential is evident by $\chi$ reaching values as large as 70 per cent. 
Due to that, we find that the dynamical invariant $J_r'$ is not properly defined by the linear approximation -- $J_r$ changes impulsively. Note that the orbital period of this particle is much longer than the total simulation time, meaning we only resolve a fraction of an orbital revolution. Within this time interval, the radial action undergoes a strongly non-linear fluctuation when the particle passes its orbital pericentre and then quickly settles to a value which is roughly half of its initial value. The linear approximation completely fails to capture this behaviour. 

\section{Diffusion formalism for radial action distributions}\label{sec_form}

Based on the first order Taylor expansion for time-dependent radial actions derived in the previous section, we now develop a diffusion formalism similar to the one in \citet{2015MNRAS.451.3537P} to analytically describe the evolution of radial action distributions in time-dependent spherical potentials. We furthermore show how this diffusion formalism can be used to differentiate regimes in integral of motion space in which the evolution of radial action distributions is linear from regimes in which it is not.  

\subsection{The diffusion equation for radial action distributions}\label{sub_diffeq}
To derive a diffusion equation for radial actions, we closely follow the formalism presented in \citet{2015MNRAS.451.3537P} based on \citet{doi:10.1002/andp.19053220806}.

Let us define $\phi(\Delta|J_{r'})d\Delta$ as the conditional probability that a particle with the dynamically invariant action $J_{r'}$ will change its action by an amount $\Delta$ in the range $d\Delta$ during the time interval $(t,t+\tau')$; $\phi$ is normalized such that 
\begin{align}
    \int \phi(\Delta|J_{r'})d\Delta = 1.
\end{align}
Subsequently, we define $p(J_r,t_0+\tau'|J_{r'},t_0)dJ_r$ as the probability that a particle with action $J_{r'}$ at $t_0$ will have an action in the interval $[J_r,J_r+d\,J_r]$ at a later time $t = t_0+\tau'$. As with the equivalent in energy space presented in \citet{2015MNRAS.451.3537P}, these two functions obey Einstein's master equation
\begin{align}
 \nonumber &p(J_r,t_0+\tau'|J_{r'},t_0)dJ_r =\\
  & d J_r \int p(J_r-\Delta,t_0|J_{r'},t_0)\phi(\Delta|J_{r'})d\Delta.\label{master_eq}
\end{align}
Now if we are in the adiabatic limit where Eq.~(\ref{linear_order}) applies, or equivalently $\chi=\Delta/J_r \ll 1$, we can expand Eq.~(\ref{master_eq})
in $\Delta$ to find 
\begin{align}
\nonumber &p(J_r,t_0+\tau'|J_{r'},t_0) \approx p(J_r,t_0|J_{r'},t_0)\\
\nonumber&-\left. \frac{\partial p}{\partial J_{r}}\right|_{\Delta = 0}\int \phi(\Delta|J_{r'})\Delta\,d\Delta\\
  &+\frac{1}{2}\left. \frac{\partial^2p}{\partial J_r^2}\right|_{\Delta = 0}\int \phi(\Delta|J_{r'})\Delta^2\,d\Delta.\label{master_eq2}
\end{align}
Since we can also expand the lhs of Eq.~(\ref{master_eq2}) around $t=t_0$ to first order
\begin{align}
    \nonumber &p(J_r,t_0+\tau'|J_{r'},t_0) \approx p(J_r,t_0|J_{r'},t_0)+\frac{\partial p}{\partial t}\tau',
\end{align}
we then have
\begin{align}
    \frac{\partial p}{\partial t} = C\left. \frac{\partial p}{\partial J_r}\right|_{\Delta=0}+D\left. \frac{\partial^2 p}{\partial J_r^2}\right|_{\Delta=0},\label{diffusion}
\end{align}
where we define the drift coefficient 
\begin{align}
    C(J_{r'},t) = -\frac{1}{\tau'}\int \phi(\Delta|J_{r'})\Delta\,d\Delta \label{drift_coefficient}
\end{align}
and the diffusion coefficient 
\begin{align}
     D(J_{r'},t) = \frac{1}{2\tau'}\int \phi(\Delta|J_{r'})\Delta^2\,d\Delta.\label{diffusion_coefficient}
\end{align}
Analogous to the discussion in \citet{2015MNRAS.451.3537P}, the initial condition to solve Eq.~(\ref{diffusion}) is $p(J_r,t_0|J_{r'},t_0) = \delta(J_r-J_{r'})$ and the general solution to this equation is a Green function 
\begin{align}
\nonumber    p(J_r,t&|J_{r'},t_0) = \frac{1}{\sqrt{4\pi \Tilde{D}(J_{r'},t)}}\\
&\times \exp\left\{ -\frac{\left[J_r-J_{r'}+\Tilde{C}(J_{r'},t)\right]^2}{4\Tilde{D}(J_{r'},t)}\right\},\label{green_solution}
\end{align}
the properties of which are discussed in detail in \citet{2015MNRAS.451.3537P}. In Eq.~(\ref{green_solution}) we have implicitly defined scaled drift and diffusion coefficients $\Tilde{C} = C\tau'$, $\Tilde{D}= D\tau'$. Note that the fact that $\Tilde{C}$ and $\Tilde{D}$ are independent of time implies that there is no divergence for short transition times in Eq.~(\ref{green_solution}). 

In the perturbative regime, we can use the transition probability given by Eq.~(\ref{green_solution}) to calculate the radial action distribution, $N(J_r,t)$, from the invariant action distribution.  
To calculate the invariant distribution $N(J_{r'})$ from the initial distribution $N(J_r,t_0)$, 
we also need $p(J_{r'}|J_r,t_0)$. The derivation of this probability works analogous to the calculation above, and one obtains:
\begin{align}
\nonumber   & p(J_{r'}|J_r,t_0) = \frac{1}{\sqrt{4\pi \Tilde{D}(J_r,t_0)}}\\
&\times \exp\left\{ -\frac{\left[J_{r'}-J_r-\Tilde{C}(J_r,t_0)\right]^2}{4\Tilde{D}(J_r,t_0)}\right\}.\label{green_solution_two}
\end{align}
We can then calculate $N(J_{r'})$ from $N(J_r,t_0)$ as
\begin{align}
    N(J_{r'}) = \int dJ_r\,p(J_{r'}|J_r,t_0)N(J_r,t_0).\label{fwd_conv}
\end{align}

Contrary to the time average of a particle's energy in a time-dependent potential, the time average of the radial action is constant according to Eq.~(\ref{linear_order}), and thus we do not need an extra convolution analogous to Eq. 28 of \citet{2015MNRAS.451.3537P}.
The radial action distribution at a time $t$ is thus obtained from
\begin{align}
    N(J_r,t) = \int dJ_{r'}\,p(J_r,t|J_{r'},t_0)\,N(J_{r'}),\label{rev_conv}
\end{align}
where the transition probability is the one defined in Eq.~(\ref{green_solution}).

\subsection{Drift and diffusion coefficients}\label{sub_coeff}
Similar to Section 2.3 of \citet{2015MNRAS.451.3537P} we now briefly discuss how to calculate the drift and diffusion coefficients defined in Eqs.~(\ref{drift_coefficient}) and (\ref{diffusion_coefficient}). To this end, we define the microcanonical distribution function $w(J_r,E)$ depending on both energy and radial action,
\begin{align}
    w(J_r,E) = \int \delta\left( J_r - X\right)\delta(E-H)\mathbf{d^3r\,d^3v},\label{DF}
\end{align}
where 
\begin{align}
    X = \frac{1}{\pi}\int_{r_{\rm{peri}}}^{r_{\rm{apo}}}dr\,\sqrt{2(E-\Phi(r))-\frac{L^2}{r^2}}.
\end{align}
We then define the drift and diffusion coefficients as microcanonical averages over the particle distribution as follows
\begin{align}
\nonumber    &\Tilde{C}(J_r,E,t) = -\frac{1}{w}\\
    &\times\int \frac{\dot{R}}{R}\left(\mathbf{r}\cdot\mathbf{v}\right)\frac{P(H,X,t)}{2\pi}\delta(J_r-X)\delta(E-H)\mathbf{d^3r\,d^3v}\label{int_drift}\\
\nonumber    &\Tilde{D}(J_r,E,t) = \frac{1}{2w}\\
    &\times \int \left(\frac{\dot{R}}{R}\left(\mathbf{r}\cdot\mathbf{v}\right)\frac{P(H,X,t)}{2\pi}\right)^2\delta(J_r-X)\delta(E-H)\mathbf{d^3r\,d^3v}.\label{int_diff}
\end{align}
Notice that here we have written the period as a function of both the energy and the radial action, which is possible in general spherical potentials where the angular momentum is a function of energy and radial action. 
To obtain drift and diffusion coefficients that depend only on the radial action, however, we have to integrate over the energies
\begin{align}
    &\Tilde{C}(J_r,t) = \frac{1}{w(J_r)}\int dE\,w(J_r,E)\,\Tilde{C}(J_r,E,t)\\
    &\Tilde{D}(J_r,t) = \frac{1}{w(J_r)}\int dE\,w(J_r,E)\,\Tilde{D}(J_r,E,t),
\end{align}
where 
\begin{align}
    w(J_r) = \int dE\,w(J_r,E)
\end{align}
is the microcanonical distribution function depending only on the radial action.

\subsection{Adiabatic, diffusive, and impulsive evolution} \label{subsec:ana}
We can use the diffusion coefficient defined in Eq.~(\ref{int_diff}) to characterize the rate at which a gravitational potential evolves. As we argued in Section \ref{sub_tay_one}, the regime in which the evolution of radial actions becomes non-perturbative corresponds to the regime in which the evolution of the gravitational potential is impulsive. Mathematically, this means that in Eq.~(\ref{linear_order}), $\Delta = |J_r-J_{r'}|\sim J_r$. 
For a given time-dependent potential, there is always some part of integral of motion space in which this condition is fulfilled on average. Whether the evolution of a dynamical system is adiabatic, diffusive, or impulsive then depends on how populated this part of integral of motion space is. Since $\Tilde{D}\sim 1/2\,\Delta^2$, a possible way to estimate whether distributions of particles occupying a particular integral of motion space volume evolve either adiabatically to diffusively (linearly) or impulsively (non-linearly) is by the ratio $\sqrt{\Tilde{D}}/J_r$. 

This is particularly useful for phase-mixed particle ensembles. In such cases $\Tilde{C} = 0$ and thus the drift cannot be used to estimate whether the evolution of such ensembles is diffusive or impulsive. The diffusion coefficient, on the other hand, can be calculated analytically provided one can make the simplifying assumption that in Eqs.~(\ref{int_drift}) and (\ref{int_diff}) the radial period is approximately constant within a small region in $J_r-\mathcal{E}$ space. In this case,
\begin{align}
  &\Tilde{C}(J_r,E,t) \approx -\frac{\dot{R}}{R}\frac{P(E,L)}{2\pi}\left<(\mathbf{r}\cdot\mathbf{v}) \right>\label{drift_theory}\\
    &\Tilde{D}(J_r,E,t) \approx \frac{1}{2}\left(\frac{\dot{R}}{R}\right)^2\left(\frac{P(E,L)}{2\pi}\right)^2\left<(\mathbf{r}\cdot\mathbf{v})^2 \right>\label{diffusion_theory}
\end{align}
where $\left<.\right>$ denotes an ensemble average. For time-dependent spherical potentials, all factors appearing in Eq.~(\ref{diffusion_theory}) can be calculated if $\Phi(r)$ is known. Appendix \ref{app3} discusses how $\dot{R}/R$ can be calculated and $\left<\mathbf{r}\cdot\mathbf{v}\right>$ can be obtained following \citet{2019MNRAS.484.5409P}:
\begin{align}
    \left<(\mathbf{r}\cdot\mathbf{v})^2\right> = \frac{1}{2P(E,L)}\int_{r_{\rm peri}}^{r_{\rm apo}}dr\,r^2\sqrt{2(E-\Phi(r))-\frac{L^2}{r^2}}. \label{eq:rvr}
\end{align}
In general potentials, equation \ref{diffusion_theory} has to be evaluated numerically. From the value of $\sqrt{\Tilde{D}}/J_r$ we can then estimate whether the dynamical evolution of tracers with a given set of integrals of motion ($E,J_r$) is likely to be diffusive or impulsive for a given evolving potential.

A special case is again the Kepler potential in which the right hand side of Eq.~(\ref{diffusion_theory}) can be evaluated analytically. We therefore illustrate the above point on our benchmark potential. 
\citet{2013MNRAS.433.2576P} shows that in a scale-free potential with a time-dependent force 
\begin{align}
    F(r,t) = -\mu(t)r^n,
\end{align}
a good analytic approximation to the scale factor is
\begin{align}
    R(t) \approx \left(\frac{\mu(t)}{\mu(0)}\right)^{-1/(n+3)}.\label{decoupled}
\end{align}
and for a phase-mixed distribution of particles orbiting in a time-dependent Kepler potential ($n=-2$), Eq.~(\ref{diffusion_theory}) yields
\begin{align}
    \Tilde{D}(J_r,E,t) = \frac{1}{4}\left(\frac{\dot{R}}{R}\right)^2\left(GM\right)^2\frac{GM J_r-\sqrt{2\mathcal{E}}J_r^2}{\left(2\mathcal{E}\right)^{\frac{7}{2}}}.\label{drift_ana}
\end{align}
We identify the linear (adiabatic) region in $\mathcal{E}-J_r$ space with the region corresponding to all combinations of energy and radial action for which $\sqrt{\Tilde{D}} \ll J_r$. 
Based on the ratio between the (theoretical) diffusion coefficient calculated according to Eq.~(\ref{drift_ana}) and the radial action we define three different areas in the $J_r-\mathcal{E}$ space, characterized by $\sqrt{\Tilde{D}} < 0.1 J_r$, $\sqrt{\Tilde{D}} < J_r$ and $\sqrt{\Tilde{D}} > J_r$.
%
\begin{figure}
    \centering
    \includegraphics[height=6cm,width=9cm,trim=0.5cm 0.5cm 0.5cm 0.4cm, clip=true]{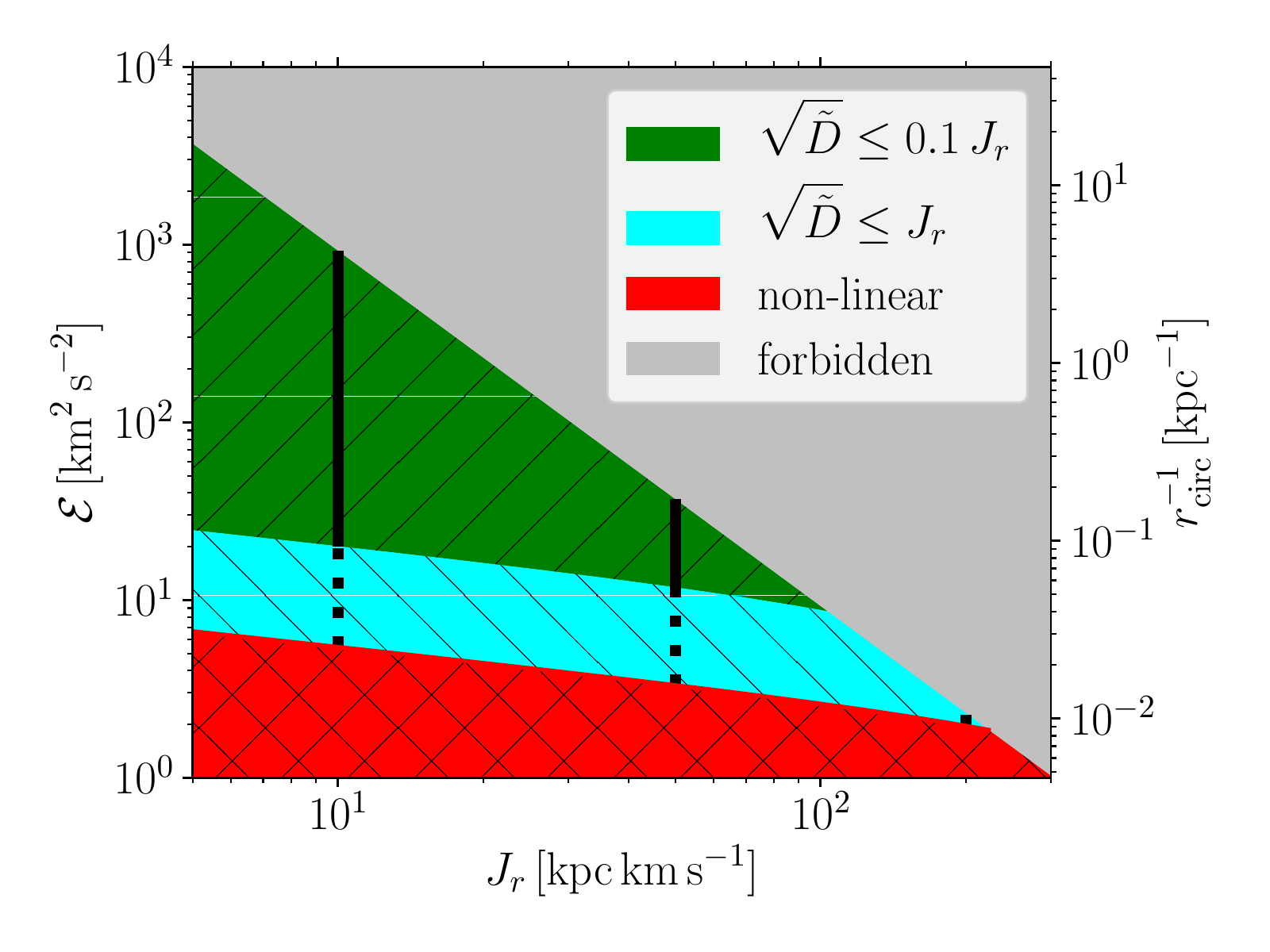}
    \caption{Areas in $\mathcal{E}-J_r$ space in which the expected dynamical evolution of tracer distributions is distinctly different, calculated for our benchmark time-dependent Kepler potential at $t=0$. The grey area is not populated, as no particles can exist there according to Eq.~(\ref{action_kepler}). The green area denotes the perturbative ``linear" regime in which $\sqrt{\Tilde{D}}\ll 0.1 J_r$. The cyan area denotes a ``transition" regime between the ``linear" regime and the ``fringe". Here, $0.1<\sqrt{\Tilde{D}}/J_r < 1$. In red, we show the non-linear ("fringe") part of the phase space. Here, the first order variation can be bigger than the action itself, meaning that the invariant action defined in Eq.~(\ref{linear_order}) is not a dynamical invariant and the diffusion formalism is not applicable. 
   On the right y-axis we show the radius of a circular orbit of a particle with energy $\mathcal{E}$ as an indication of the scales involved.}
    \label{fig:keyplot}
\end{figure}

Fig. \ref{fig:keyplot} shows a plot of $\mathcal{E}-J_r$ space where the different correspond to the above-defined regimes in our benchmark potential
at $t = 0$. For reference, we show the radius of a circular orbit with energy $\mathcal{E}$ on the right y-axis.

The grey area is dictated by Eq.~(\ref{action_kepler}) and its border corresponds to the line at which $L=0$. Any allowed orbits are located to the left in one of the coloured areas. 

Particles inhabiting the green area have an average diffusion coefficient which is smaller than one per cent of the squared radial action. In this (linear, i.e. adiabatic to diffusive) area, the diffusion formalism is applicable. 

For particles in the cyan area, the average diffusion coefficient is larger than one per cent of the squared radial action, yet smaller than the squared radial action itself. This represents a ``transition" regime between the diffusive, perturbative regime and the impulsive ``fringe". 

The red area is the ``fringe" of the potential. Here, the evolution of radial actions is highly non-linear and the diffusion formalism is not applicable. 

The exact locations of the areas vary with time, as the mass of the Kepler potential directly impacts all the three boundaries shown in this picture.
Furthermore, as not all relevant distributions of tracer particles are phase-mixed and in virial equilibrium, using Eq.~(\ref{drift_ana}) along with $\Tilde{C} = 0$ is not always a valid approximation.  
Nonetheless, the areas defined in Fig.~\ref{fig:keyplot} give a good indication as to which radial action - energy combinations imply a linear, diffusive evolution 
and which do not. 

In Section \ref{sec_results} we validate the predictions of Fig. \ref{fig:keyplot}. We then apply the above analysis to investigate whether the flattening of the tail in Fig. \ref{act_dist_evolution} is a linear or a non-linear phenomenon. Given that the tail is comprised mainly of particles inhabiting the ``fringe" area in Fig. \ref{fig:keyplot}, our expectation is that the effect is a non-linear one. To investigate this further, we construct five restricted simulations and then compare their evolution to the evolution predicted by the diffusion formalism.

\section{Tests on a time-dependent Kepler potential}\label{sec_results}

In this section, we test the diffusion formalism developed in Sections \ref{section_one} and \ref{sec_form} in a series of numerical simulations performed using AREPO. The simulations follow the evolution of five different initial tracer particle populations in our benchmark time-dependent Kepler potential. We start by describing the initial conditions of each run. We also test the analysis outlined in Section \ref{subsec:ana}, using it to forecast whether the diffusion formalism will adequately describe the evolution of the radial action distribution of the tracers for each simulation. Then we compare the evolved radial action distributions to the predictions of the diffusion formalism and evaluate if our prior assessment of whether the evolution of each distribution will be diffusive or impulsive was accurate. Finally, we use our results to discuss whether the net drift observed in Fig. \ref{act_dist_evolution} is a linear or a non-linear effect. 


\subsection{Initial Conditions}\label{sub_inidist}

\begin{table*}
	\begin{center}
		\begin{tabular}{cccccccccc}
			\hline
			Distribution    &  central $J_r$ & $\sigma_{J_r}$ & $\mathcal{E}_{\rm{min}}$ & $r_{\rm{peri,min}}$  &  $r_{\rm{ini}}$  & $L_{\rm{ini}}$   \\
			& 	[$\rm{kpc\,km\,s}^{-1}$]	&	[$\rm{kpc\,km\,s}^{-1}$] &  $[\mathrm{km\,s}^{-1}]$          & [kpc]    & [kpc] &   [$\rm{kpc\,km\,s}^{-1}$]              \\
			\hline     
			\hline 
			10-transition     & 10  & 2 & 5.44   & 0.1  & 0.1-15     & 5-150 \\
			50-transition     & 50 & 3 & 3.32 & -  & 0-15  & 5-150   \\
			200-transition     &200  &  12 & 1.96 & -  & 0-15& 5-150     \\
			\hline
			10-linear     & 10 & 2&  20.06 & 0.1    & 0.1-15   &  2-50 \\
			50-linear & 50 & 3 & 11.78 & 0.1    & 0.1-15   & 5-150 \\
			\hline
		\end{tabular}
	\end{center}
	\caption{The relevant parameters of the five initial particle distributions 
	All distributions consist of 20000 particles which are sampled from a Gaussian distribution in radial action. ``Linear" refers to distributions which are confined to the green area of Fig. \ref{fig:keyplot} (i.e. the area in which the linear approximation of $J_r$ is accurate). ``Transition" refers to distributions which are confined to the green and cyan area, implying that the linear correction can be of the same order as the action itself for some particles. In column 2 we show the mean of the Gaussian distribution in each of the cases, whereas the spread of the distribution is given in column 3. The energy limit which is applied to confine the particles to the integral of motion space areas specified above is shown in column 4. The last three columns show various additional cuts that have been imposed for numerical reasons. Column 5 shows minimum values for the pericenter radii, column 6 the range of initial radii at which we set up the tracers. The last column shows the initial range of angular momenta. Notice that in some cases, these collide with other cuts such as the pericenter radius cut. In fact, the reason for the ``10-linear" cut to be different from the others is that with the imposed energy cut, no angular momenta larger than 50 ${\rm kpc\,km\,s^{-1}}$ are allowed in the sampled range of radial actions. } 
	\label{tab_distributions}
\end{table*}
We set up five different Gaussian distributions in radial action to follow their evolution in our benchmark Kepler potential. The black lines in Fig. \ref{fig:keyplot} indicate the mean of the different distributions, highlighting the respective 
values of 10,50 and 200 ${\rm kpc\,km\,s}^{-1}$. The solid black lines indicate two initial configurations in which we confine the particles to the integral of motion space area in which the evolution of the radial action is expected to be linear. The dotted lines indicate three simulations that are confined to the ``linear" and the ``transition" regimes. The simulation whose central value is 200 ${\rm kpc\,km\,s}^{-1}$ is a special case, as here only a tiny area of non-impulsive integral of motion space is available which is entirely part of the ``transition" area. 
As a consequence, the average diffusion coefficient is rather large. As the occupied integral of motion space is in close vicinity to the ``fringe", we anticipate that the diffusion formalism may fail for this simulation. In table \ref{tab_distributions} we show the parameters defining the initial conditions.
\begin{figure*}
    \centering
    \includegraphics[width=0.48\linewidth]{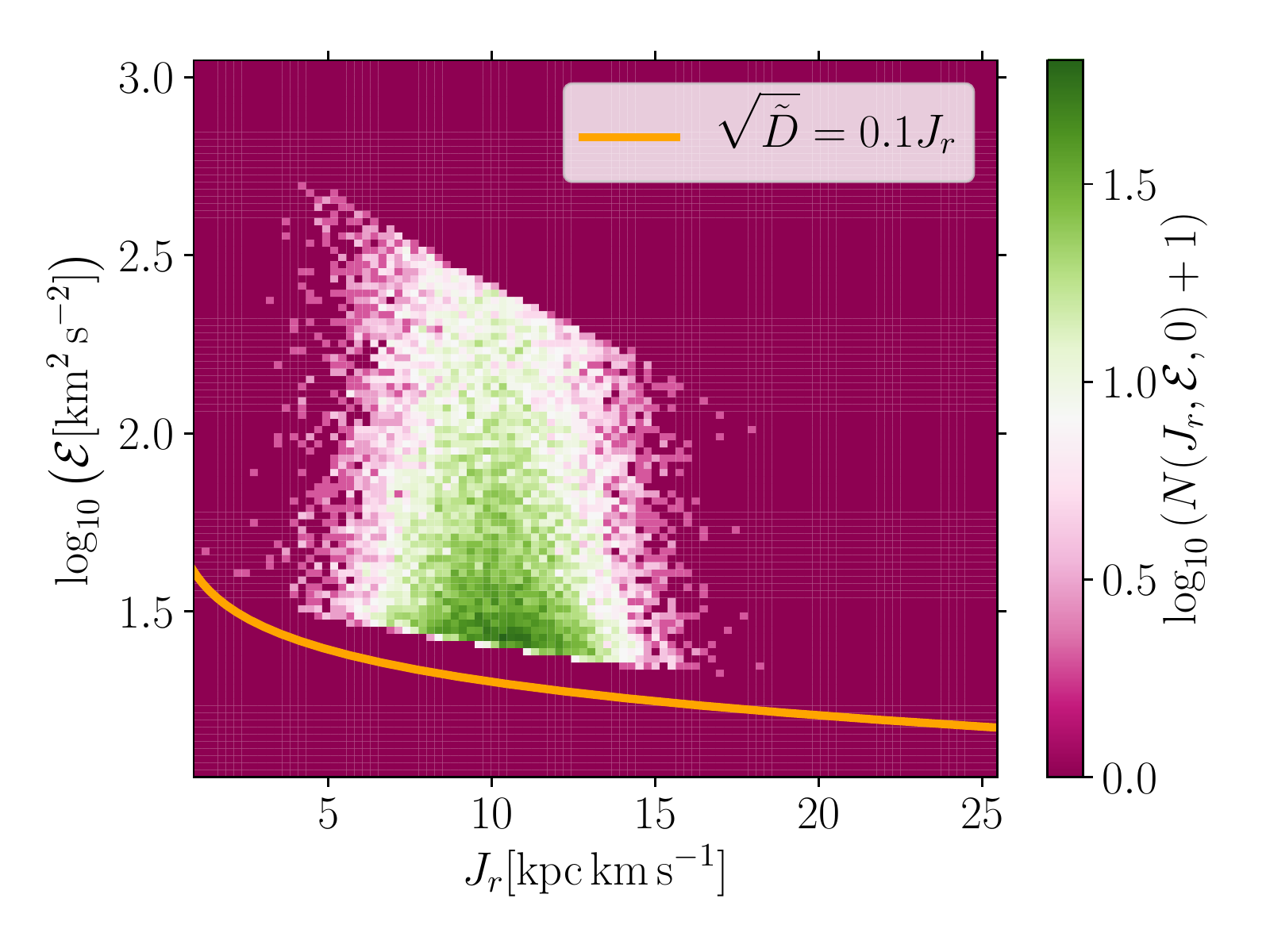}
    \includegraphics[width=0.48\linewidth]{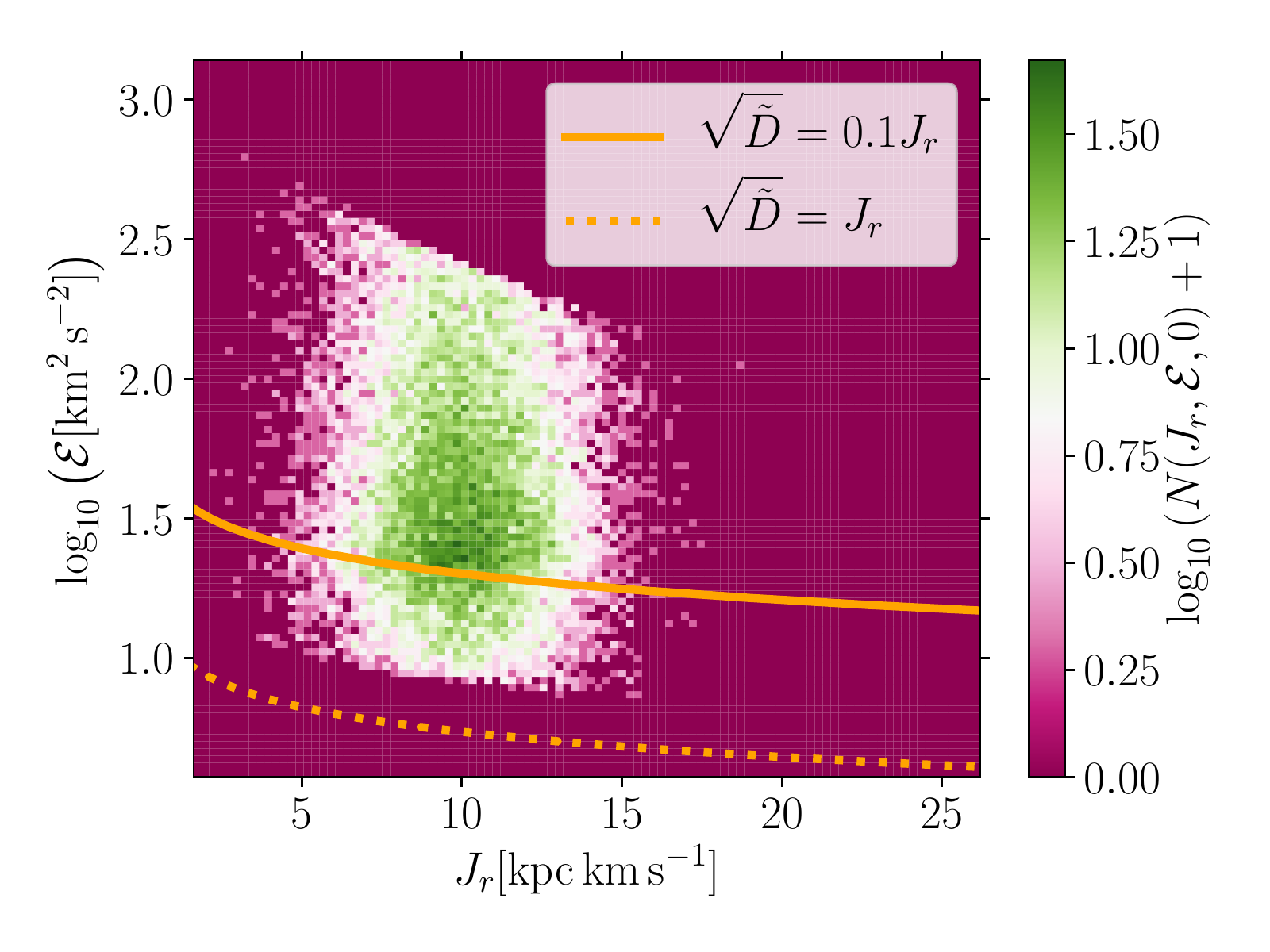}
    \caption{Comparison of the initial integral of motion space distributions in the 10-linear (left) and the 10-transition (right) cases. The lines defining the limits between different areas in Fig. \ref{fig:keyplot} are shown as orange lines here. The solid orange line is the border between the ``linear" and the ``transition" regime, whereas the dashed orange line is the limit to the ``fringe" regime. 
    The logarithmic colour scale indicates the number of particles.}
    \label{fig:10_comp}
\end{figure*}

\begin{figure*}
    \centering
    \includegraphics[width=0.48\linewidth]{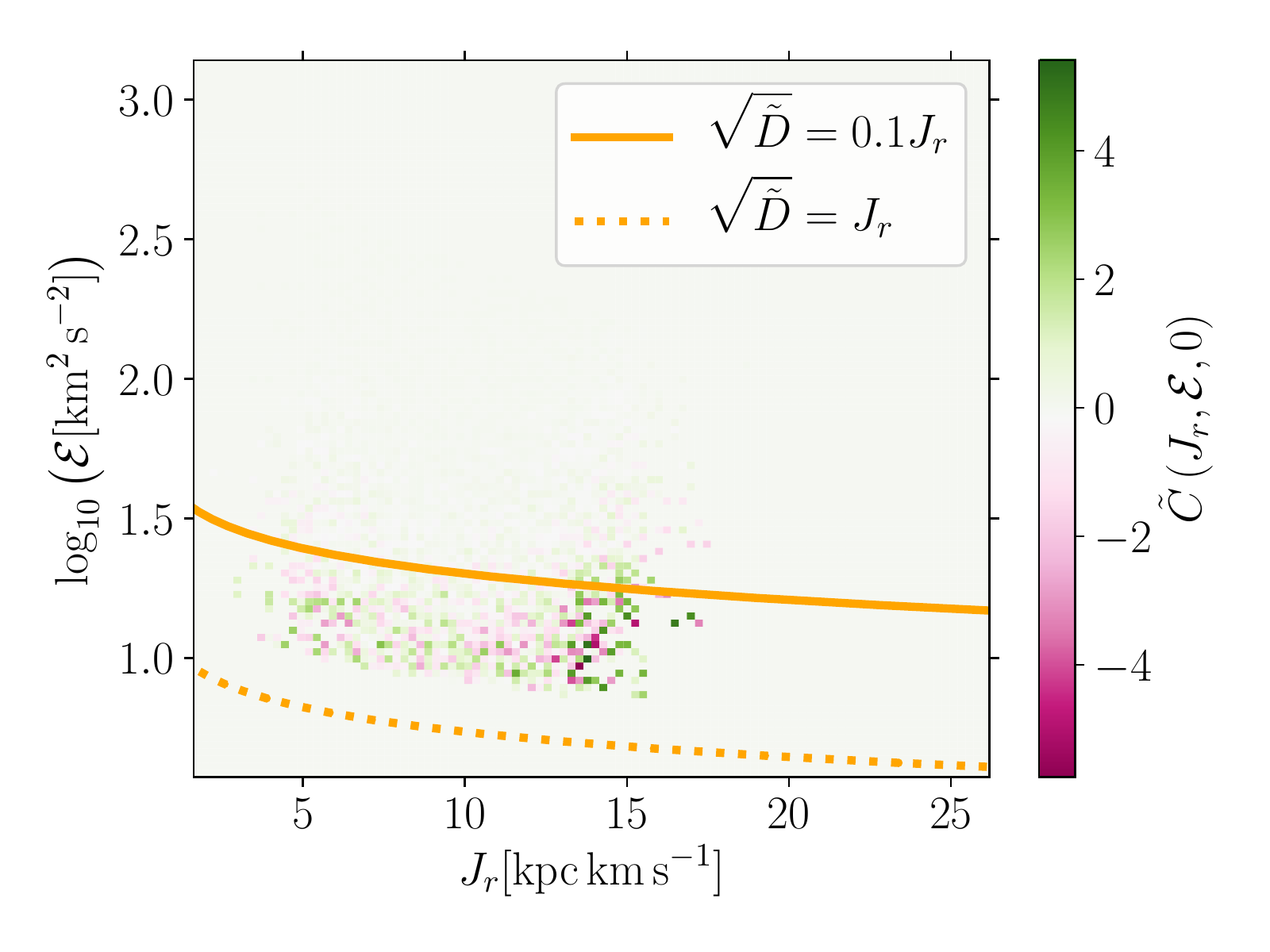}
    \includegraphics[width=0.48\linewidth]{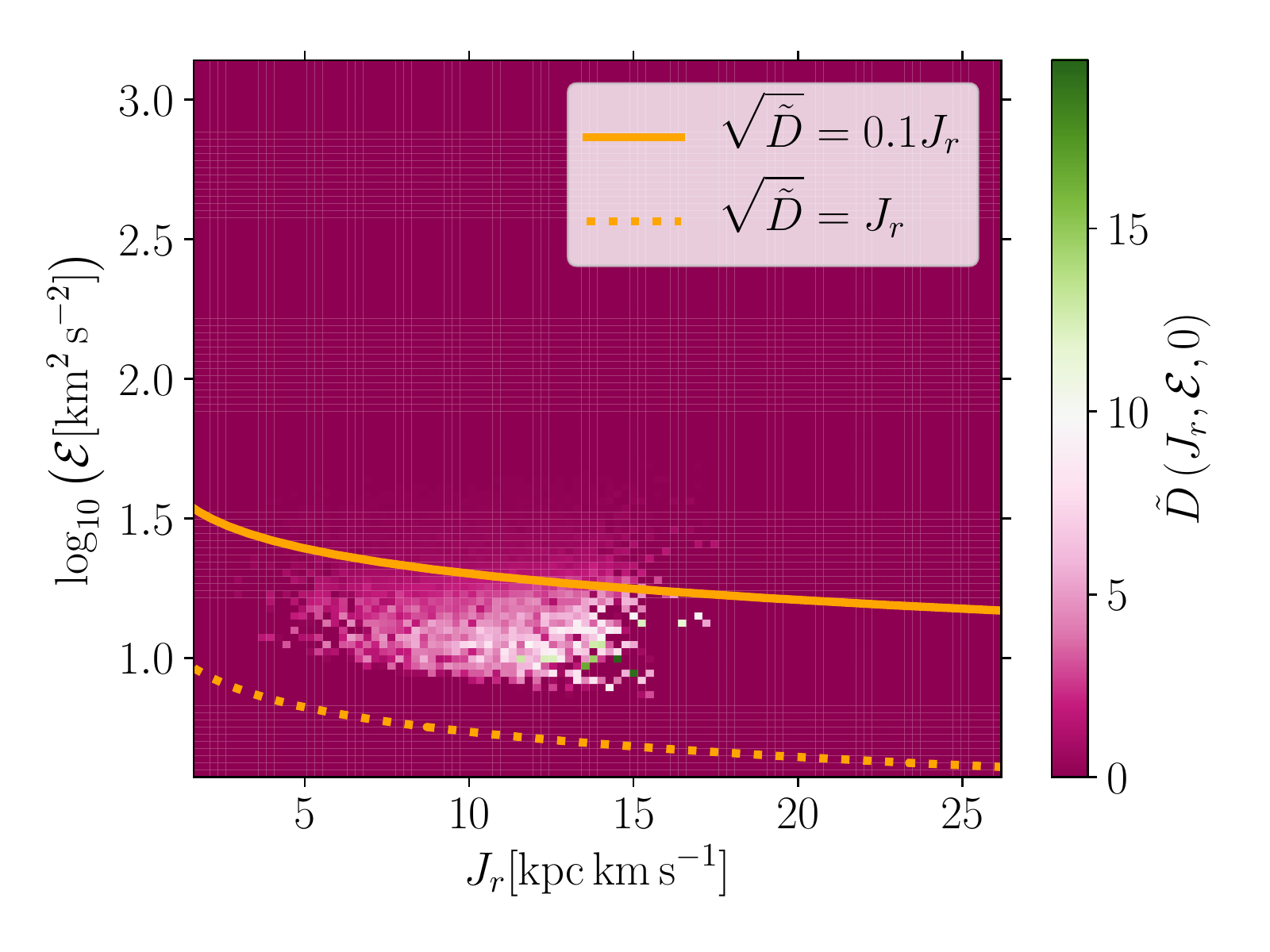}
    \caption{Initial drift (left) and diffusion (right) coefficients in the ``10-transition" case as functions of energy and radial action. The colour scales indicate the drift (diffusion) coefficient in units of ${\rm kpc\,km\,s^{-1}}$ ($\left({\rm kpc\,km\,s^{-1}}\right)^2$) in the left (right) panel. The orange lines are the same as in Fig. \ref{fig:10_comp}. Both the drift and diffusion coefficients increase in magnitude for larger energies and radial actions. In the ``transition" area, the drift coefficients can reach values which are of the same order of magnitude as the radial action at which they are measured.}
    \label{fig:10_cmp_2}
\end{figure*}

Fig. \ref{fig:10_comp} shows a comparison of the initial coarse-grained distribution functions in $\mathcal{E}-J_r$ space between the ``10-linear" case on the left panel and the ``10-transition" case on the right panel. The Gaussian shape in the direction of $J_r$ is apparent in both cases and reflects the input parameters given in Table \ref{tab_distributions}. The distribution in energy is largely determined by the additional cuts we impose (see Table \ref{tab_distributions}). In particular, the lower limit on the pericentre radius effectively introduces a lower limit on the angular momentum. Hence, some theoretically possible combinations of $\mathcal{E}$ and $J_r$ which correspond to low angular momenta are forbidden. The energy cuts stated in table \ref{tab_distributions} are clearly reflected in the high energy end of the distribution functions.

The distribution functions presented in Fig. \ref{fig:10_comp} are coarse-grained versions of Eq.~(\ref{DF}), which we calculate as
\begin{align}
\nonumber    &\Tilde{N}_{i,j}(J_{r,i},E_j) = \sum_{k=0}^{N_{tot}}\left[\theta(\log(E_k)-\log(E_{j,min}))\right.\\
\nonumber    &\times \theta(\log(E_{j,max})-\log(E_k))\theta(J_{r,k}-J_{r,i,min})\\
    &\left.\times\theta(J_{r,i,max}-J_{r,k})\right].
\end{align}
The coarse-grained versions of the drift and diffusion coefficients defined in Eqs. (\ref{drift_theory}) and (\ref{diffusion_theory}) are given by  
\begin{align}
\nonumber    \Tilde{C}_{i,j}(J_{r,i},E_j) &= \frac{1}{N_{i,j}}\left. \sum_{k=0}^{N_{tot}}\right|_{E_{j,min}<E_k<E_{j,max}}^{J_{r,i,min} < J_{r,k}<J_{r,i,max}}\\
    &\left[-(\mathbf{r}_k\cdot\mathbf{v}_k)\frac{\dot{R}_k}{R_k}\frac{P(E_k)}{2\pi}\right]\label{coarse_drift}
\end{align}
and 
\begin{align}
\nonumber    \Tilde{D}_{i,j}(J_{r,i},E_j) &= \frac{1}{2\,N_{i,j}}\left. \sum_{k=0}^{N_{tot}}\right|_{E_{j,min}<E_k<E_{j,max}}^{J_{r,i,min} < J_{r,k}<J_{r,i,max}}\label{coarse_diff}\\
    &\left[(\mathbf{r}_k\cdot\mathbf{v}_k)\frac{\dot{R}_k}{R_k}\frac{P(E_k)}{2\pi}\right]^2 
\end{align}
In Fig. \ref{fig:10_cmp_2} we show the coarse-grained drift and diffusion coefficients of the "10-transition" distribution in $\mathcal{E}-J_r$ space at $t = 0$. 
The drift coefficients (in units of ${\rm kpc\,km\,s^{-1}}$) are shown in the left panel, while the diffusion coefficients (in units of $\left({\rm kpc\,km\,s^{-1}}\right)^2$) are shown on the right panel. Both the drift and the diffusion coefficients show the same trend with energy and radial action. The largest drift and diffusion coefficients are well within the ``transition" area in phase space. 
Moreover, the largest drift coefficients are quite substantial in magnitude, up to the same order of magnitude as $J_r$ itself. Furthermore, the average drift and diffusion coefficients increase for larger values of $\mathcal{E}$ and $J_r$. Since drift and diffusion coefficients are larger in the ``transition" area than in the ``linear" area, the analysis of Section \ref{subsec:ana} suggests that the diffusion formalism will perform better at predicting the evolution of the ``10-linear" distribution than that of the ``10-transition" distribution.

\begin{figure}
    \centering
    \includegraphics[width=\linewidth]{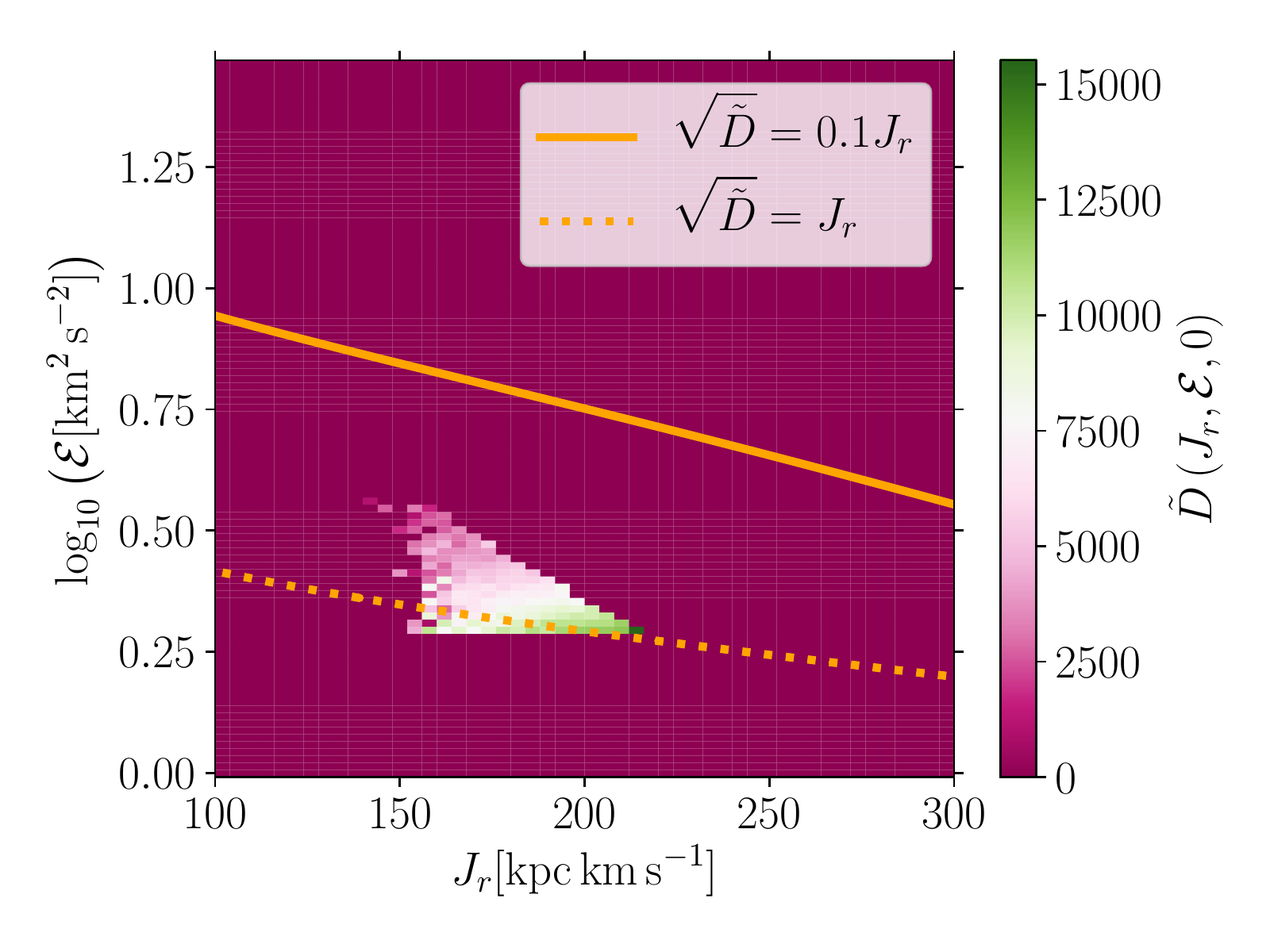}
    \caption{Diffusion coefficients for the ``200-transition" distribution (see Table \ref{tab_distributions}) at $t=0$ (units as in Fig. \ref{fig:10_cmp_2}).
    Only a narrow area in integral of motion space is populated, which is very close to the border of the ``fringe" (dotted orange curve here). The upper limit of the populated integral of motion space area in the $\mathcal{E}$-direction is given by the $L=0$ curve. The measured diffusion coefficients are very large over the entire populated integral of motion space area.
    } 
    \label{fig:200_dists}
\end{figure}

Fig. \ref{fig:200_dists} shows the initial diffusion coefficients of the ``200-transition" distribution. 
Since the populated integral of motion space area lies in close vicinity to the "fringe", the measured diffusion coefficients are very large. Given the magnitude of the averaged diffusion coefficients, we expect the evolution of the ``200-transition" radial action distribution to be impulsive and non-perturbative. 
We thus do not expect our diffusion formalism to give an accurate prediction. Across all five simulations, we expect the predictions of the diffusion formalism to deteriorate when a larger fraction of particles occupies the increasingly non-linear ``transition" area in integral of motion space.

\subsection{Evolution of the ``10-linear" distribution}\label{sub_10lin}

In this Section we compare the evolved ``10-linear" distribution to predictions from the diffusion formalism. 
We calculate the drift and diffusion coefficients in two different ways. In the first (numerical) method, we calculate them directly from the particle's phase space coordinates using Eqs.~(\ref{coarse_drift}) and (\ref{coarse_diff}). To obtain coefficients that depend solely on radial action, we marginalize over the energy as
\begin{align}
&\Tilde{C}_i(J_{r,i}) = \frac{1}{N_i}\sum_{j=0}^{N_{bin}}\Tilde{C}_{i,j}(J_{r,i},E_j)\label{coarse_C}\\
&\Tilde{D}_i(J_{r,i}) = \frac{1}{N_i}\sum_{j=0}^{N_{bin}}\Tilde{D}_{i,j}(J_{r,i},E_j)\label{coarse_D},
\end{align}
where we have used 
\begin{align}
    N_i = \sum_{j=0}^{N_{bin}}N_{i,j}.
\end{align}
In the second (analytic) method, we assume the particle distribution to be phase-mixed, set the drift coefficient to zero, and calculate the diffusion coefficient using Eq.~(\ref{drift_ana}).

\begin{figure}
    \centering
    \includegraphics[width=\linewidth]{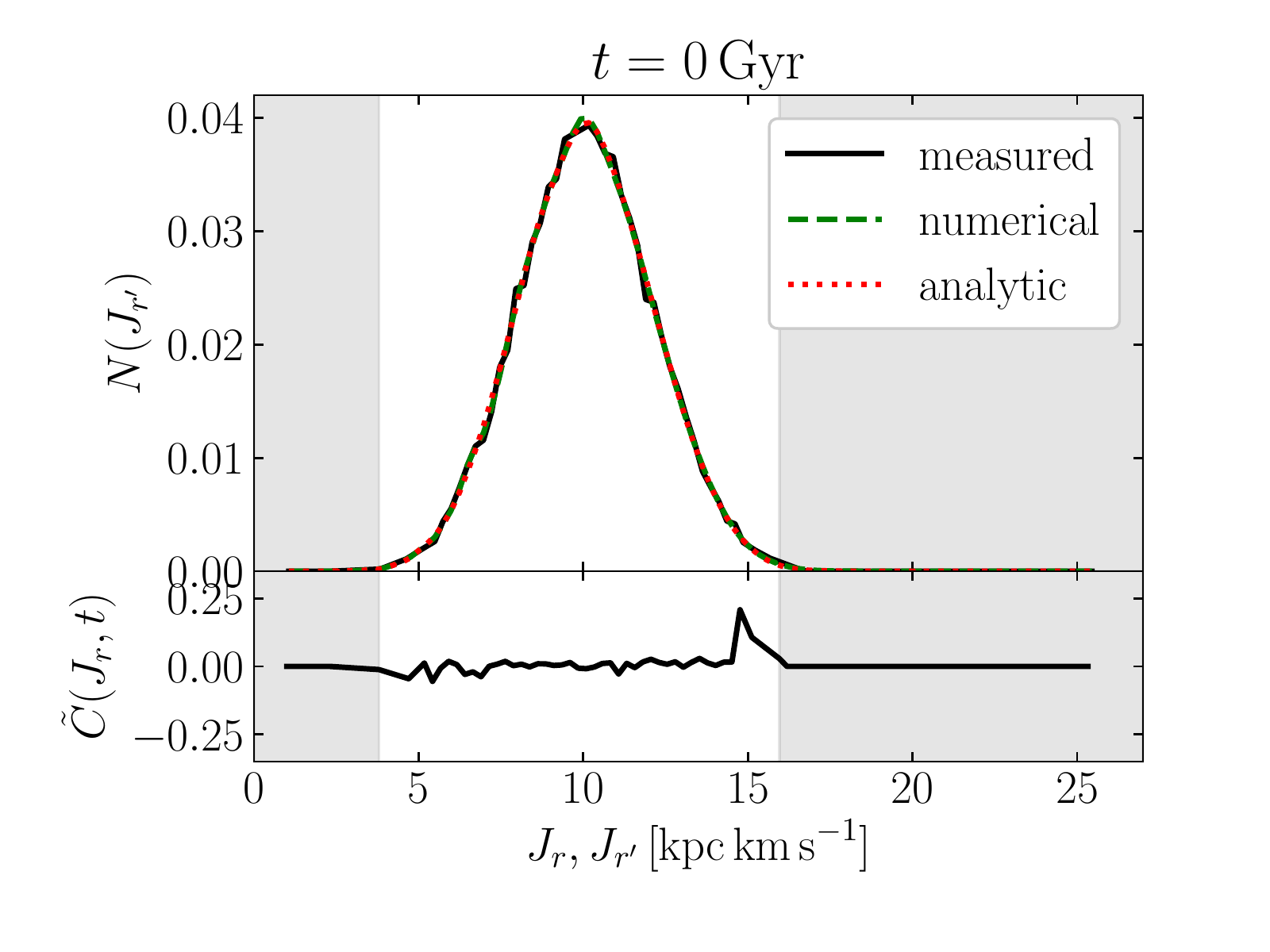}
    \caption{Invariant distribution $N(J_{r'})$ in the upper panel, and the drift coefficient $\Tilde{C}(J_r,t)$ in the lower panel for the ``10-linear" simulation at $t=0$ (see Table \ref{tab_distributions}). The black solid line is the measured distribution, obtained by calculating $J_{r'}$ for each particle and binning the results. The green dashed line is the result of the convolution (Eq.~\ref{fwd_conv}) using drift and diffusion coefficients obtained numerically from the data. The drift coefficient shown in the lower panel is the one depending on the instantaneous action $J_r$, i.e., the one that was used to do the convolution resulting in the green dashed line of the upper panel. The red dotted line in the upper panel shows the result of Eq.~(\ref{fwd_conv}) using the analytic diffusion coefficient from Eq.~(\ref{drift_ana}). In order to avoid sampling noise in the data we combine bins with less than 50 particles in the tails of the distribution. In the grayed-in areas $\Tilde{C}(J_r)$ is too poorly sampled.}
    \label{fig:invariant_10}
\end{figure}

The upper panel of Fig. \ref{fig:invariant_10} shows the invariant distribution $N(J_{r'})$ obtained at the start of the ``10-linear" simulation, while the lower panel shows the drift coefficient $\Tilde{C}(J_r,t)$ that was used to calculate $N(J_r')$ using Eq.~(\ref{fwd_conv}). 
We combine sparsely populated adjacent bins until the combined bin contains more than 50 particles. The center of the new bin is calculated as a weighted mean of the centres of the combined bins, using the particle number as a weight. The grey area marks the resulting radial action range in which bins are empty, 
suggesting that the drift obtained in this area is not a robust measurement due to insufficient particle sampling.
The prediction of the diffusion formalism using drift and diffusion coefficients calculated according to the first (second) method is shown as a green dashed (red dotted) line. 
The ``measured" distribution (black line) is obtained by calculating $J_r'$ for each particle individually and then calculating a histogram in $J_r'$.
Comparing the ``measured" distribution with the results of the diffusion formalism, we find that they all coincide remarkably well. The agreement between the two results of the diffusion formalism reveals that our initial sampling algorithm created a fully phase-mixed distribution in radial action with no net drift. This is confirmed on the lower panel, where we show  $\Tilde{C}(J_r,0)$ and find it to be consistent with zero over the full radial action range. 

\begin{figure*}
    \centering
    \includegraphics[width=0.48\linewidth]{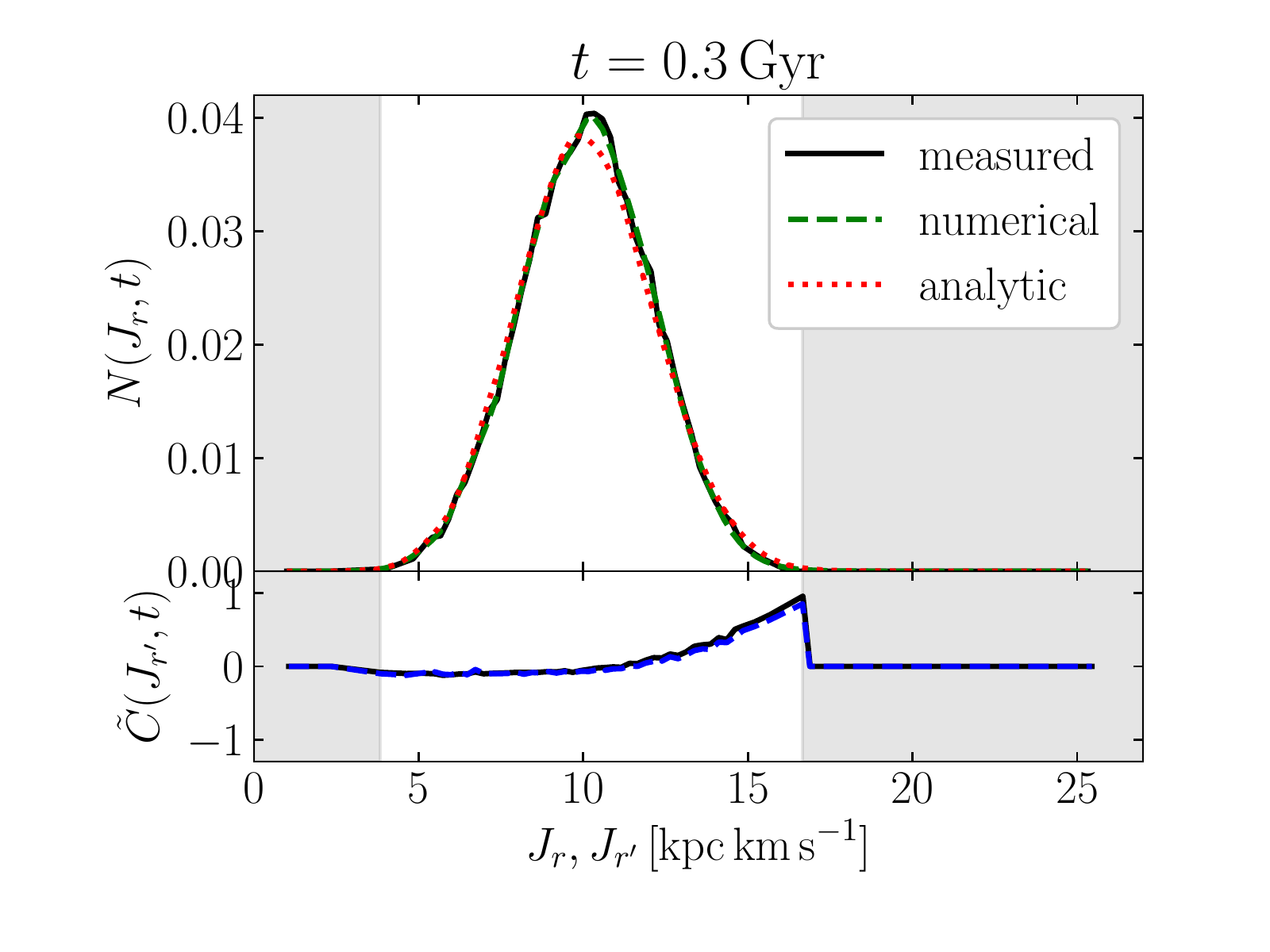}
    \includegraphics[width=0.48\linewidth]{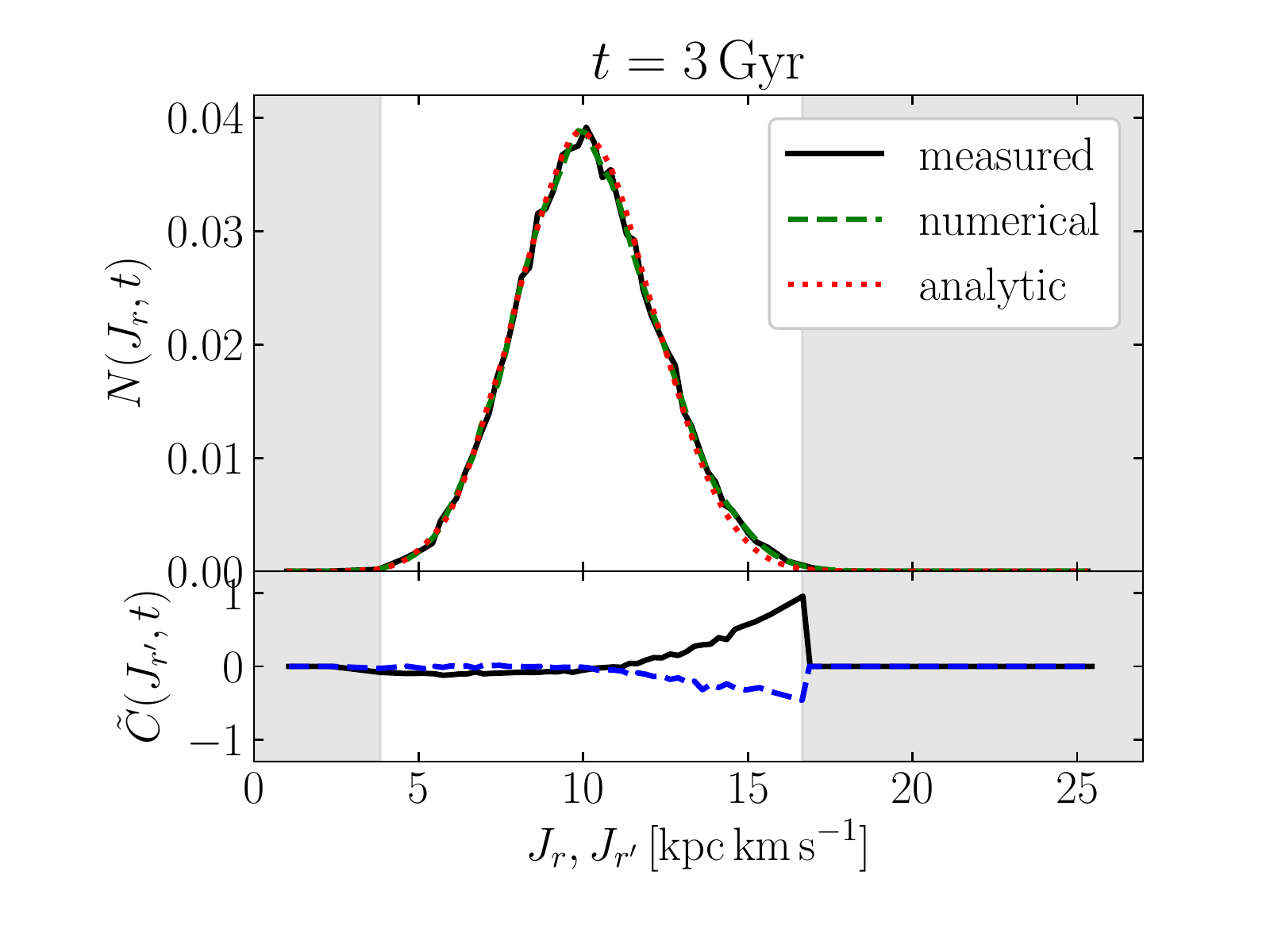}
    \caption{$N(J_r,t)$ at $t=0.3\,\rm{Gyr}$ on the left panel and at $t=3\,\rm{Gyr}$ on the right panel for the ``10-linear" simulation (see Table \ref{tab_distributions}). As in Fig. \ref{fig:invariant_10}, the black line in the upper panels is the measured distribution, the green dashed line is obtained using the diffusion formalism with coefficients calculated from the simulation directly and the red dotted line is calculated using the analytic diffusion approach assuming phase mixing (no drift). In the lower panels, we show the drift coefficient used in the convolution (Eq. \ref{rev_conv}) to obtain the green dashed line in the upper panels. The black line in the lower panels is the drift coefficient $\Tilde{C}(J_{r'},t)$ at $t=0$, whereas the blue dashed line is the one calculated at time $t$ (shown in the upper legend). We combine bins as in Fig. \ref{fig:invariant_10}.
    Initially, the drift coefficient is non-zero, especially towards the tail of the distribution to the right, but it decreases considerably with time. In line with this, we find that the analytic approximation is not a perfect match at initial times, but improves significantly as phase mixing in $N(J_r')$ progresses. }
    \label{fig:10_green_comp}
\end{figure*}
In Fig. \ref{fig:10_green_comp} we $N(J_r,t)$ at  $t=300\,$Myr on the left panel and at $t = 3\,$Gyr on the right panel. The colours used are the same as in Fig. \ref{fig:invariant_10}. In the lower part of each panel we show the drift coefficients as functions of the invariant action, $\Tilde{C}(J_{r'},t)$. Black solid lines denote the drift at the initial time, whereas blue dashed lines indicate the drift at the time displayed in the top panels. We combine bins as in Fig. \ref{fig:invariant_10}. 

Although at $t=300\,{\rm Myr}$, $\Tilde{C}(J_r',t)$ deviates strongly from zero at large radial actions, 
by $t=3\,{\rm Gyr}$, it
has decreased considerable over the entire range of invariant actions. The impact of this evolution in the drift is evident when comparing between the two results of the diffusion formalism. If numerically calculated drift and diffusion coefficients are used, the result of the diffusion formalism is in perfect agreement with the `measured" distribution at both times. Using the ``analytic" method to calculate the diffusion coefficients results in a worse match at $t=300\,{\rm Myr}$, as this method assumes $N(J_{r'})$ to be phase-mixed.



Overall, we conclude that the diffusion formalism outlined in Section \ref{sec_form} provides a remarkably good description of the adiabatic time evolution of $N(J_r,t)$ in the ``10-linear" simulation. In the next section, we will investigate if and how the accuracy of this formalism deteriorates when analyzing the more impulsive evolution of the other distributions from Table \ref{tab_distributions}.

\subsection{The diffusion formalism in different regimes}\label{sub_var_diff}

In this section we present results obtained when applying the diffusion formalism  to the ``10-transition", ``50-linear" and ``50-transition" distributions from Table \ref{tab_distributions}. For each of those cases, we show a comparison between the directly measured invariant distribution and the result of Eq.~(\ref{fwd_conv}) at $t = 0$. We furthermore show a measurement of $N(J_r,t)$ after $3\,{\rm Gyr}$ of simulation time, as well as the result of the diffusion formalism using Eq.~(\ref{rev_conv}) and the initial radial action distribution $N(J_r,0)$. The drift and diffusion coefficients used in Eqs.~(\ref{fwd_conv}) and (\ref{rev_conv}) are calculated numerically from the phase space coordinates of the tracers.  

\begin{figure*}
    \centering
    \includegraphics[width=0.48\linewidth]{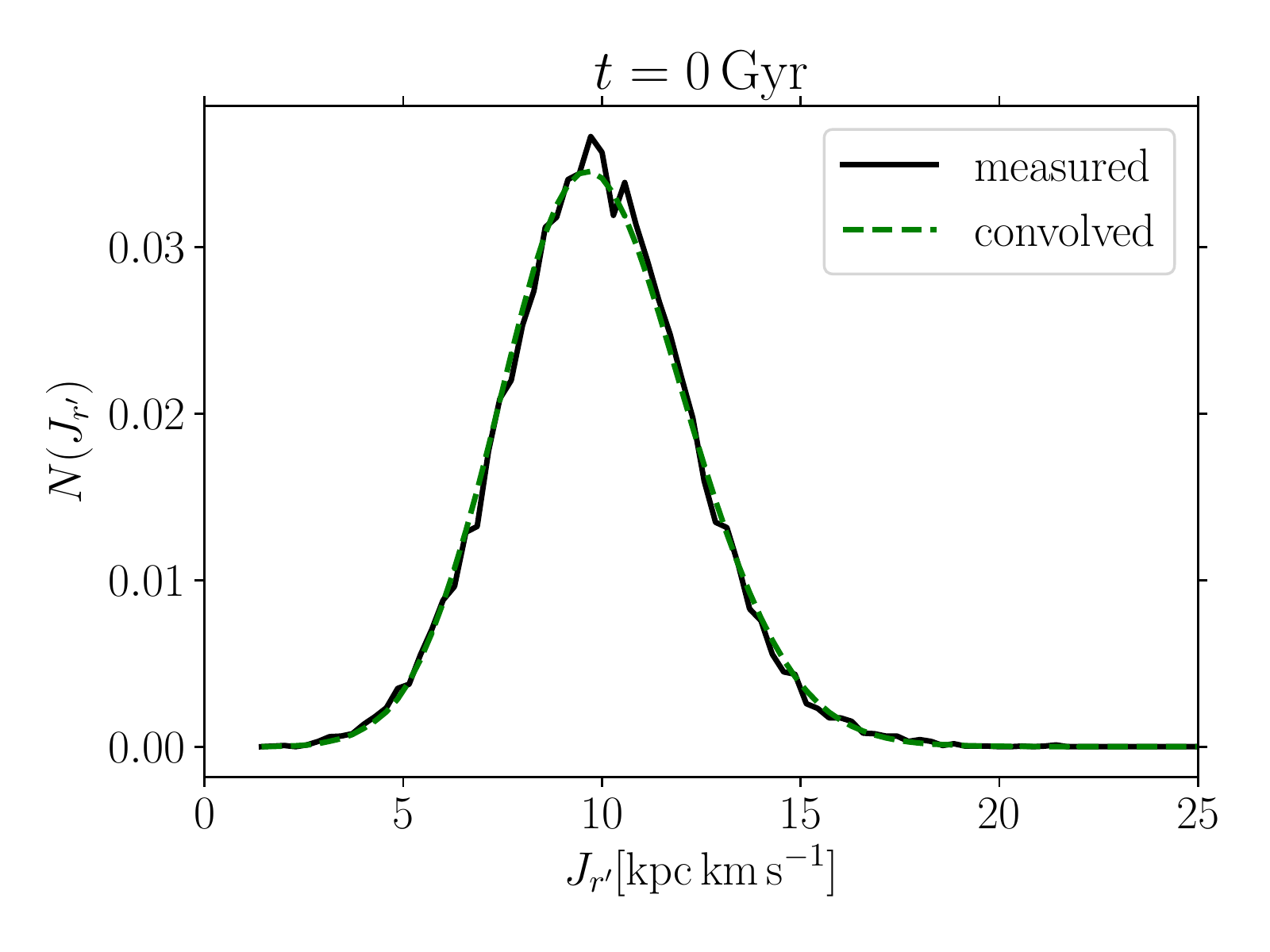}
    \includegraphics[width=0.48\linewidth]{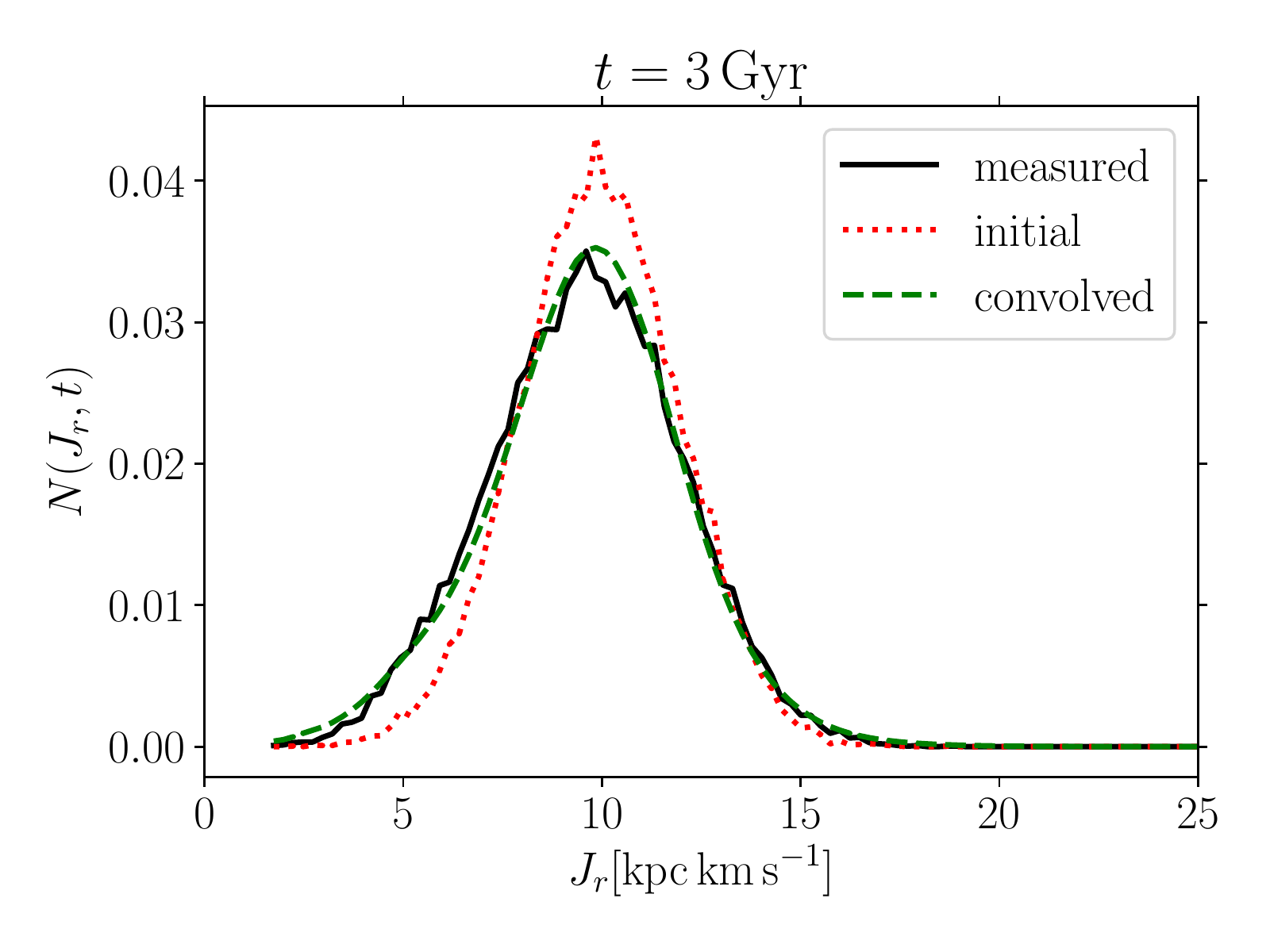}\\
    \includegraphics[width=0.48\linewidth]{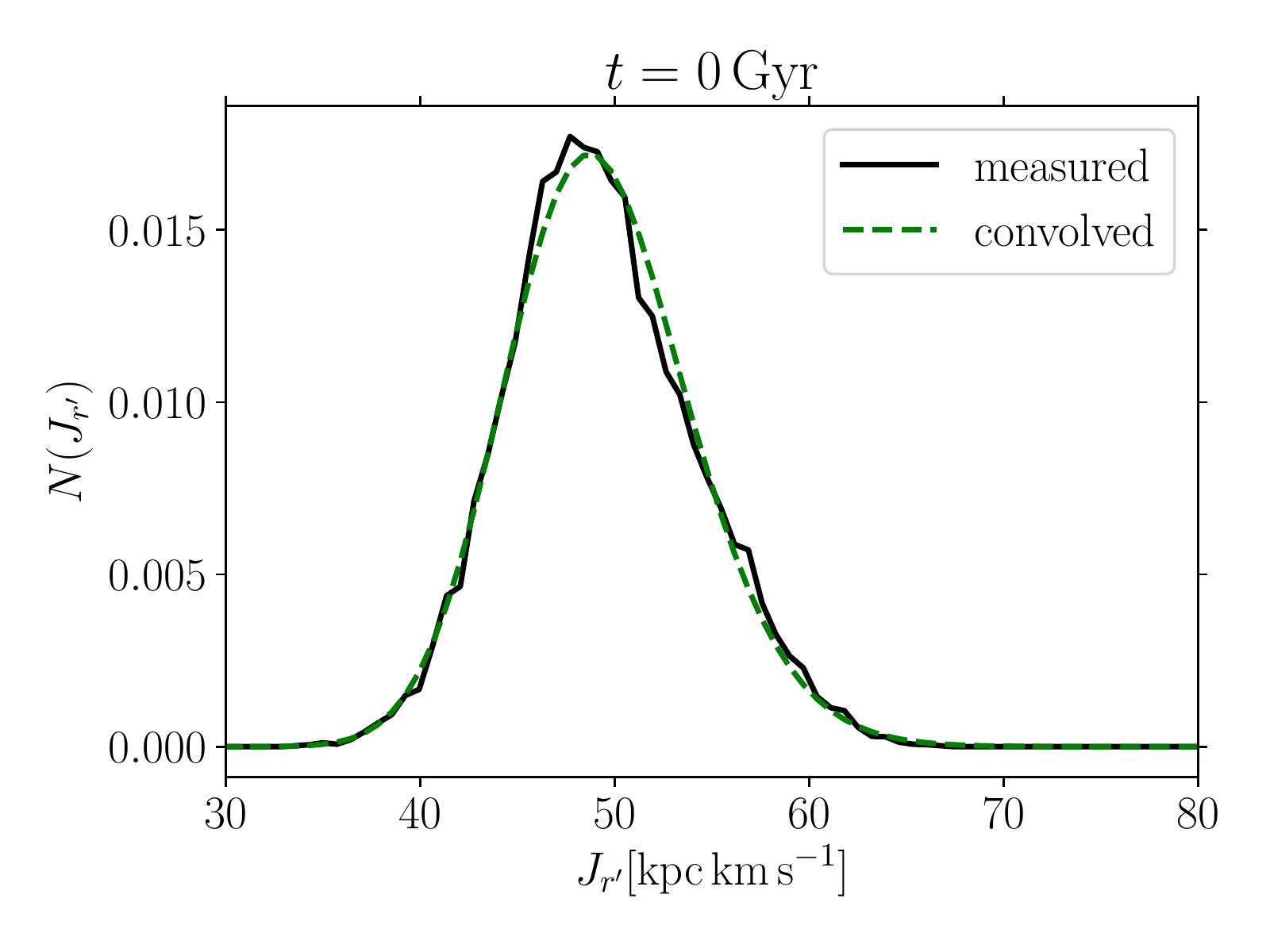}
    \includegraphics[width=0.48\linewidth]{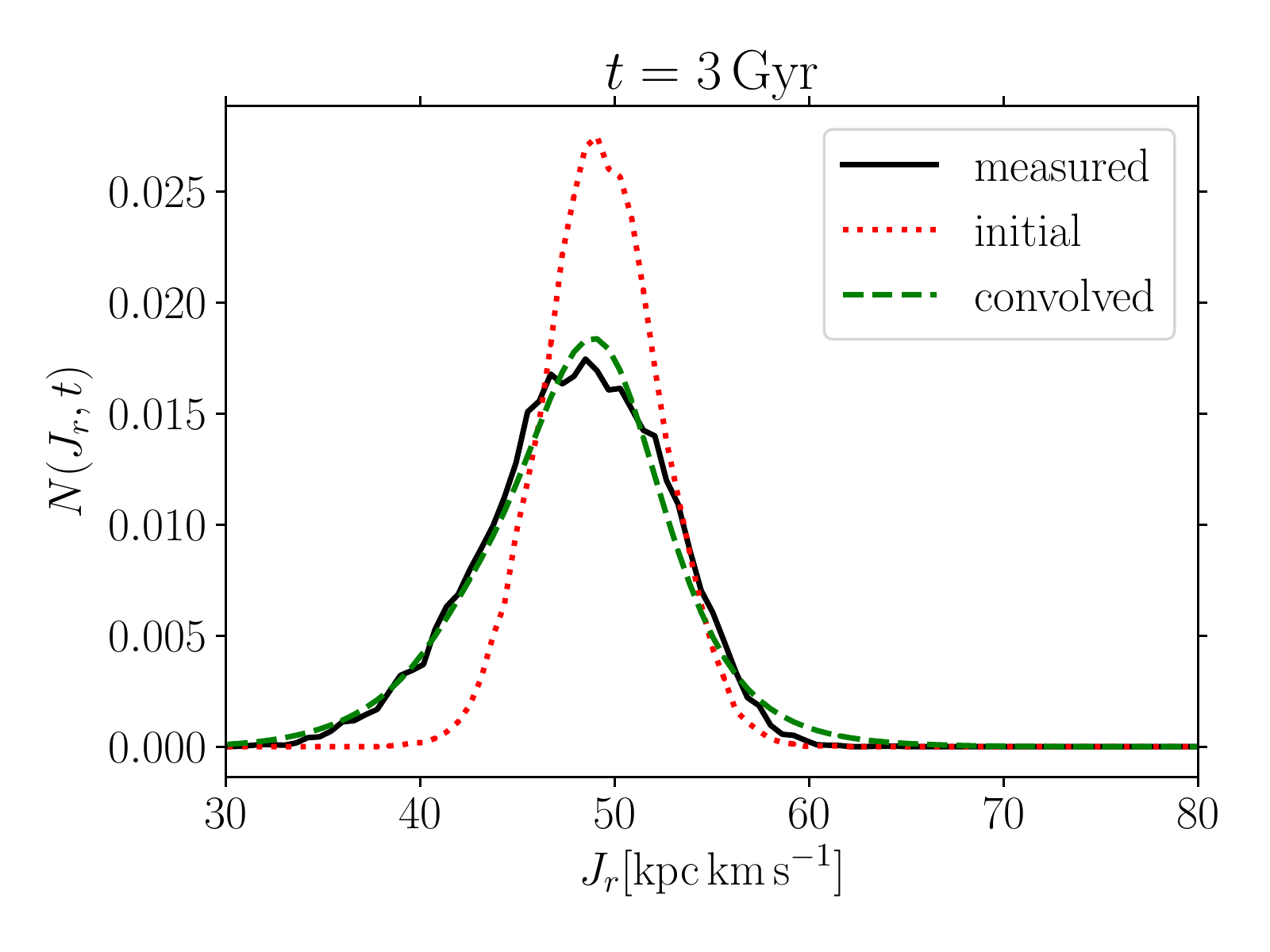}\\
    \includegraphics[width=0.48\linewidth]{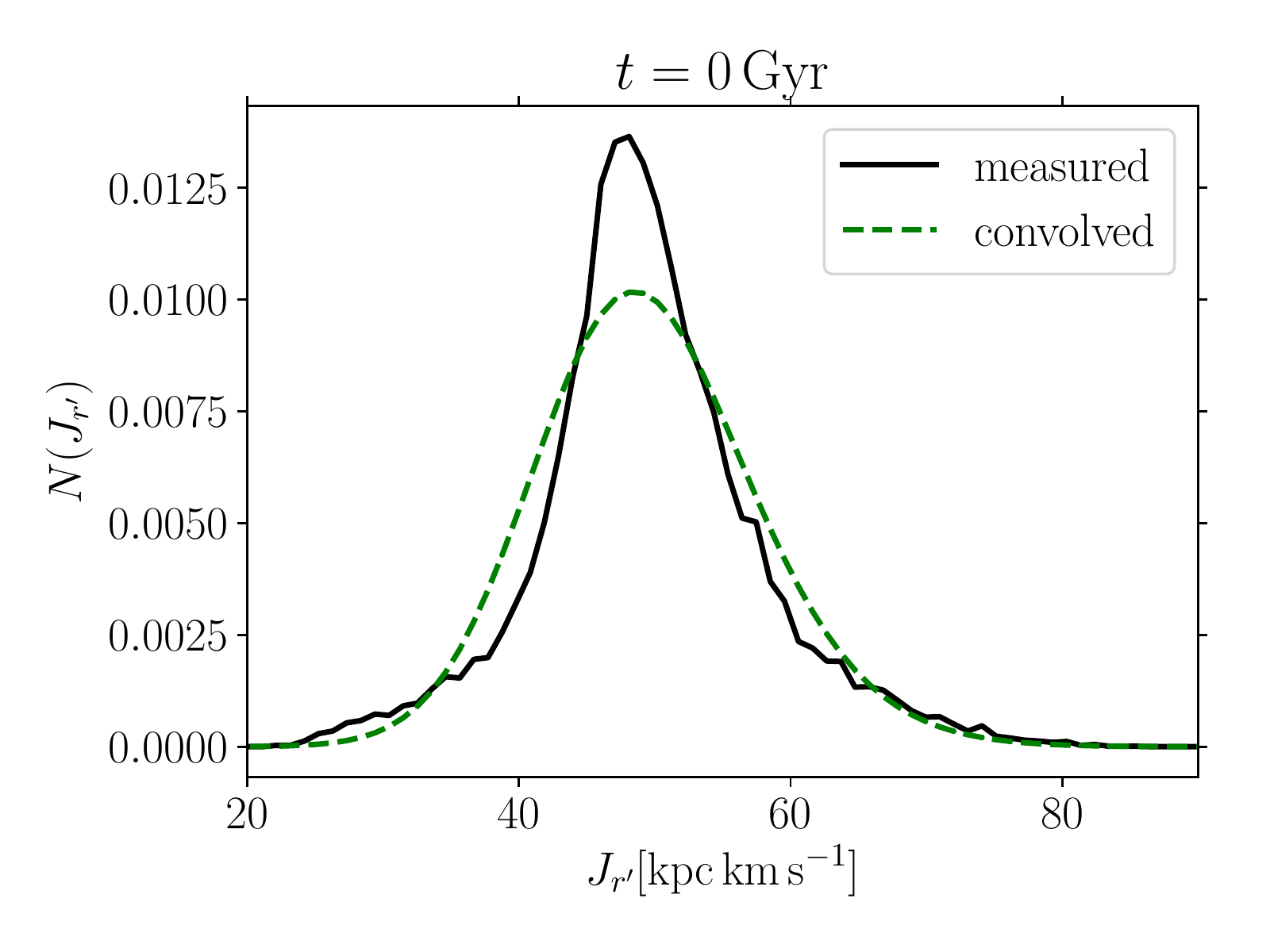}
    \includegraphics[width=0.48\linewidth]{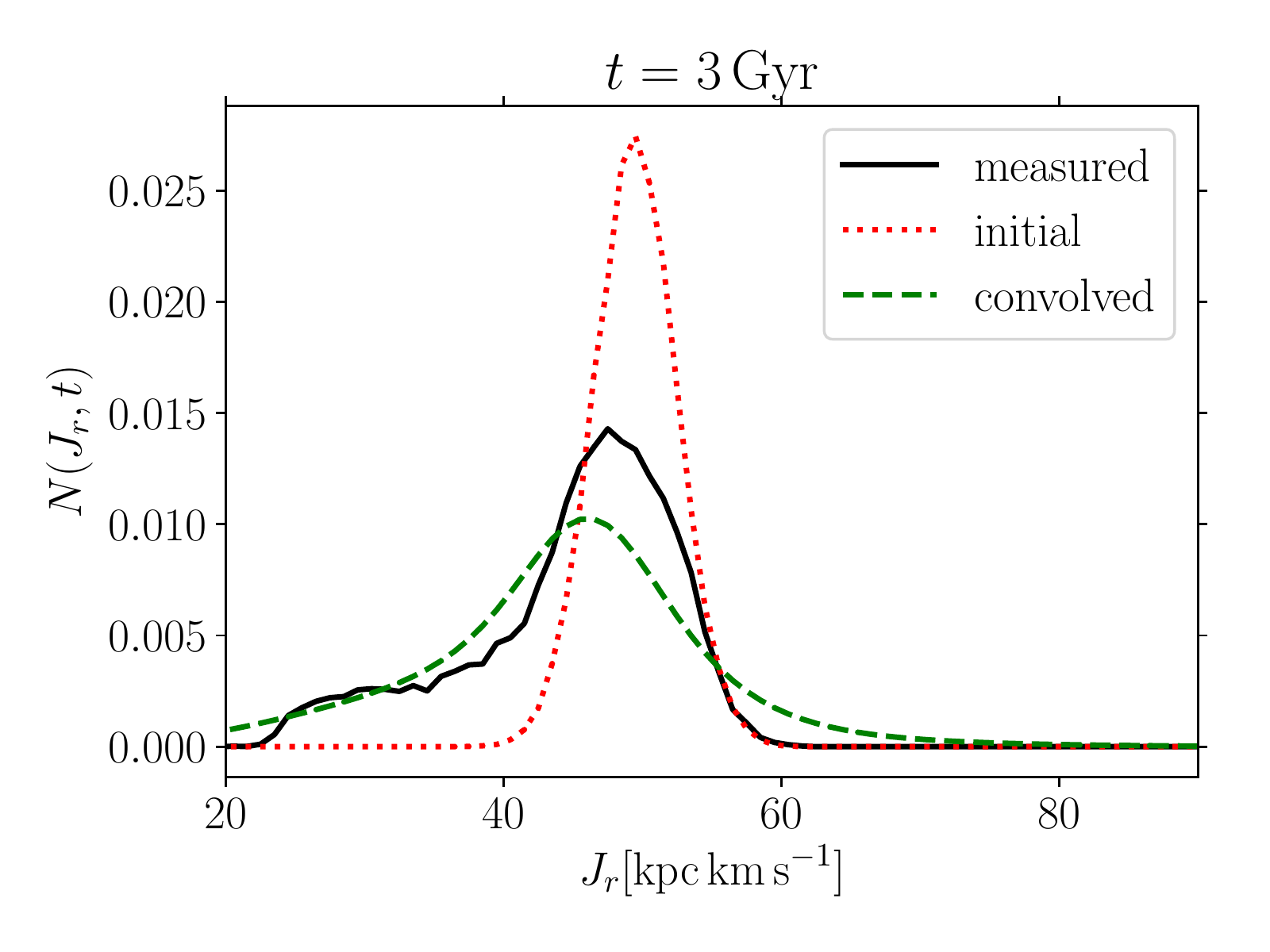}
    \caption{Invariant distributions (on the left) and radial action distributions after a simulation time of $3\,{\rm Gyr}$ (on the right) of the ``10-transition" simulation (top row), the ``50-linear" simulation (middle row) and the ``50-transition" simulation (bottom row); see Table \ref{tab_distributions}. The green dashed lines marked ``convolved" are the result of applying the diffusion formalism (Eqs. \ref{fwd_conv} and \ref{rev_conv}) using coarse-grained drift- and diffusion coefficients (see Eqs.~\ref{coarse_C} and \ref{coarse_D}). 
    The black lines marked ``measured" are direct measurements of the distributions. On the right panels, we also show $N(J_r,t=0)$ as red dotted lines.}
    \label{fig:invariant_distributions}
\end{figure*}
Fig. \ref{fig:invariant_distributions} shows the invariant distributions at time $t=0$ in the ``10-transition" (top), ``50-linear" (middle) and ``50-transition" (bottom) cases in the left column. In solid black lines, we show the ``measured" distribution obtained by calculating the invariant action using Eq.~(\ref{linear_order}) for each particle. The green dashed lines are the ``convolved" distributions obtained from Eq.~(\ref{fwd_conv}). The match between the measured and the calculated distributions is accurate in the ``10-transition" simulation (top left) and only marginally worse in the ``50-linear" case (mid left).
However, in the ``50-transition" case (bottom left), the match between the 
result of Eq.~(\ref{fwd_conv}) and the direct 
measurement of $N(J_r')$ is substantially worse. 
In particular, the result of the convolution is does not resolve the peak in the measured invariant distribution.

The right column of Fig. \ref{fig:invariant_distributions} displays $N(J_r,t)$ after $3\,{\rm Gyr}$ 
in the ``10-transition" (top), ``50-linear" (middle) and ``50-transition" (top) cases. Black solid lines are direct measurements, green dashed lines are the results of the convolutions (Eqs. \ref{fwd_conv} and \ref{rev_conv}) and red dotted lines show $N(J_r,t=0)$ for comparison. 
In the "10-transition" simulation, shown in the top right panel of Fig. \ref{fig:invariant_distributions}, we find a very good agreement between the measured distribution and the result of the diffusion formalism. Furthermore, we find that there is only relatively little evolution in the shape of $N(J_r,t)$, with the most obvious effect being the formation of a tail towards smaller values of $J_r$.  

In the middle right panel, we show the ``50-linear" distribution at the end of the simulation. We find good agreement between the result of the diffusion formalism and the measured distribution, despite the rather substantial evolution with respect to $N(J_r,0)$. A slight mismatch can be observed in the tail of the distribution towards large values of $J_r$, where the convolution  overpredicts the true measured distribution. Note that the diffusion formalism nonetheless provides a substantial improvement over the assumption that radial actions are invariant. 

To explain why the evolution of the ``10-transition" distribution appears to be better captured by the diffusion formalism than the ``50-linear" distribution, we consult Fig. \ref{fig:keyplot}. While some particles in the ``10-transition" simulation inhabit the cyan ``transition" area, most particles are far within the ``linear" regime (see right panel of Fig.~\ref{fig:10_comp}). Furthermore, the linear energy range is much larger for $J_r  = 10\,{\rm kpc\,km\,s^{-1}}$ than it is for  $J_r  = 50\,{\rm kpc\,km\,s^{-1}}$. In the former case particles are on average more bound and have shorter periods (see Eq.~\ref{Kepler_period}), and thus the 
evolution of $N(J_r,t)$ is closer to adiabatic    
and the diffusion formalism is more accurate. 

In the bottom right panel of Fig. \ref{fig:invariant_distributions} we show the ``50-transition" case. Fig. \ref{fig:keyplot} suggests that the fraction of ``transition" integral of motion space 
available to the particles is much larger here than in the ``10-transition" case. 
The consequences of this can immediately be seen in the mismatch between the measured and convolved invariant distributions in the bottom left panel of Fig.~\ref{fig:invariant_distributions}. 
In the bottom right panel, we find that $N(J_r,t)$ has evolved substantially at the end of the simulation, with an extended tail towards lower actions being present in the final distribution. 
While the peak of the distribution remains around its initial value of $50\,{\rm kpc\,km\,s^{-1}}$, its mean shifts towards smaller radial actions and the resulting distribution is non-Gaussian. 
The diffusion formalism does not fully capture the evolution of the radial action distribution. Most notably, it resolves neither the peak of the distribution nor the tail at large $J_r$ values. 
This is to some extent expected, given that the calculated invariant distribution already deviates significantly from the measured invariant distribution (bottom left panel of Fig.~\ref{fig:invariant_distributions}).
Interestingly, the tail of $N(J_r,t)$ towards smaller radial actions is captured fairly well by the diffusion approximation, indicating that its formation is a linear effect. We note that the diffusion formalism still yields a considerable improvement over the assumption that $N(J_r,t)$ is invariant, yet the growing mismatch between the predictions of the formalism and the actual measurement is a strong hint that non-linear effects are becoming increasingly significant. 

Finally, we note that the ``50-transition" distribution drifts towards smaller radial actions. Since this drift is roughly captured by the diffusion formalism, it appears to be a linear effect. Given that the number of particles around $J_r=50\,\rm{kpc\,km\,s}^{-1}$ grows with time in Fig. \ref{act_dist_evolution}, a more substantial drift towards smaller actions must occur at initially larger actions, where the available integral of motion space becomes increasingly non-linear according to Fig. \ref{fig:keyplot}.

\subsection{Evolution of the ``200-transition" distribution}\label{sub_200tran}

According to Fig. \ref{fig:keyplot}, there is no ``linear" integral of motion space available for particles with $J_r = 200\,{\rm kpc\,km\,s^{-1}}$. We thus expect the evolution to be impulsive. Indeed, we find that the diffusion formalism fails and that
the ``invariant" distribution measured using Eq.~(\ref{linear_order}) evolves strongly with time. To understand this better, we look at the time evolution of both $N(J_r,t)$ and $N(J_r',(t))$, where $J_r'$ is defined by Eq.~(\ref{linear_order}) -- 
and cannot be considered a dynamical invariant.
\begin{figure}
    \centering
    \includegraphics[width=\linewidth]{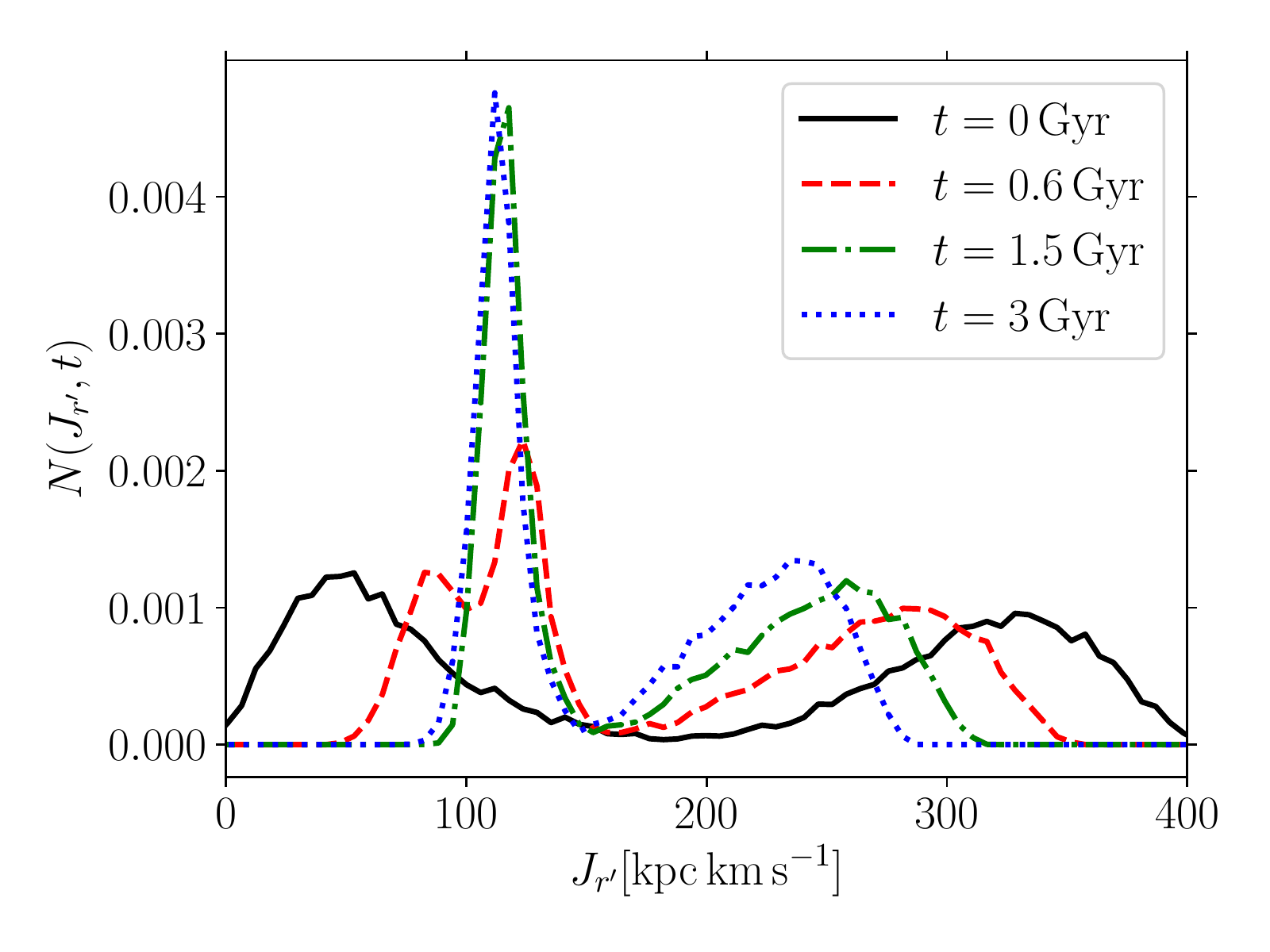}\\
    \includegraphics[width=\linewidth]{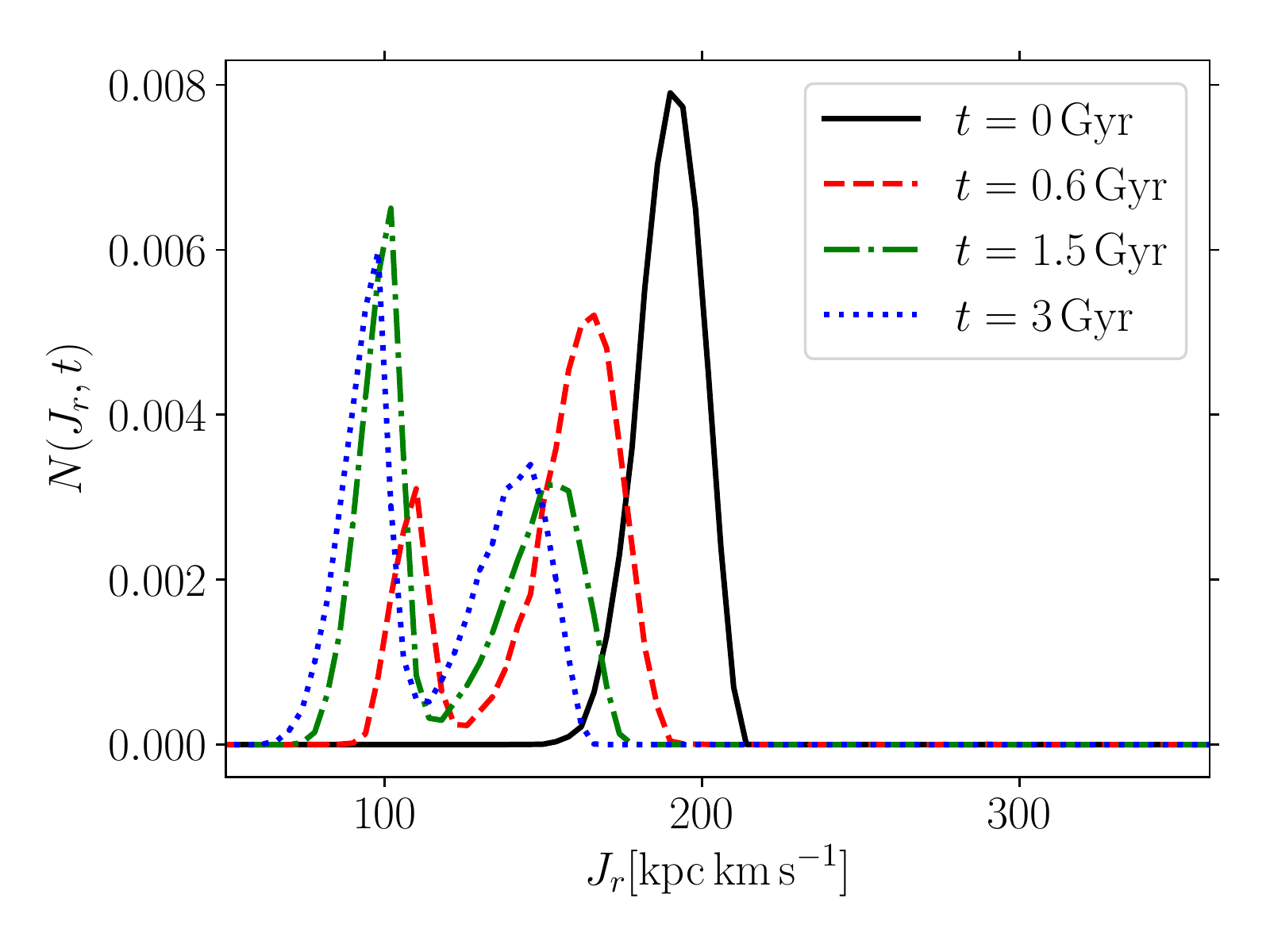}
    \caption{Time evolution of $N(J_r',(t))$ as defined in Eq.~(\ref{linear_order}) (upper panel) and of $N(J_r,t)$ (lower panel) for the ``200-transition" case (see Table~\ref{tab_distributions}). Different lines correspond to distributions at different times as described in the legends.} 
    \label{fig:200_cyan_evolution}
\end{figure}
Fig. \ref{fig:200_cyan_evolution} shows $N(J_r,t)$ (lower panel) and $N(J_r')$ as calculated using the linear approximation (Eq. \ref{linear_order}) (upper panel) at different times. Both evolve strongly with time. 
The fact that $N(J_r')$ is not time-invariant indicates that the evolution of the ``200-transition" distribution is highly non-adiabatic (and clearly non-linear). However, the evolution of the ``invariant" distribution explains some of the time evolution of $N(J_r,t)$ in the bottom panel of Fig. \ref{fig:200_cyan_evolution}.
While $N(J_r,t)$ is initially a Gaussian distribution set up as in Table \ref{tab_distributions}, the initial ``invariant" distribution is bimodal, with one peak around $50\,\rm{kpc\,km\,s}^{-1}$ and a second peak around $350\,\rm{kpc\,km\,s}^{-1}$.

This initial bimodal shape of the distribution can be explained by Fig. \ref{fig:200_dists}. A large fraction of the particles in the ``200-transition" distribution populates an integral of motion space region with diffusion coefficients of the order of $J_r^2$. This implies that for most individual particles, the linear variation is of the order of the central radial action itself. Since its sign depends on the direction of the particle's radial velocity, the linear correction can be either positive or negative. Therefore, particles with $J_r = 200\,\rm{kpc\,km\,s}^{-1}$ can have a linear ``invariant" action which is either larger or smaller by an amount roughly equal to $J_r$ itself. 
As a result the diffusion formalism breaks and we here attempt to qualitatively explain the evolution of the radial action distribution. 

Right after the start of the simulation, $N(J_r,t)$ develops a bimodal structure. This happens because the initial 200-transition distribution consists mainly of tracers at their orbital apo- and pericentres, respectively. Their subsequent evolution splits the radial action distribution into the populations corresponding to the two peaks of the ``invariant" distribution, which can be understood from Eq.~(\ref{linear_order}). 
To explain the subsequent evolution, we calculate $N(J_r')$ at various times keeping the second order terms of Eq.~(\ref{radial_action_expansion})
(see Appendix \ref{app2} for the detailed calculation). 
\begin{figure}
    \centering
    \includegraphics[width=\linewidth]{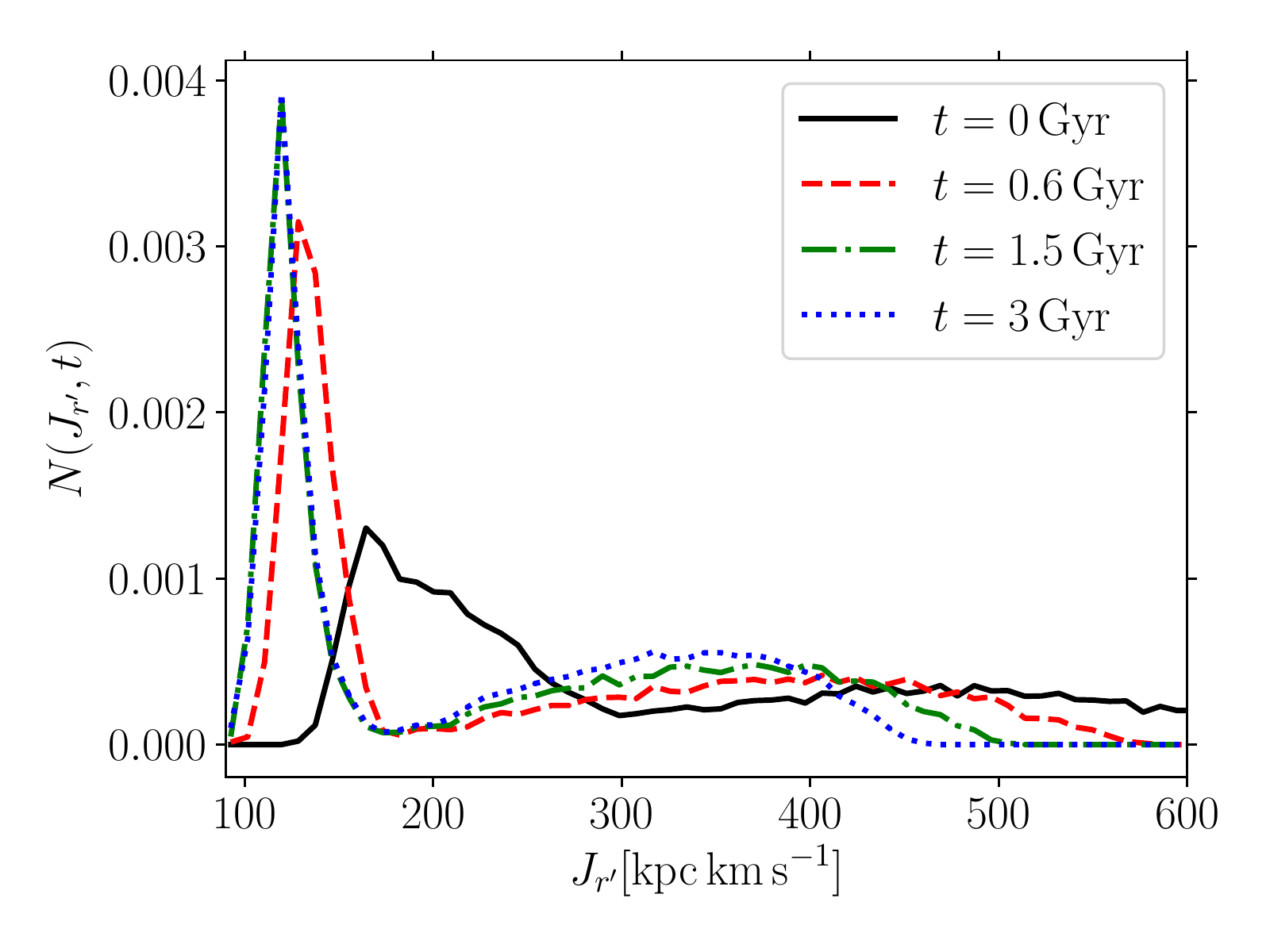}
    \caption{Time evolution of the ``invariant" distribution for the ``200-transition" case (see Table~\ref{tab_distributions}) calculated using a second order expansion of the radial action (Eq.~\ref{2nd_order}). Different lines correspond to distributions at different times as indicated in the legend.} 
    \label{fig:2o200}
\end{figure}
%
The resulting distributions are shown in Fig. \ref{fig:2o200} where $J_r'$ is now defined by Eq.~(\ref{2nd_order}).
Initially, the second order version of $N(J_r',t)$ is more populated around $J_r' = 200\,{\rm kpc\,km\,s^{-1}}$ than the first order (linear) version shown on the upper panel of Fig.~\ref{fig:200_cyan_evolution}. The bimodal shape of the linear version of $N(J_r')$ vanishes when including second order corrections. Instead, the second order $N(J_r')$ obtains an extended tail at large values of $J_r'$, which demonstrates that higher order corrections are rather significant for the ``200-transition" distribution. 

Both $N(J_r,t)$ and the second order version of $N(J_r')$ exhibit a net drift towards smaller radial action values over the course of the simulation. As the second order ``invariant" distribution is still not a true invariant, we conclude that the evolution of the ``200-transition" distribution is truly non-linear and impulsive. 
Therefore, the drift observed in Fig. \ref{act_dist_evolution} is a non-linear effect which cannot be captured by the diffusion formalism developed in Section \ref{sec_form}. 
Moreover, we conclude that the analysis presented in Section \ref{subsec:ana} is a fairly good indicator of whether the radial action distribution of a set of tracer particles will evolve linearly or non-linearly.

\section{A couple of consequences for dark matter haloes} \label{sec_applications}

So far, we have calculated the evolution of radial actions of tracer particles in generic time-dependent spherical potentials and developed a diffusion formalism for radial action distributions. We then tested our formalism in a time-dependent Kepler potential and found that it accurately describes the time-evolution of radial action distributions in a large part of integral of motion space but breaks down near the ``fringe". 
Hereafter, we 
discuss a couple of examples where the
diffusion formalism developed in Section \ref{sec_form}, and its limitations (i.e. in the impulsive regime), provide new physical insights. These examples are i) the mass accretion history of a Milky-Way (MW) like halo and ii) core formation in a self-interacting dark matter (SIDM) halo. For both of these cases, we discuss whether or not the rate of change of the self-gravitating potential of the halo implies an adiabatic evolution of the radial action distributions of tracers. Notice that our discussion is solely focused on the evolution that arises due to a global time-dependence of the gravitational potential. Potential resonant diffusion that arises due to local perturbations is not included in our current formalism, but could easily be included, either using Hamiltonian perturbation theory or an extension of the diffusion theory presented here into two dimensions, similar to e.g. \cite{2015MNRAS.451.3537P,2019MNRAS.484.5409P}. The key point of the calculations here is to assess whether conservation of radial actions is a plausible assumption on average, given the rate at which i) the MW accretes mass or ii) an SIDM halo forms a core. 

\subsection{Mass accretion in dark matter haloes}\label{hma}

\citet{2013MNRAS.430..121P} developed a theoretical formalism to derive the distribution function of a DM halo by maximizing the entropy of an ensemble of collisionless self-gravitating particles, while imposing the extra constraint that the ensemble average of the particles' radial actions is conserved, $\langle J_r\rangle\approx const$. This was motivated by the observation that the evolution of $\langle J_r\rangle$ in simulated haloes is much slower than the variation of the radial action of individual DM particles, $J_r$. The authors find that the distribution function of simulated haloes is accurately reproduced over several orders of magnitude in $J_r$. However, the number of particles with low angular momentum 
is underpredicted in the central regions of the halo (see their Fig.~4), and in turn \citet{2013MNRAS.430..121P}'s formalism does not reproduce the ubiquitous CDM cusps. The authors show that this mismatch can be alleviated by including a second, dynamically decoupled population of particles into the formalism, the so-called "cusp" particles. 

Collision-less N-body simulations show that centrally-divergent cusps arise during the early build up of DM haloes, 
when the gravitational potential of these systems undergo impulsive changes. After that epoch, central DM cusps are retained throughout the hierarchical accretion history of the parent halo. In galactic haloes that are initialized with a cored DM profile, a DM cusp can re-grow through dry mergers with massive cuspy dark matter subhalos that fall into the central regions of the host via dynamical friction \citep{lp15}.

Here, we use the analysis presented in Section \ref{subsec:ana} to assess if a MW-like halo today -- and at a characteristic early time -- contains a significant sub-population of DM particles whose radial actions are not conserved on average -- which would be a plausible explanation for the mismatch between \citet{2013MNRAS.430..121P}'s analysis and the results of simulations. In our formalism, this sub-population consists of particles that inhabit the ``fringe".


To this aim, we apply Section \ref{subsec:ana}'s analysis to the mass accretion of MW-like haloes as follows. We self-consistently construct an idealised MW-size halo following a Hernquist density profile at two different redshifts. We use the mean accretion history reported in \citet{10.1111/j.1365-2966.2010.16774.x} to determine the mass of the halo at a given redshift and use the rejection sampling scheme described in \citet{2019MNRAS.485.1008B} based on Eddington's formalism (\citealt{Eddington1916}, \citealt{2008gady.book.....B}) to sample a particle representation of the DM halo in dynamical equilibrium. We then calculate the (theoretical) diffusion coefficient of each individual DM (simulation/sampled) particle from Eq.~(\ref{diffusion_theory}). The MW-like halo we consider here has a virial mass of $M_{\rm 200} = 1.43\times 10^{12}M_{\odot}$ and a concentration\footnote{With $r_{-2}$ being the radius at which the logarithmic slope of the halo's density profile equals $-2$ and $r_{200}$ the radius at which the enclosed density equals 200 times the critical density of the Universe.} of $c_{200} = r_{200}/r_{-2} = 12$ at redshift $z=0$. These are typical values for MW-like haloes (e.g. \citealt{10.1111/j.1365-2966.2010.16774.x}).\\
For the scale factor $R(t)$ (formally given by Eq.~\ref{scalefactor}), we use an approximate formula derived in analogy to the scale-free case (Eq.~\ref{decoupled})\footnote{For a discussion of the validity of this approximation see Appendix \ref{app3})}:
\begin{align}
    R(t) = \frac{1}{2+\alpha(r,t)}\left(\frac{\Psi(r,t)}{\Psi(r,0)}\right)^{-\frac{1}{2+\alpha(r,t)}},\label{gen_scale}
\end{align}
where 
\begin{align}
    \alpha(r,t) = \frac{d\log\left(\Psi(r,t)\right)}{d\log(r)}\label{general_R}
\end{align}
is the logarithmic slope of the potential at a given radius and
\begin{align}
    \Psi(r,t) = \Phi(r,t)-\Phi(0,t)
\end{align}
is the potential shifted such that $\Psi(0,t) = 0$. To calculate the diffusion coefficient, we need the time derivative of the scale factor, namely
\begin{align}
 \nonumber   \frac{{\rm d}R}{{\rm d}t} &= -\frac{1}{\left(2+\alpha(r,t)\right)^2}\times \left(\frac{\Psi(r,t)}{\Psi_0(r)}\right)^{-\frac{1}{2+\alpha(r,t)}}\\
    &\times \left\{\left[1-\frac{1}{2+\alpha(r,t)}\ln\left(\frac{\Psi(r,t)}{\Psi_0(r)}\right) \right]\times\frac{{\rm d}\alpha}{{\rm d}t} + \frac{\dot{\Psi}}{\Psi} \right\}.\label{uno}
\end{align}
In our model scenario of mass accretion into a MW-size halo, we assume that mass accretes primarily into the outer parts of the halo, leaving the inner mass content unchanged. 
This is in line with the simple picture of cosmological halo mass assembly in layers/shells in which the concentration parameter $c_{200} = r_{200}/r_{-2}$ evolves with redshift only due to the evolution of $r_{200}\propto(1+z)$ in an expanding Universe. This simple picture is approximately validated by full cosmological $N$-body simulations (e.g.
\citealt{2014MNRAS.442.2271S}, \citealt{Ludlow2014}). Under these assumptions, the scale radius in a Hernquist halo can be written as a function of redshift as
\begin{align}
    r_s(z) = 2 r_{-2}(z) = \frac{2(1+z)}{c_{200,0}}\left(\frac{GM(z)}{100 H^2(z)}\right)^{1/3},
\end{align}
where $c_{200,0}$ is the concentration parameter at the current time, $M(z)$ is the redshift-dependent halo mass and $H(z)$ is the Hubble rate. In a Hernquist halo, 
\begin{align}
\alpha(r,z) &= 1-\frac{r}{r+r_s(z)}\\
\Psi(r,z) &= \frac{G M(z) r}{r_s(z)(r+r_s(z))}
\end{align}
and we can thus write the relevant time derivatives in equation \ref{uno} as 
\begin{align}
    \frac{\dot{\Psi}}{\Psi} &= \frac{\dot{M}}{M} - \frac{r+2r_s}{r_s(r+r_s)}\dot{r}_s\\
    \frac{{\rm d}\alpha}{{\rm d}t} &= \frac{r}{(r+r_s)^2}\dot{r}_s,
\end{align}
where we suppress the redshift-dependence and derivatives are taken with respect to cosmic time. In our model of mass accretion, the density profile is that of a Hernquist halo at all redshifts. Therefore, for most times and radii, we expect that the evolution of the potential's logarithmic slope is negligible. For the cases we study in the following, we have explicitly verified that $\dot{\Psi}/\Psi \gg \dot{\alpha}$. Independent of our definition of the static frame in which we define the action invariant, and hence independent of the definition of $\Psi_0(r)$, we therefore find that   
\begin{align}
    \frac{\dot{R}}{R} = -\frac{1}{2+\alpha(r)}\left(\frac{\dot{M}}{M}-\frac{r+2r_s}{r_s(r+r_s)}\dot{r}_s\right).\label{mass_frac}
\end{align}
In order to evaluate Eq.~(\ref{mass_frac}) at different times, we adopt the median mass accretion history for MW-like haloes reported in \citet{10.1111/j.1365-2966.2010.16774.x}
\begin{align}
    M(z) = M_0\left(1+z\right)^{\eta}\exp\left(-\kappa'\left(\sqrt{1+z}-1\right)\right),
\end{align}
with $\kappa'= 4.9$ and $\eta = 2.23$. Thus 
\begin{align}
    \frac{\dot{M}}{M} &= \left(\frac{\eta}{1+z}-\frac{\kappa'}{2\sqrt{1+z}}\right)\times\frac{{\rm d}z}{{\rm d}t}\\
    &= \left(\frac{\eta}{1+z}-\frac{\kappa'}{2\sqrt{1+z}}\right)\times \left(-(1+z)H(z)\right)\label{MAH}
\end{align}
where $H(z)$ is the Hubble rate at redshift $z$. At the current time ($z=0$) we find that for $M(z=0) = 1.43\times 10^{12}M_{\odot}$ and $c_{200,0} = 12$, 
\begin{align}
   \frac{\dot{M}}{M}(z=0) \approx 1.57\times 10^{-2}\,{\rm Gyr}^{-1}.
\end{align}
In the past, however, the amplitude of the specific accretion rate may have been very different. As an example, let us look at the redshift $z_{\rm max}$ maximizing ${\rm d}M/{\rm d}z/M$, which is easily shown to be
\begin{align}
  1+z_{\rm max} = \frac{16\eta^2}{\kappa'^2}.  
\end{align}
At this redshift, we find that $M(z=z_{\rm max})\approx 3.7\times 10^{11}M_{\odot}$ and \begin{align}
    \frac{\dot{M}}{M}(z=z_{\rm max}) \approx 0.35\,{\rm Gyr}^{-1}.
\end{align}
The second term in equation \ref{mass_frac} can be written as 
\begin{align}
    \dot{r}_s = -(1+z)H(z)\frac{{\rm d}r_s}{{\rm d}z}.
\end{align}
Throughout our calculations, we use $\Omega_{\Lambda,0} = 0.7$ and \\
$\Omega_{m,0}= 0.3$. 
Using Eq.~(\ref{mass_frac}), we can now calculate the theoretical diffusion coefficients at different redshifts using Eq.~(\ref{diffusion_theory}).   
%
To that end, we calculate energy, angular momentum, radius and radial velocity of each (simulation) particle in our halo. We then numerically calculate the radial action, the radial period, and $\left<\mathbf{r}\cdot\mathbf{v}\right>$ (according to Eq.~\ref{eq:rvr}) of each particle.
%
From the ratio $\sqrt{\Tilde{D}}/J_r$ we then construct the ``linear", ``transition" and ``fringe" areas introduced in Fig. \ref{fig:keyplot}.

\begin{figure}
    \centering
    \includegraphics[width=\linewidth]{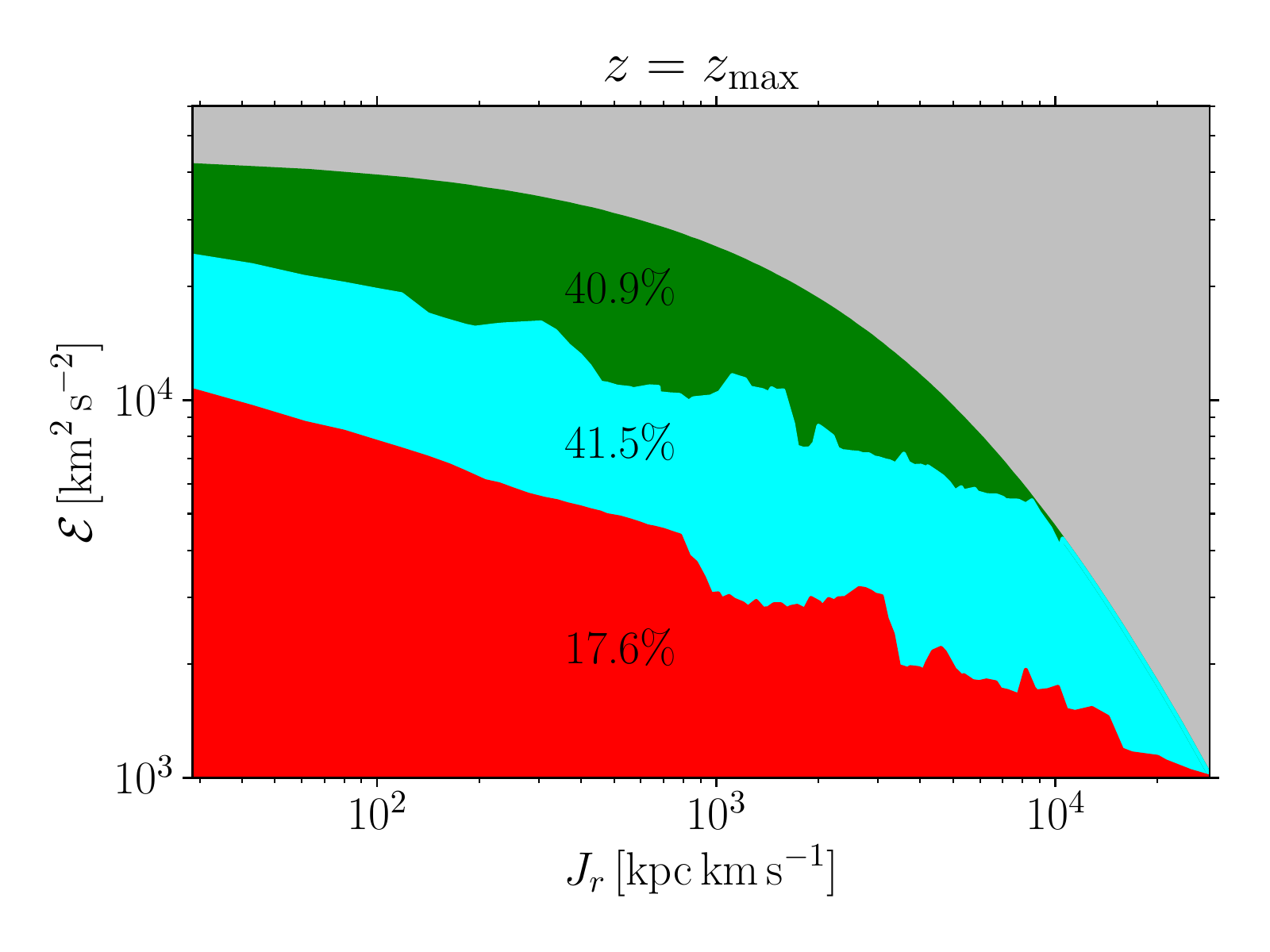}\\
    \includegraphics[width=\linewidth]{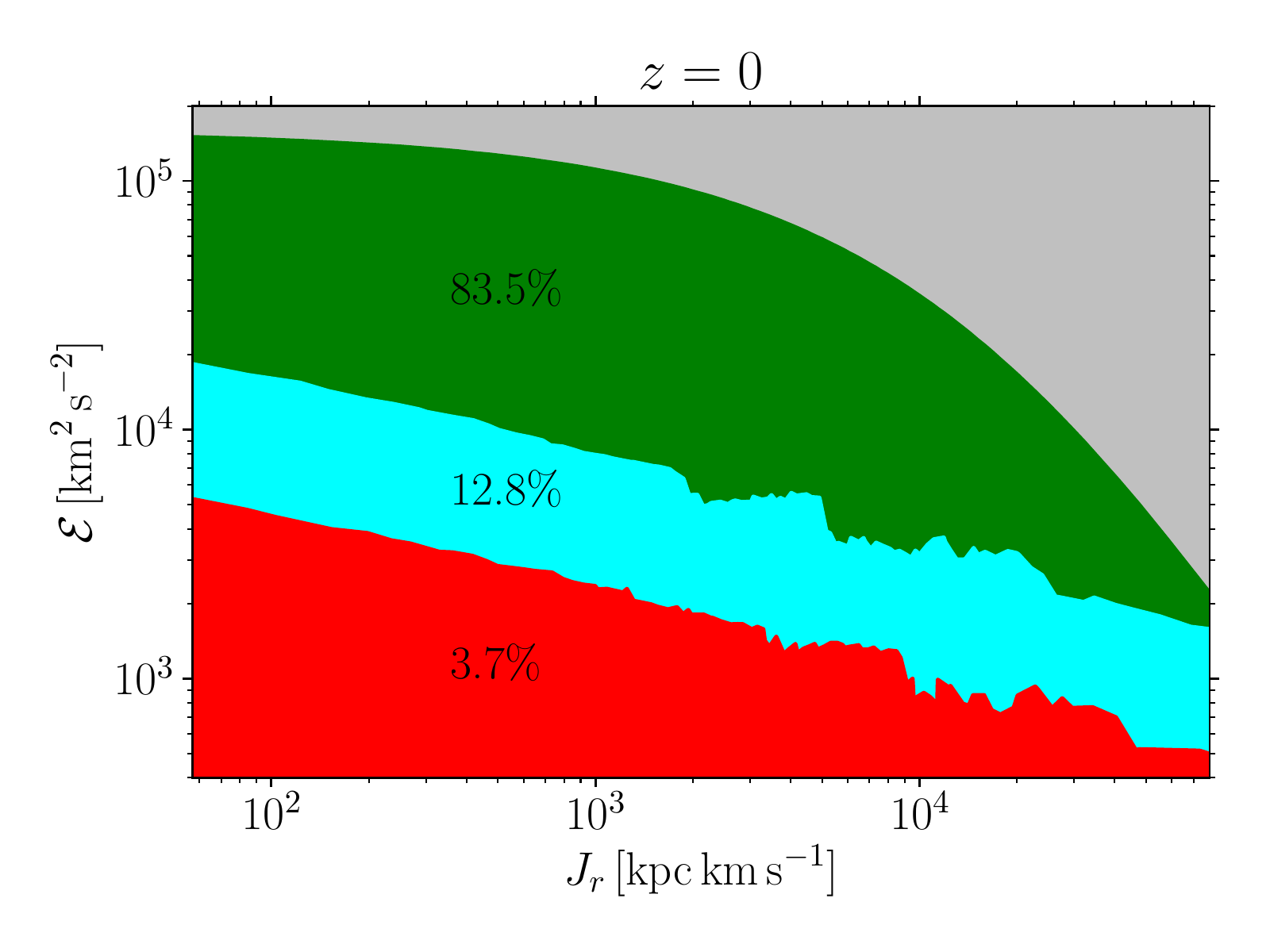}
    \caption{Occupancy of the $\mathcal{E}-J_r$ space in an idealised MW-size halo accreting mass according to the mass assembly history obtained in cosmological simulations (see Eq.~\ref{MAH}). The silver area is not populated, the green area is the ``linear" regime in which $\sqrt{\Tilde{D}}/J_r < 0.1$. The area coloured in cyan is the ``transition" area in which $\sqrt{\Tilde{D}}/J_r < 1$ and the red area is the ``fringe". The fraction of particles in the different regions is shown as a percentage of the total amount of particles. The top panel shows the integral of motion space occupancy at $z=z_{\rm max}$ (see text), the bottom panel at $z=0$. 
    }
    \label{fig:mass_accretion}
\end{figure}

Fig. \ref{fig:mass_accretion} shows the integral of motion space distribution of one million particles at different times. In the top (bottom) panel, we show the distribution at the redshift $z_{\rm max}$ (today).
In each panel, we show the fraction of particles inhabiting the different integral of motion space areas, which correspond to the ones introduced in Fig. \ref{fig:keyplot} for the time-dependent Kepler potential. To determine the boundaries between the different regions, we sort the particles within different radial action bins by their respective values of $\sqrt{\Tilde{D}}/J_r$.

In the upper panel of Fig.~\ref{fig:mass_accretion}, we see that at $z = z_{\rm max}$ only $\sim41\%$ of the particles in the DM halo inhabit the ``linear" integral of motion space area in which the diffusion formalism applies, while
$\sim18\%$ are located in the ``fringe", where we expect impulsive evolution. 
Just as in the Kepler case, at large values of $J_r$, we find that the integral of motion space area of the ``linear" regime shrinks and eventually disappears with all particles at very large radial actions being in the ``fringe", where the evolution of the radial action is non-linear.
We note that in the Kepler case, particles with these large actions tend to drift towards smaller values of $J_r$ and get "locked in" on short to intermediate time-scales (see Fig. \ref{act_dist_evolution}).
Since such drift does not seem to occur in the opposite direction in the Kepler case, 
radial actions which are too large to satisfy a ``linear" evolution are effectively erased from the distribution. Assuming that this quantitative behaviour will be the same in a DM halo (with a potential closer to that of a NFW distribution)
an integral of motion space distribution of DM particles such as the one seen at $z = z_{\rm max}$ could have significant consequences. In particular, if the ``fringe" is as populated as in the upper panel of Fig. \ref{fig:mass_accretion}, the radial action distribution is expected to drift towards smaller radial action at later times, i.e., the amount of particles with smaller radial actions increases and the tail of the distribution is erased. 

The lower panel of Fig. \ref{fig:mass_accretion} shows the integral of motion space occupancy today. 
In contrast to the upper panel, the evolution of radial actions in a MW-size halo today is on average significantly more adiabatic, with 
83 per cent of particles inhabiting the ``linear" regime, as opposed to only 3.7 per cent that inhabit the ``fringe".
This implies that {\it at the current time}, it is fair to assume that the ensemble average of radial actions is conserved in a MW-size halo, as shown in the bottom panel of Fig. \ref{fig:mass_accretion}.
The results of our simple analysis confirm that the assumptions made by \citet{2013MNRAS.430..121P} are well founded.  Yet, the top panel suggests that the (mean) evolution of the gravitational potential becomes gradually more impulsive the further we go back in time. Since we know that cusps form immediately after the gravitational collapse of the parent halo, this likely points to a link between the formation of the cusp and the impulsive evolution of the gravitational potential. Fig.~\ref{act_dist_evolution} shows that an impulsive evolution leads to a net evolution of radial action distributions towards smaller actions. A possible explanation for that may come from Eqs.~(\ref{linear_energy}) and (\ref{linear_order}). In fast evolving potentials, infalling particles have lower energies and thus shorter ``instantaneous radial periods". Such infalling tracers may thus be ``locked" into the regime in which $\Delta/J_r\lesssim 1$ and  Eq.~(\ref{linear_order}) applies. Fig. \ref{fig:mass_accretion} demonstrates that at a fixed radial action, the linear oscillation amplitude (see also Eq.~\ref{linear_order}) is (on average) smaller for particles with smaller energies. Comparison between the two panels also reveals that faster evolution of the gravitational potential shifts the ``linear'' integral of motion space towards lower radial actions and energies. At a fixed radial action, particles with smaller energies also have smaller angular momenta. In very fast evolving potentials, this lock-in mechanism is therefore most efficient for particles with low angular momenta. This way, impulsively evolving potentials themselves would create populations of particles with small radial actions and low angular momenta. These may be the ``cusp particles" of \citet{2013MNRAS.430..121P}. However, since this is an impulsive/non-linear effect, a complete understanding cannot be obtained using the diffusion formalism presented here. A first step towards an overall better understanding will be to expand our formalism beyond the analysis of tracers and to self-consistently evolve the radial action distribution and the gravitational potential of an ensemble of gravitating particles at subsequent time-steps. 
Finally, we note that diffusion formalism may be useful to improve the analysis of tidal streams in the MW. \citet{2015A&A...584A.120B} have analyzed the evolution of radial action distributions of tidal streams in a time-dependent Aquarius potential. In tidal streams whose orbits are not adiabatic, they find an increased spread in radial action between the stream particles when compared to the final distribution in a static potential. In the bottom panel of Fig. (\ref{fig:mass_accretion}), we find that $\sim 17\%$ of DM particles today inhabit a region in integral of motion space in which diffusion of radial actions due to the time-dependence of the MW is a relevant phenomenon. Stars in tidal streams are essentially tracers of the gravitational potential and thus our results suggest that the observed increased spread in radial action is the result of a diffusion process in radial action space caused by the time-dependence of the potential. This diffusion in radial action space constitutes an additional baseline error source when using the clustering of tidal streams in action space to constrain the potential of the Galaxy, as was done, e.g, by \citet{2015ApJ...801...98S} and \citet{2017ApJ...836..234S}. In particular, our results suggest that the accuracy of the assumption that the true potential implies the most tightly clustered radial action distribution depends on both the accretion history of the MW and the orbit of the tidal stream's progenitor.  

\subsection{Cusp-core transformation in SIDM}\label{sec_sidm}

\citet{2019MNRAS.485.1008B} looked at the evolution of a set of tracer particles with a Gaussian distribution in radial action orbiting in the potential of a dwarf-sized DM halo developing a constant density core in one of two different ways\footnote{The halo used was a dwarf-sized halo which initially has a Hernquist density profile with mass $M_{200} = 1.43\times 10^{10}M_{\odot}$; see Table 1 of \citet{2019MNRAS.485.1008B} for the relevant simulation parameters.}, through elastic self-scattering between the DM particles (SIDM) 
or through impulsive energy injections into the system akin to supernova feedback. Fig. 7 in \citet{2019MNRAS.485.1008B} shows a comparison of the final radial action distributions of the tracer particles in these two reference scenarios, as well as a third baseline scenario in which the host halo retains its cusp. 
From this figure, it is clear that
$N(J_r,t)$ at the end of the SIDM simulation, which has a cusp-core transformation, is very close to the final distribution in the benchmark simulation (without a cusp-core transformation). 
Presumably, the difference between those two distributions can be explained by some small amount of diffusion in radial action space. Hence, we expect that a large part of the  $\mathcal{E}-J_r$ space area occupied by particles orbiting in a SIDM halo belongs to the ``linear" regime of radial action evolution. 

To test this hypothesis, 
we perform the analysis introduced in Section \ref{subsec:ana} in a SIDM halo similar to the one used in \citet{2019MNRAS.485.1008B} following Eq.~(\ref{diffusion_theory}).
The only difference is that 
we re-run the SIDM simulation 
with a smaller self-interaction cross section of $\sigma/m_\chi = 1{\rm cm^2\,g^{-1}}$. We save snapshots (simulation outputs) every $0.14\,{\rm Gyr}$ and calculate the potential $\Psi(r)$ by averaging the potential of tracer particles in logarithmically spaced spherical shells. To obtain $\Psi(r,t)$  we interpolate $\Psi(r)$ between snapshots.
The scale factor itself is calculated according to Eq.~(\ref{gen_scale}). We calculate $\alpha(r,t)$ numerically by interpolating the function $\log(\Psi)(\log(r))$ and 
calculating its derivative at each (simulation) particle's radius. 
Since the SIDM halo changes shape we cannot neglect the
time-dependence of $\alpha(r,t)$
and the derivative of the scale factor at a given radius given by Eq.~(\ref{uno}). $\dot{\alpha}$ and $\dot{\Psi}$ are calculated numerically at each particle's position.
We then calculate $\Tilde{D}(J_r,\mathcal{E})$ from Eq.~(\ref{diffusion_theory}) as in Section \ref{hma}, but using $\Phi(r,t)$ of the SIDM halo. 
Core formation in a SIDM halo is a gradual process. Initially, self-interactions are most efficient in the halo's centre and cause a relatively fast decrease of the central density. 
Subsequently, the forming core slowly thermalizes. In this latter stage, due to the prior mass redistribution, the density is completely flat (isothermal core) in the centre followed by a small region outside the inner core (but still within the scale radius) where the density rises slightly above that of the original profile
(see e.g \citealt{Vogelsberger2012,Vogelsberger2014}). At the end of our simulation, the halo's core is fully thermalized and in a transient quasi steady state. 
Notably, including baryons into the simulation can change the phenomenology of SIDM (see \citealt{2020MNRAS.495...58S} or \citealt{2017MNRAS.472.2945R}). We briefly note that there is an additional phase called gravothermal collapse (e.g. \citealt{2019PhRvD.100f3007Z} and \citealt{2020arXiv201002924T}) in which the core collapses to very large densities. However, this phase is only relevant for very large cross sections. 



\begin{figure}
    \centering
    \includegraphics[width=\linewidth]{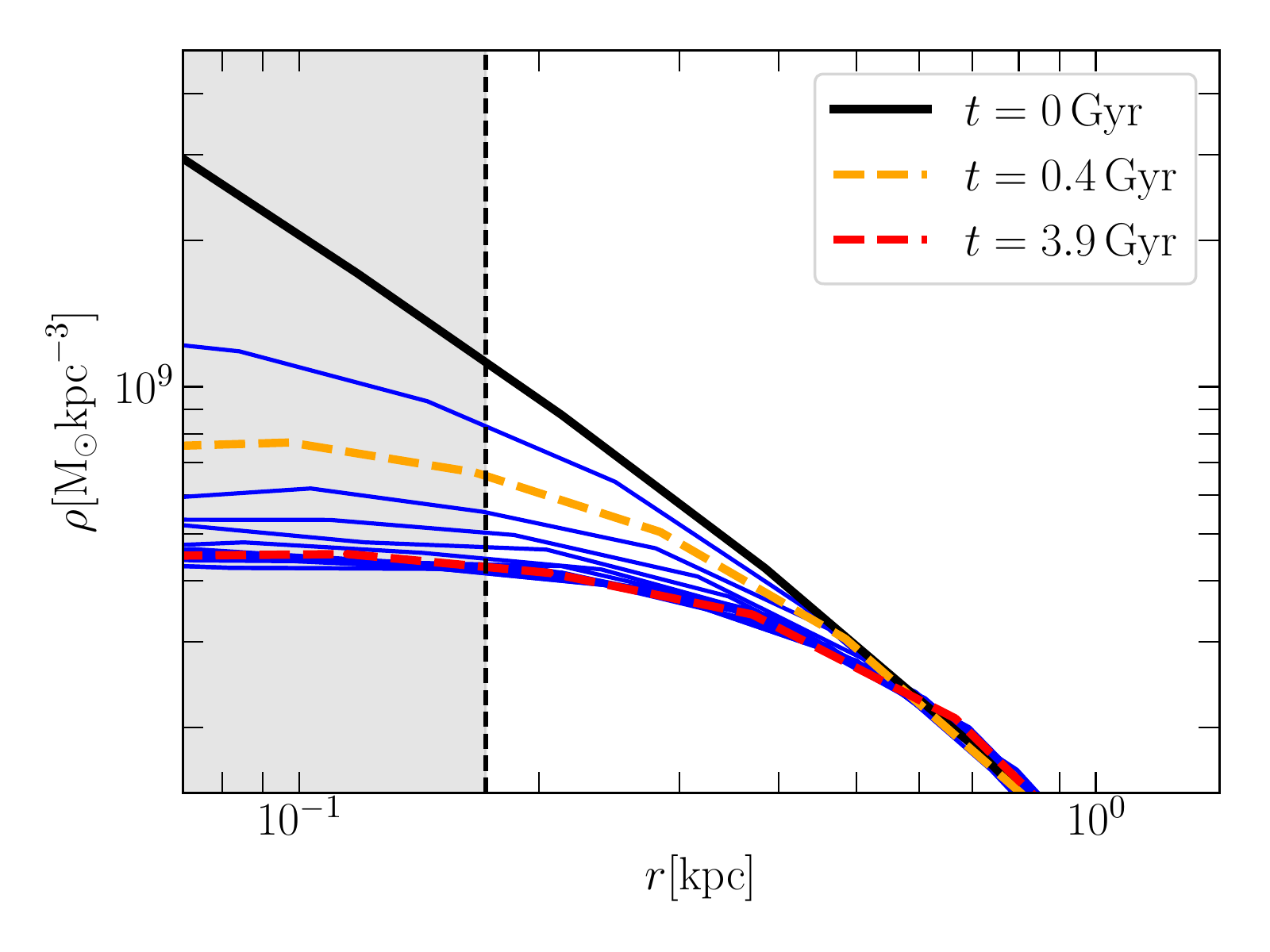}
    \caption{Evolution of the density profile in the inner region of a dwarf-size SIDM halo with $\sigma/m_\chi = 1{\rm cm^2g^{-1}}$. Lines denote the halo density profile at different times. The black solid line marks the density profile at the beginning of the simulation , whereas the orange and red dashed lines are the density profiles after $0.4\,{\rm Gyr}$ and $3.9\,{\rm Gyr}$ of simulation time respectively. Blue lines indicate the evolution in between other snapshots. The gray shaded area denotes the resolution limit of the equivalent CDM-only simulation (see \citealt{2019MNRAS.485.1008B}).}
    \label{fig:sidm_profiles}
\end{figure}

Fig. \ref{fig:sidm_profiles} shows the evolution of the central part of the SIDM-halo's density profile in our simulation. The black solid line is the initial density profile. In dashed orange we show the density profile after $0.4\,{\rm Gyr}$ and the red dashed line is the density profile after $3.9\,{\rm Gyr}$. Blue lines denote snapshots between or after the ones highlighted in the legend. The grey shaded area shows the resolution limit for CDM simulations (\citealt{Power:2002sw}). We note that SIDM profiles are usually converged to well within this so-called Power radius \citep{Vogelsberger2012,Rocha2013,Vogelsberger2014}. 
The orange density profile corresponds to a time during the initial stage of core formation when the mass density in the innermost part of the halo decreases rapidly. The red line, on the other hand, corresponds to a time at which the core is fully formed. 
We have chosen those two times because they represent different stages of the core formation process in a SIDM halo.  

\begin{figure}
    \centering
    \includegraphics[width=\linewidth]{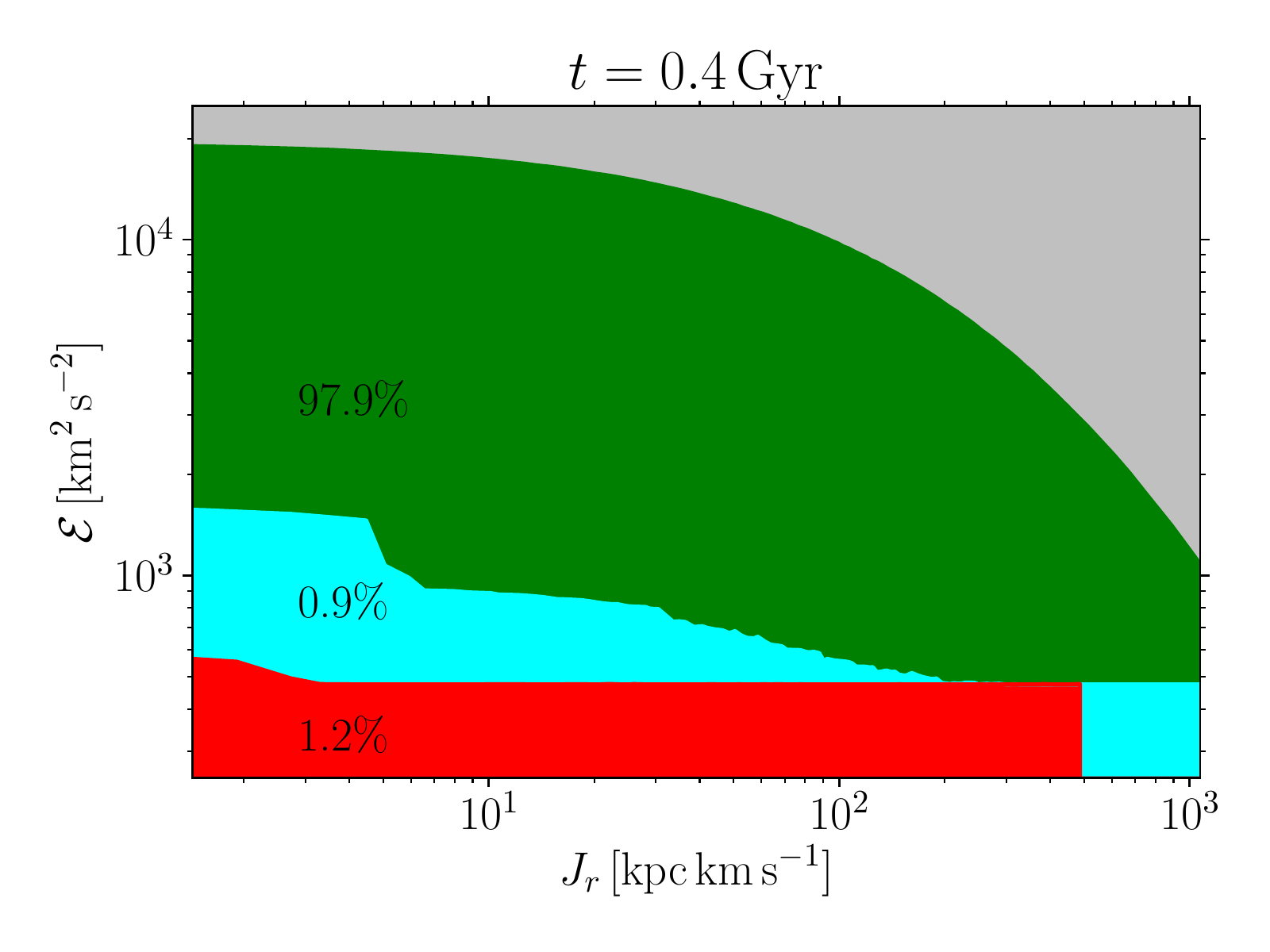}\\
    \includegraphics[width=\linewidth]{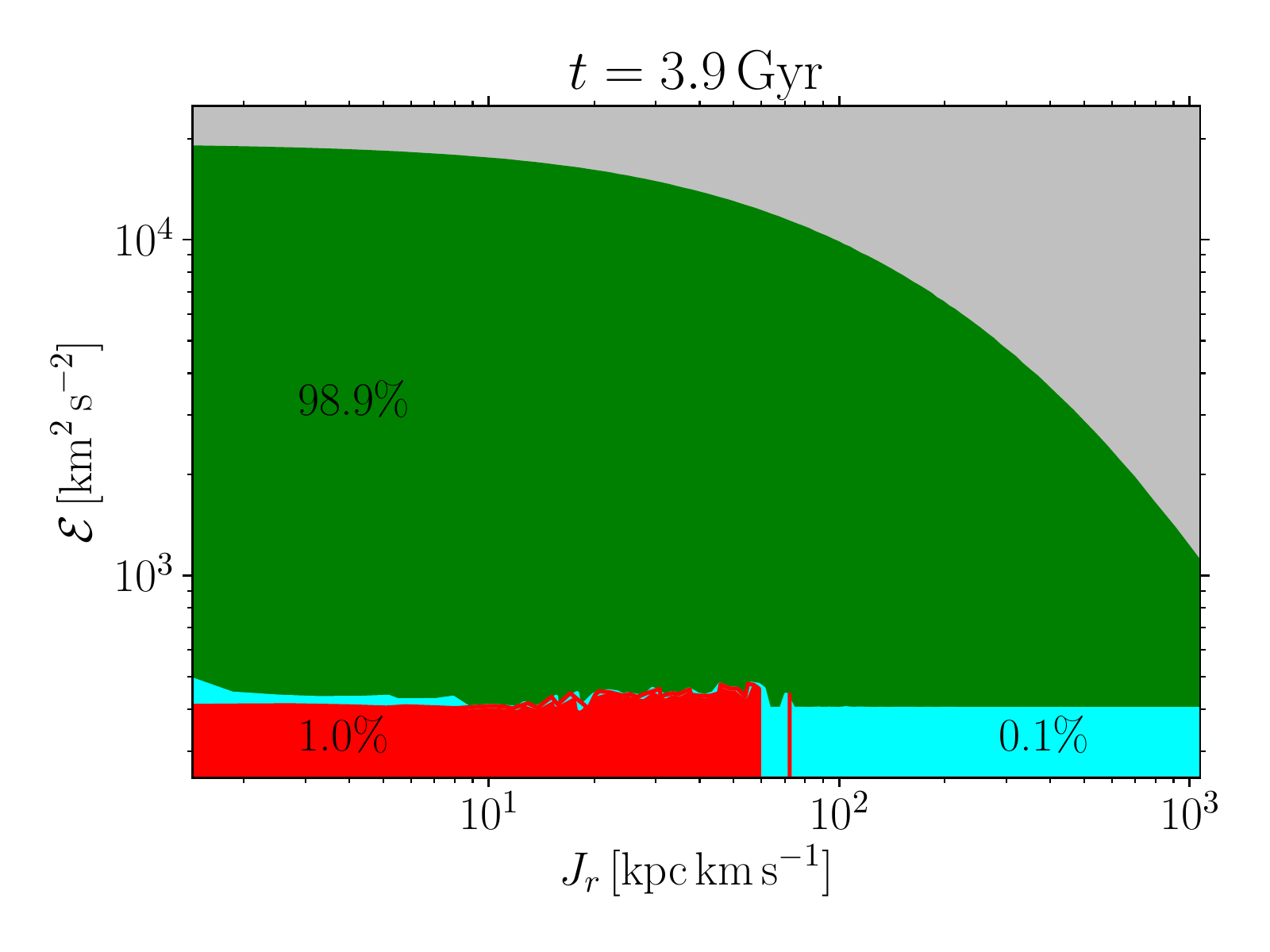}\\
    \caption{The $\mathcal{E}-J_r$ space occupancy in a dwarf-size halo forming a core due to self-scattering between its DM particles. The different curves and areas are equivalent to those in Fig. \ref{fig:mass_accretion}. The fraction of DM particles populating each area of integral of motion space is shown as a percentage in the respective areas. The upper panel corresponds to the orange dashed line in Fig. \ref{fig:sidm_profiles} (early-stage of core formation), while the lower panel corresponds to the red dashed line in Fig. \ref{fig:sidm_profiles} (late-stage of core formation).}
    \label{fig:sidm_phasespace}
\end{figure}

Fig. \ref{fig:sidm_phasespace} shows the integral of motion space occupancy for the SIDM halo at the times corresponding to the orange dashed line ($0.4\,{\rm Gyr}$) and the red dashed line ($3.9\,{\rm Gyr}$) in Fig. \ref{fig:sidm_profiles}. The areas in the plot correspond to the equivalent areas in Fig. \ref{fig:mass_accretion}.
The change in gravitational potential due to core formation (induced by DM self-interactions) causes an almost adiabatic evolution of radial action distributions over the entire simulation time. In both panels of Fig. \ref{fig:sidm_phasespace}, more than 97 per cent of DM particles are in the ``linear" integral of motion space area and only around 1 per cent inhabit the ``fringe". 
As a consequence, we expect the radial action distribution of tracer particles to be altered very little, if at all, by the change in gravitational potential triggered by SIDM, provided that the numerical values of the self-interaction cross section lies within the range that is allowed by astrophysical constraints.

We remark that the statement that the evolution of radial action distributions in an SIDM halo is expected to be linear and can be described by our diffusion formalism applies only to kinematic tracers. For the DM particles themselves, SIDM-induced elastic collisions are a different matter as the radial action of both collision partners changes in a random fashion, since their post-collision orbits are entirely different from the pre-collision orbits. 

We apply the diffusion formalism presented in Sections \ref{sec_form} and \ref{sec_results} to describe the evolution of a Gaussian radial action distribution comprised of 20000\footnote{To enable us to derive smoother drift and diffusion coefficients, we have added 18000 tracers to the original 2000 used in \citet{2019MNRAS.485.1008B}. They are set up in exactly the same way as the initial 2000 tracers and thus obey the same Gaussian distribution.} tracer particles orbiting in the core-forming SIDM halo. 
We use the approximations given by Eqs.~(\ref{gen_scale}) and (\ref{uno}) for $R(t)$ to calculate the drift and diffusion coefficients. In Appendix \ref{app3}, we demonstrate that the evolution of radial actions of kinematic tracers in a potential with a time-dependent shape is accurately described by this approximate scale factor. Since the rate at which the shape of the potential in Appendix \ref{app3} changes is larger than the rate at which the potential of the SIDM halo changes, we conclude that we can use Eqs.~(\ref{gen_scale}) and (\ref{uno}) to obtain accurate predictions of the evolution of $N(J_r,t)$ in a core-forming SIDM halo. 



\begin{figure}
    \centering
    \includegraphics[width=\linewidth]{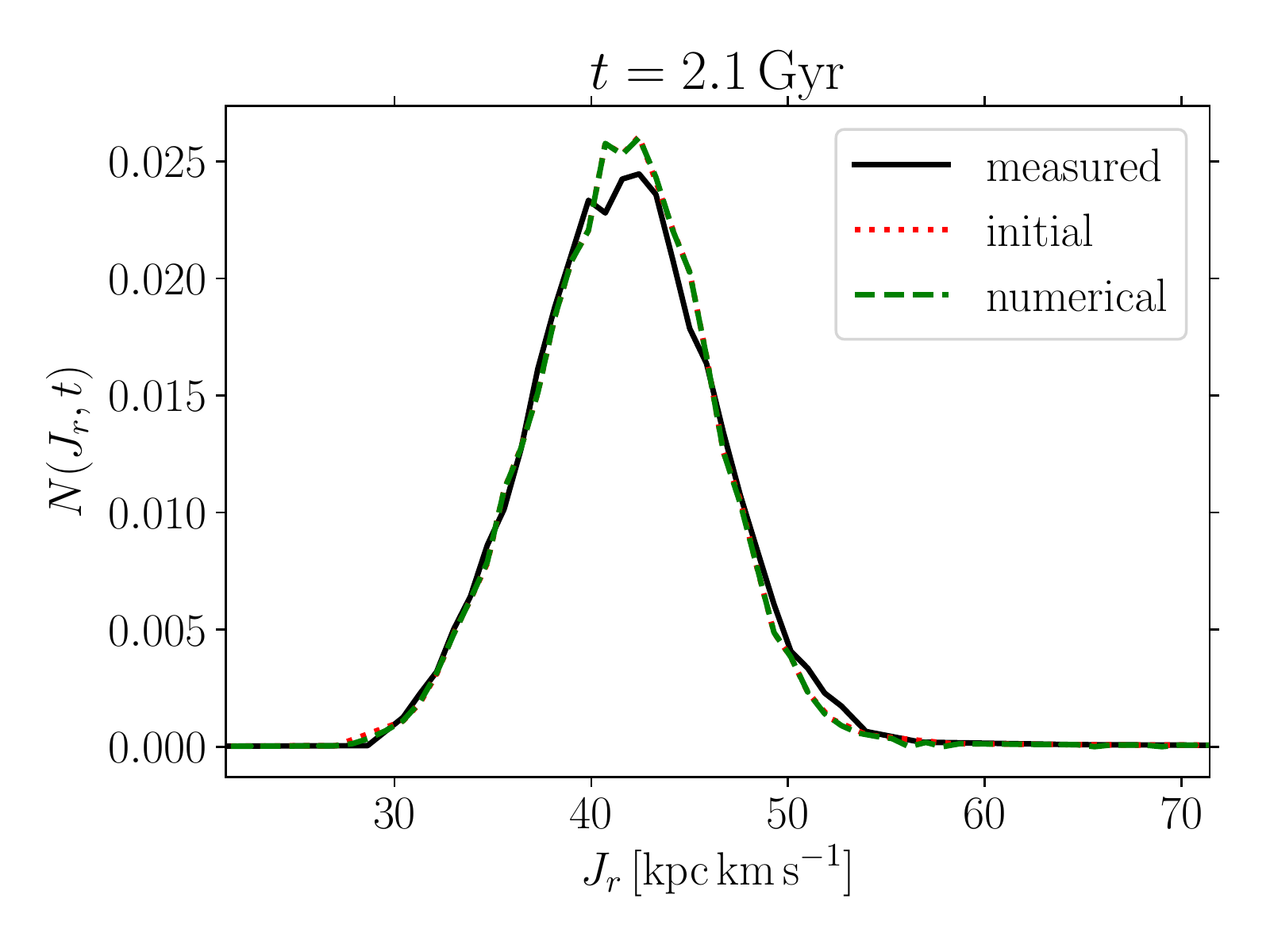}\\
    \includegraphics[width=\linewidth]{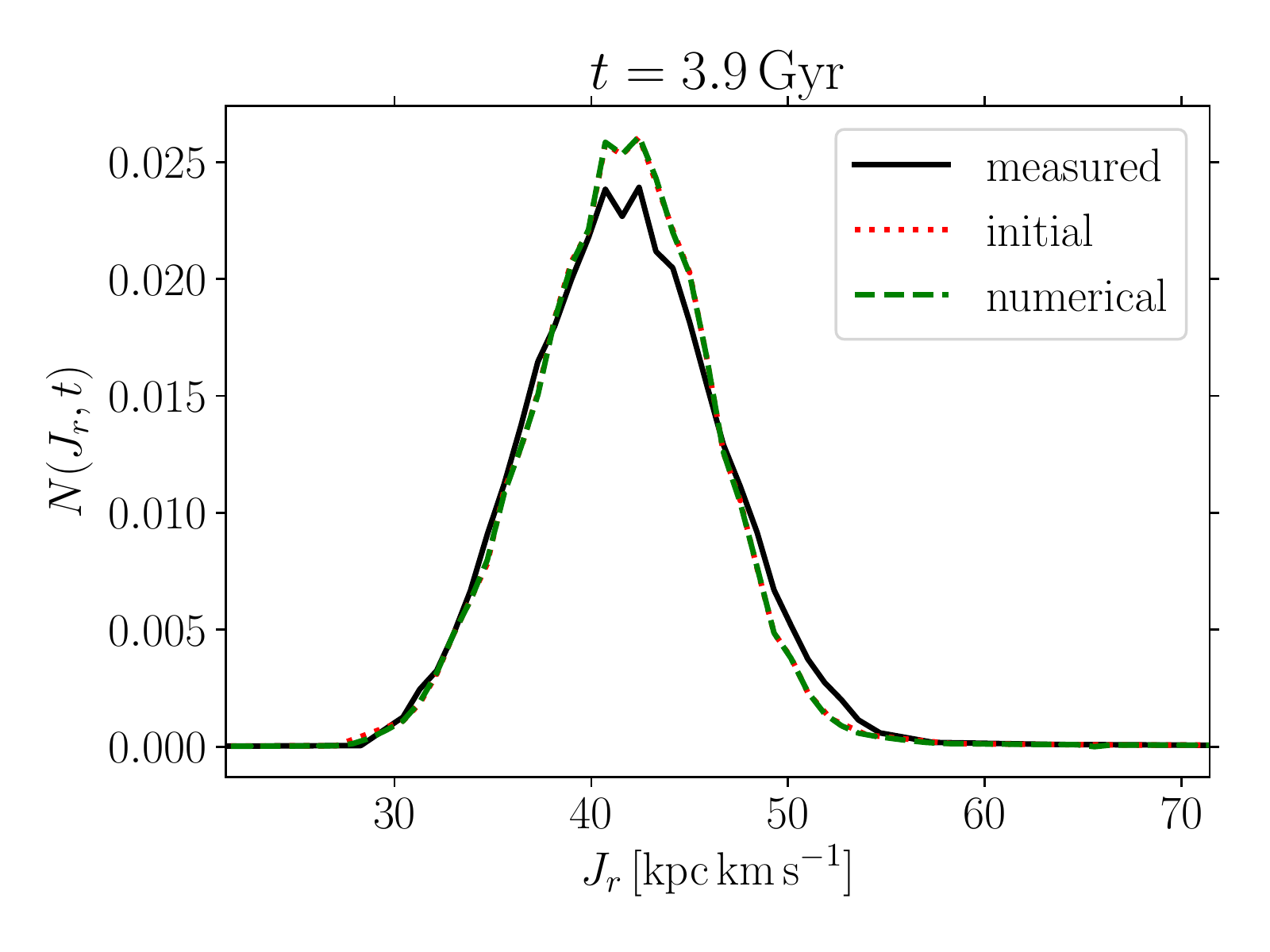}
    \caption{Radial action distribution of 20000 tracer particles after $2.1\,{\rm Gyr}$ of simulation time (top panel) and after $3.9\,{\rm Gyr}$ of simulation time (bottom panel) in the core-forming SIDM halo shown in Fig.~\ref{fig:sidm_profiles}. The black lines show the measured distribution. The red dotted line is the initial distribution and the green dashed line denotes the result of the diffusion formalism.}
    \label{fig:sidm_dist}
\end{figure}

Fig. \ref{fig:sidm_dist} shows the ``measured" radial action distribution (black solid line) after $2.1\,{\rm Gyr}$ of simulation time in the top panel and after $3.9\,{\rm Gyr}$ in the bottom panel, along with the result of the diffusion formalism (green dashed line) and the initial distribution (red dotted line). 
Since the calculation of the drift and diffusion coefficients involves numerical derivatives of time-dependent measured quantities, we take $N(J_r,t)$ at $t=0.4\,{\rm Gyr}$ as the initial distribution in order to assure that the time-derivatives are numerically stable. 
%
The result of the diffusion formalism is overall in good agreement with the measured distribution in Fig. \ref{fig:sidm_dist} (particularly in the early stages of core formation; see upper panel). 
However, in the late stage of core formation (between $2.1\,{\rm Gyr}$ and $3.9\,{\rm Gyr}$) the measured distribution gradually forms a slightly more extended tail towards larger radial actions. This evolution, albeit not very sizeable, is not predicted by our formalism. In fact, our diffusion formalism predicts hardly any evolution at all, which might imply that this is a significant deviation. In Appendix \ref{app3}, however, we show that the amplitude of the oscillation of radial actions in a potential developing a core is in general well approximated by our diffusion formalism. We thus surmise that the small but gradually increasing mismatch between the measured radial action and the result of the diffusion formalism is caused by numerical effects arising due to the discrete sampling of the SIDM halo. In fact, since the DM simulation particles are more massive than the tracers, individual close encounters between tracers and DM simulation particles can occasionally lead to a small sudden increase in the tracer's energy, leading to a small increase in radial action as well.  

\begin{figure}
    \centering
    \includegraphics[height=6cm,width=8.5cm,trim=0.5cm 0.5cm 0.5cm 0.4cm, clip=true]{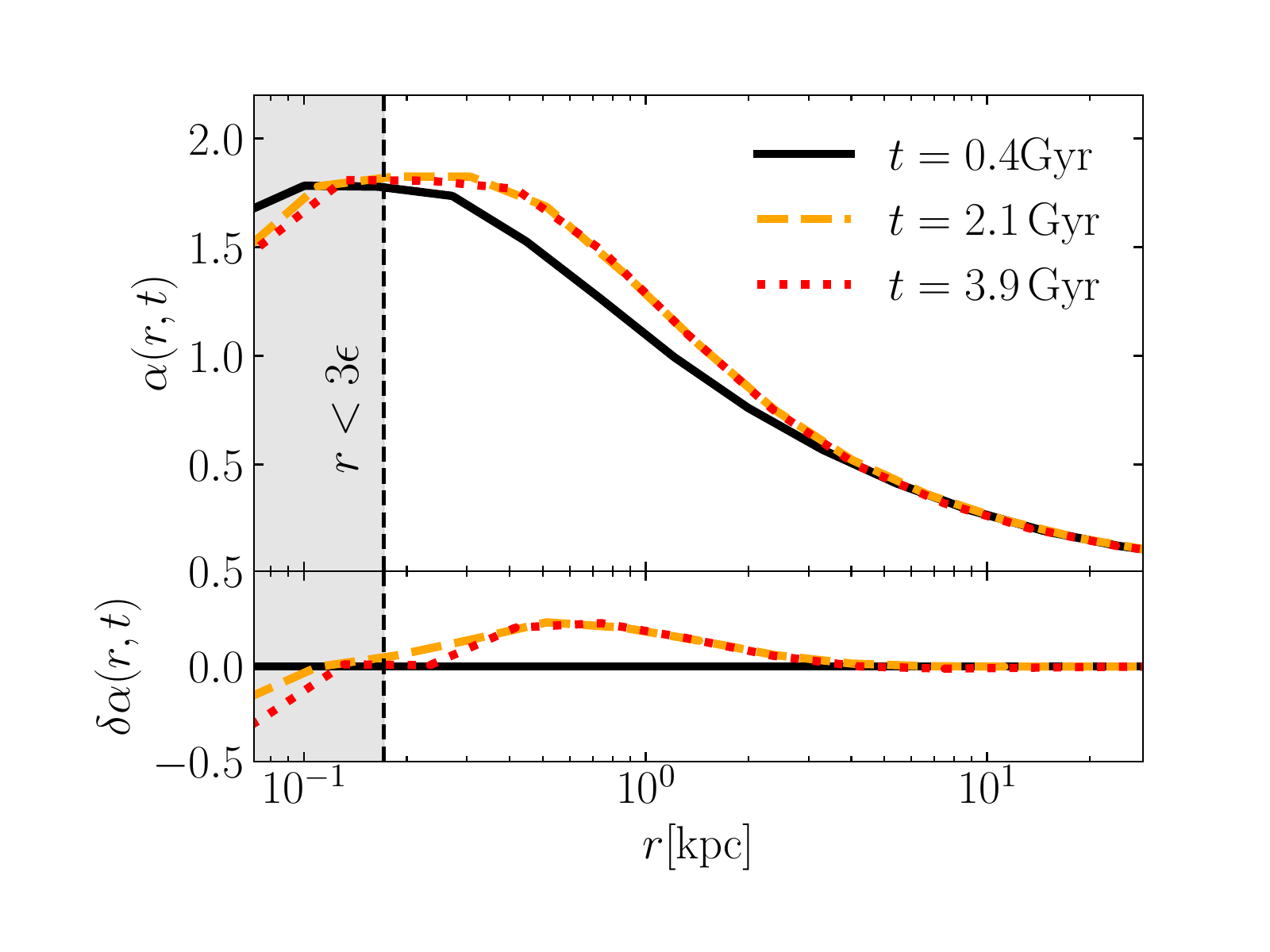}
    \caption{Logarithmic slope $\alpha(r,t)$ of the potential of the SIDM halo in Fig~\ref{fig:sidm_profiles} as a function of radius at different times (upper panel) and change of the logarithmic slope with respect to $t = 0.4\,{\rm Gyr}$, also as a function of radius (lower panel). The times shown match the ones in Fig. \ref{fig:sidm_dist}. The grey shaded area highlights the resolution limit as in Fig.~\ref{fig:sidm_profiles}.}
    \label{fig:sidm_alpha}
\end{figure}

Nonetheless, as we find in Appendix \ref{app3}, a shape-changing host potential can cause sizeable oscillations of the radial action values if the change in shape of the potential occurs fast enough. 
The fact that our diffusion theory does not predict any sizeable evolution of $N(J_r,t)$ implies that the change of the shape of $\Psi(r,t)$ is rather slow between $t = 0.4\,{\rm Gyr}$ and $t=3.9\,{\rm Gyr}$. 
In Fig. \ref{fig:sidm_alpha}, we look at the time evolution of the logarithmic slope $\alpha(r,t)$ of the SIDM halo's potential as a function of radius. We show the logarithmic slope of the potential at three different times, $t = 0.4\,{\rm Gyr}$ (solid black line), $t = 2.1\,{\rm Gyr}$ (dashed orange line) and $t = 3.9\,{\rm Gyr}$ (dotted red line). In the lower panel, we furthermore show the change of the logarithmic slope with respect to $t_0 = 0.4\,{\rm Gyr}$, defined as 
\begin{align}
    \delta\alpha(r,t) = \alpha(r,t)-\alpha(r,t_0).
\end{align}
We find that the evolution of $\alpha(r,t)$, which directly affects the ratio $\dot{R}/R$ through the time-derivative $\dot{\alpha}(r,t)$, is most significant at earlier times and at smaller radii. In fact, in Fig. \ref{fig:sidm_alpha} we show that most of the evolution in $\alpha(r,t)$ occurs between $0.4\,{\rm Gyr}$ and $2.1\,{\rm Gyr}$. Afterwards, the shape of the potential is essentially constant, implying that at that point the core is fully formed. Considering that Fig. \ref{fig:sidm_dist} hardly shows any evolution of both predicted and measured $N(J_r,t)$ between $t=0.4\,{\rm Gyr}$ and $t=2.1\,{\rm Gyr}$, we conclude from Fig. \ref{fig:sidm_alpha} that there should not be any physical evolution afterwards either.
We take this as confirmation that the small observed drift in this time interval is due to discreteness effects, i.e., it occurs because the sampled potential is not perfectly smooth. 

In conclusion, we find that SIDM with a cross section of $\sigma/m_{\chi} = 1\,{\rm cm^2g^{-1}}$ causes virtually no evolution in $N(J_r,t)$ and can thus be considered a fully adiabatic process. 
In a cosmological SIDM halo the impact of mass accretion on the radial action distribution of tracers will thus far outweigh the impact of SIDM. 

\section{Conclusion}\label{sec_conclusions}

Cosmological simulations of CDM yield concordant results on all resolved scales. An example is the inner density profile of collapsed DM haloes, which is found to be universal over 20 orders of magnitude in halo mass (\citealt{2020Natur.585...39W}). Simulations of the dynamics of virialized self-gravitating systems have also provided valuable insight into the long-term evolution of such systems, modeling, for instance, the shapes of observed galaxies or the formation of tidal streams in the galactic halo. 

However, many of the results of N-body simulations have not yet been understood from fundamental principles. An approach that is frequently used to model the secular evolution of bound systems is Hamiltonian perturbation theory \citep[e.g.][]{1972MNRAS.157....1L,1984MNRAS.209..729T}. In this method, the (time-dependent) Hamiltonian is written as a power series expansion in some small parameter $\epsilon$ and the zero'th order term is a time-independent Hamiltonian that can be written in terms of canonical action variables (see Eq.~\ref{eq:ham}). Hamiltonian perturbation theory can be formulated into a kinetic theory for the evolution of self-gravitating systems (\citealt{2010MNRAS.407..355H}), an approach that has proven quite successful in recent years (e.g. \citealt{2015A&A...584A.129F}). However, a crucial assumption behind Hamiltonian perturbation theory is that actions are integrals of motion of the zero'th order term $H_0$ and that they evolve very slowly. The theory performs well for potentials in which these conditions are fulfilled, but breaks once actions evolve significantly with time. 

An alternative to Hamiltonian perturbation theory is to use statistical mechanics. Historically, however, deriving a thermodynamic model for the evolution of self-gravitating systems has proven to be difficult for a variety of reasons. 
Negative specific heat and non-ergodicity of gravitating systems forbid the use of canonical or grand canonical ensembles. On top of that, \citet{1990PhR...188..285P} shows that in gravitating systems the energy is not an extensive parameter, thus the system cannot be divided into non-interacting macrocells and the laws of standard thermodynamics do not apply. Despite these challenges, several attempts have been made to describe the evolution of self-gravitating systems within the framework of statistical physics. \citet{2013MNRAS.430..121P} attempted to derive the distribution function of a virialized CDM halo, using a maximum entropy approach and the additional constraint that the ensemble average of the DM particle's actions be conserved. Their formalism accurately recovers the distribution function of simulated haloes over several orders of magnitude, but does not predict the centrally-divergent density cusps without adding a second, dynamically decoupled particle population with low angular momenta.

\cite{2013MNRAS.433.2576P} developed a coordinate transformation relating the equations of motion in a frame with a time-dependent potential to the dynamics in a different frame in which the potential is static. 
The integrals of motion in this new reference frame expressed in the coordinates of the original frame are so-called dynamical invariants. Subsequently, \cite{2015MNRAS.451.3537P} developed a statistical theory for the diffusion in energy space of a particle ensemble in the time-dependent potential based on those dynamical invariants.


In this paper, we have developed a similar theory for the diffusion of particle ensembles in radial action space. For now, our theory is restricted to spherically symmetric potentials but we note that a modified theory that includes diffusion along resonant spatial directions can be obtained as discussed in \citet{2015MNRAS.451.3537P}. In our theory, the diffusion is derived by treating the tracers as a microcanonical ensemble and relating the drift and diffusion coefficients to the evolution of the radial actions of individual tracers. This makes our approach fundamentally different from Hamiltonian perturbation theory, in which perturbed systems do not depart strongly from a state of dynamical equilibrium. To first order, we show that the time-evolution of the radial action of a particle moving in a time-dependent spherical potential is given by an oscillation around the invariant action, with an amplitude that is fully determined by the phase space coordinates of the particle and the transformation between the static and the time-dependent frames. Since actions are traditionally considered adiabatic invariants, we can use the ratio between the amplitude of the linear oscillation and the radial action to determine whether the evolution takes place in an adiabatic or impulsive regime. 
We illustrate this theory with two examples i) an idealised Milky-Way (MW) size galaxy with a time-dependent potential that grows according to the average mass accretion history of MW-like $\Lambda$CDM haloes, and ii) the  time-dependent potential that of a dwarf-size self-interacting DM (SIDM) halo forming a constant density core. A summary of our main methods and key results is as follows:
\begin{enumerate}
    \item {\it Calculating the linear variation of $J_r$.-} 
    The starting point is to find a numerical (or approximate) solution
    to the differential equation for the scale factor $R(t)$ which defines the coordinate transformation introduced in \citet{2013MNRAS.433.2576P}. We then write the action invariant $J_r'$ defined in the new time-independent reference frame in terms of the phase space coordinates of the particle in the original frame. 
    Under the assumption that the potential evolves slowly, we make a first order Taylor expansion of $J_r'$ in 
    $\dot{R}/R$ (see Section \ref{sub_tay_one}). We find that to zero'th order, the dynamical invariant $J_r'$ and the radial action $J_r$ in the original reference frame are the same. 
    The first order correction to this equality is an oscillation with an amplitude proportional to $\dot{R}/R$ -- and thus to $\dot{\Phi}/\Phi$ -- around $J_{r'}$. This oscillation happens in phase with the radial motion and is proportional to the radial period of the particle's orbit. We have here for the first time presented an analytical calculation of the linear oscillation amplitude of radial actions in time-dependent potentials (Eq.~\ref{linear_order}). 
    As a numerical test of this linear model, we follow the orbits of three tracer particles in a time-dependent analytic Kepler potential using the N-body code AREPO \citep{Springel:2009aa} (see Section~\ref{sec_3particles}). 
    In case the oscillation amplitude is small compared to $J_r$ itself our linear model provides an excellent fit to the time-evolution of $J_r$ (see left panel of Fig. \ref{3particles}). For orbits in the ``fringe" of the potential ($E\approx 0$), such as the one shown on the right panel of Fig. \ref{3particles}, the linear oscillation amplitude is of the order of $J_r$ itself and the radial action's evolution becomes non-perturbative. 
    \item {\it Formulating a diffusion formalism in radial action space. -} Having calculated the linear variation of $J_r$ for individual particles in a time-dependent potential, we then aim to statistically predict the evolution of the radial action distribution $N(J_r,t)$ of a particle ensemble. In Section \ref{sub_diffeq}, we show that if the time evolution of the gravitational potential is sufficiently slow, $N(J_r,t)$ is related to the invariant distribution of dynamical action invariants $N(J_r')$ through a diffusion equation. The drift and diffusion coefficients $\Tilde{C}(J_r,t)$ and $\Tilde{D}(J_r,t)$ which appear in the diffusion equation are proportional to the ensemble average and the variance of the first order Taylor correction, respectively. Having computed $\Tilde{C}(J_r',t)$ and $\Tilde{D}(J_r',t)$, we can calculate $N(J_r,t)$ from the invariant distribution $N(J_r')$ through a simple Gaussian convolution. 
    We define a numerical criterion based on $\Tilde{D(J_r,t)}$ to determine whether the expected evolution of radial action distributions can be described as a diffusion process.
     Dependent on the ratio between $\sqrt{\Tilde{D}}$ and $J_r$ we define three different regions in integral of motion space:
    \begin{itemize}
        \item In the \textbf{linear} regime, $\sqrt{\Tilde{D}}<0.1\,J_r$. We expect that the diffusion formalism will provide a good prediction of the evolution of $N(J_r,t)$ for particle distributions inhabiting this area in phase space.
        \item In the \textbf{transition} regime, $0.1 < \sqrt{\Tilde{D}}/J_r < 1$ we expect (average) deviations from linear evolution, and thus our expectation is that the diffusion formalism will yield only rough results when applied to particle distributions in this part of phase space. 
        \item In the \textbf{fringe}, $\sqrt{\Tilde{D}} > J_r$, and thus we expect that the diffusion formalism does not apply. 
    \end{itemize}
    To confirm these predictions, we perform a set of 5 restricted simulations of 20000 tracer particles orbiting in the same time-dependent Kepler potential as the three individual tracers before and covering different integral of motion space regions (see Table \ref{tab_distributions} and Fig. \ref{fig:keyplot} for the simulation parameters and the restrictions in phase space). A theoretical calculation of the diffusion coefficients in energy-action space agrees well with the coefficients measured from the initial tracer distributions (see Figs. \ref{fig:keyplot}$-$\ref{fig:200_dists}). For the ``10-linear" simulation
    we show that if the drift and diffusion coefficients are measured directly from the particle distribution, the predictions of the diffusion formalism are in perfect agreement with direct measurements of both the invariant distribution $N(J_r')$ and $N(J_r,t)$ at the end of the simulation (see Figs. \ref{fig:invariant_10}, \ref{fig:10_green_comp}). Moreover, we find that a theoretical calculation of the diffusion coefficient yields equally sound results provided the drift coefficient is negligible, i.e., for fully phase-mixed particle ensembles (see Section \ref{sub_10lin} for a discussion). In general, we find that the agreement between the measured distribution and the prediction of the diffusion formalism deteriorates the larger the fraction of particles populates the ``transition" region in integral of motion space as opposed to the ``linear" region (Section \ref{sub_var_diff}, Fig. \ref{fig:invariant_distributions}). As a consequence
    , the evolution of $N(J_r,t)$ in the `200-transition" simulation is highly non-linear and cannot be modelled as a diffusion process (see Section \ref{sub_200tran}). We attempt to explain the observed evolution using higher-order perturbation theory (see Fig. \ref{fig:2o200} and Appendix \ref{app2}) but we have to concede that while we manage to obtain an improved understanding of the shortcomings of the first order approximation, a full understanding of the non-linear evolution of $N(J_r,t)$ in the ``fringe" remains beyond our reach. Phenomenologically, however, we observe that on average, there is a clear trend for particles whose evolution is non-linear to drift towards considerably smaller radial actions on the simulation time-scale (see Figs. \ref{act_dist_evolution}, \ref{3particles}, \ref{fig:200_cyan_evolution}). In fact, particles whose actions are initially so large that there is no ``linear" integral of motion space available (see Fig. \ref{fig:keyplot}) tend to quickly drift towards values of $J_r$ where the evolution becomes linear. We conclude that fast changes in the gravitational potential occurring over a short period of time can permanently alter the radial action distribution of particle ensembles in a non-trivial way. This effect is particularly important if the fraction of particles in the ``fringe" area is significant while the change in gravitational potential occurs.   
    \item {\it Implications for halo mass accretion and cusp-core transformation. -} Following our arguments from Section \ref{subsec:ana} we can decide from the ratio $\sqrt{\Tilde{D}}/J_r$ whether the evolution of a particle ensemble's radial action distribution is linear, and hence whether actions are approximately conserved on average. 
    For illustration, we analyze whether radial action distributions evolve linearly during the accretion history of a MW-size halo and during the process of cusp-core transformation in a dwarf-size SIDM halo.
    \\
    
\noindent     
    \textbf{Mass accretion in a MW-size halo}\\ 
    To assess whether actions evolve linearly during the mass accretion history of a typical MW-size halo
    we calculate the relative accretion rate $\dot{M}/M$ as a function of redshift from the mass-redshift relation reported in \citet{10.1111/j.1365-2966.2010.16774.x} and determine the redshift $z_{\rm max}$ that maximizes $dM/dz/M$. We then self-consistently sample the phase space coordinates of $10^6$ particles from the distribution function of a Hernquist halo with mass $M(z)$ and a scale radius $r_s(z)$  - both today and at $z_{\rm max}$ (see Section \ref{hma}). Calculating $\Tilde{D}(J_r,t)$ for each particle, we determine the fraction of particles inhabiting the ``linear" part, the ``transition" part and the ``fringe" of the available integral of motion space (see Fig. \ref{fig:mass_accretion}). We find that at the current time, the radial action distribution of particles in a MW-size halo evolves rather linearly, with 83 per cent of particles being in the ``linear" regime of phase space. However, at $z=z_{\rm max}$, we find that radial actions could not be considered conserved quantities on average. In fact, with 18 per cent of particles inhabiting the ``fringe" and another 41 per cent being in the ``transition" area in phase space, we expect a highly non-linear evolution on average. Furthermore, we find that at $z=z_{\rm max}$ no linear integral of motion space is available at the largest radial actions. 
    
    From our discussion of the Kepler case, we expect that the radial action distributions of DM particles in haloes drift strongly towards smaller actions at early times (high redshifts). This impulsive drift is likely related to the formation of primordial cusps. At later times, our results support \citet{2013MNRAS.430..121P}'s statement that the ensemble average of radial actions is approximately conserved. We note, however, that there is a sparsely populated volume in phase space where the evolution of radial actions is not completely linear, even at late times. Our results suggest that the action distribution of tidal streams in Milky-Way like galaxies is not invariant, and that the hierarchical growth of the host DM halo necessarily causes diffusion and drift in radial actions. This may explain the results of \citet{2015A&A...584A.120B}, who show that streams in time-dependent potentials have a systematically larger spread of radial actions than streams in static potentials. Treating the evolution of radial actions as a diffusive process may thus be a way to improve the analysis of tidal streams in the MW and constrain the time evolution of the Galaxy potential. 
    
    It is important to bear in mind that our diffusion formalism provides approximate results. 
    The most significant caveat of our theory is that we consider diffusion in only one dimension. While this is sufficient to describe the impact of a global time-dependence of the potential, we cannot model resonant diffusion that may arise due to local perturbations caused by mergers, for instance. Taking these effects into account requires modeling the diffusion in multiple dimensions (see \citealt{2019MNRAS.484.5409P} for how this may be done). For now, our main achievement is that by modeling the diffusion of radial action distributions in time-dependent potentials we can successfully describe ensembles of tracer particles in potentials whose rate of evolution lies between adiabatic and impulsive. Since we do not resolve the effect of localized perturbations, the diffusion of stars in tidal streams predicted by our formalism would constitute a lower limit to the observed diffusion. 

   \noindent \\
   \\
    \textbf{Cusp-core transformation in a dwarf-size SIDM halo}\\
    Finally, we use the analysis presented in Section \ref{subsec:ana} to investigate whether cusp-core transformation in a dwarf-size SIDM halo 
    leads to linear or non-linear evolution of radial action distributions of tracer particles (see Section \ref{sec_sidm}). We re-run the SIDM simulation of table 1 in \citet{2019MNRAS.485.1008B}, increasing the number of tracers by a factor of ten and changing the self-scattering cross section to $\sigma/m_\chi = 1\,{\rm cm^2g^{-1}}$, a value which is around the current constraint on SIDM at dwarf scales \citep{Read2018}. We take snapshots every $143\,{\rm Myr}$ and closely follow the orbits of the tracers at each time step. We use the same functional form for the scale factor 
    as for the MW-size halo above. However, we now take the changing halo shape into account when calculating its time derivative. At each snapshot, we calculate the fraction of DM particles in the ``linear" and the ``transition" regime as well as the ``fringe" by calculating $\Tilde{D}$ and $J_r$ for each particle. The fraction of particles in the linear regime is never below 97 per cent (see Fig. \ref{fig:sidm_phasespace}) and we conclude that cusp-core transformation due to SIDM leads to a very slow change in the halo's potential, implying that the evolution of radial action distributions of particles orbiting in the SIDM halo is adiabatic.
    
     Our diffusion formalism predicts that $N(J_r,t)$ remains essentially constant during core formation. However, the measured radial action distribution gradually develops a tail towards larger radial actions that extends beyond the prediction made by the diffusion formalism (see Fig. \ref{fig:sidm_dist}). For that reason, we test in Appendix \ref{app3} whether the use of the approximate scale factor $R(t)$ is valid in potentials with a time-dependent shape. Modeling core formation in a smooth external potential rather than a live halo, and using the approximate scale factor, we find the linear Taylor expansion of the time evolution of $J_r$ to be in very good agreement with the directly measured evolution. Furthermore, we show that during the time in which the extended tail forms in the measured distribution $N(J_r,t)$, the shape of the potential is largely constant (see Fig. \ref{fig:sidm_alpha}), implying that radial actions should be conserved during that time. This leads us to conclude that the observed evolution of $N(J_r,t)$ is caused by individual close encounters between tracers and DM (simulation) particles, and is thus a numerical effect. Therefore, we take the prediction of our diffusion formalism at face value and deduce that core formation in dwarf-sized haloes due to SIDM with $\sigma/m_\chi \sim 1\,{\rm cm^2g^{-1}}$ has no detectable impact on the radial actions of kinematic tracers. In other words, it is fully adiabatic. 
    
\end{enumerate}

In summary, we have here for the first time analytically calculated the linear time evolution of radial actions in time-dependent spherical potentials. Based on that, we have developed a diffusion theory for radial action distributions of tracers in time-dependent gravitational potentials. This diffusion theory relates the evolution of the distribution $N(J_r,t)$ to an invariant distribution $N(J_r')$ in a frame which is related to the time-dependent frame via a coordinate transformation defined by a scale factor. We have demonstrated the validity of our formalism, and furthermore provided a discussion of its limitations, by performing restricted $N$-body simulations of tracer particles with different initial distributions in radial action and applying the diffusion formalism to model how these distributions evolve with time. We have shown that the size of the diffusion coefficient $\Tilde{D}(J_r,t)$ relative to the square of the radial action itself is a good indicator for whether radial action distributions will evolve linearly or not, and thus for whether the diffusion formalism we developed is applicable. We have furthermore shown that highly non-linear evolution causes a drift of radial action distributions. We have then applied our mechanism to two distinct cases of interest in astrophysics. 

First we have argued that while the radial actions of DM particles in a MW-size halo are likely conserved on average at late times, 
this was not the case shortly after the gravitational collapse of these objects. At the current time, we expect the evolution of radial action distributions of gravitational tracers to be mildly diffusive. Diffusion becomes increasingly important at earlier times, and we predict a significant asymmetric drift towards smaller radial actions for $z\gtrsim 2$. 
The impulsive evolution of haloes at very early times may be linked to the formation of primordial cusps, and taking into account diffusion due to the time-dependence of the MW halo's potential can potentially improve the analysis of tidal streams.
As a second application, we have demonstrated that radial actions are conserved in core-forming dwarf-size SIDM haloes.



In future work, we aim to extend our diffusion theory into two dimensions to take into account spatial resonances and deviations from spherical symmetry. Moreover, applying our diffusion theory to massive particles instead of tracers is an important future step towards developing a fully self-consistent statistical theory for the dynamical evolution of self-gravitating systems


\section*{Data Availability}
The data underlying this article were accessed from the Garpur supercomputer. The derived data generated in this research will be shared on reasonable request to the corresponding author.


\section*{Acknowledgements}

We thank Volker Springel for giving us access to the {\scriptsize AREPO} code. We thank Andrew Pontzen for helpful comments and we thank the anonymous referee for suggestions that have helped us clarify some important aspects of our theory. JB and JZ acknowledge support by a Grant of Excellence from the Icelandic Center for Research (Rann\'is; grant number 173929). The simulations in this paper were carried out
on the Garpur supercomputer, a joint project between the University of Iceland and University of Reykjav\'ik with funding from Rann\'is. 



\bibliographystyle{mnras}
\bibliography{main}

\appendix

\section{Numerical calculation of the scale factor}\label{app1}

In this section we discuss how we calculate the scale factor $R(t)$ numerically using Eq.~(\ref{scalefactor}). We start by recalling that Eq.~(\ref{scalefactor}) has an analytic solution if the evolution is adiabatic. In this case, we can neglect the term proportional to the second derivative of the scale factor to have 
\begin{align}
    R^3\mathbf{F}(R\mathbf{r}',t) \approx \mathbf{F}'(\mathbf{r'}).\label{A1}
\end{align}
Substituting Eqs.~(\ref{ind_pot}) and (\ref{td_kepler_pot}) into Eq.~(\ref{A1}), we then obtain
\begin{equation}
    -R^3\frac{GM_0(1+\epsilon t)}{(R r')^2} = -\frac{G M_0}{(r')^2}
\end{equation}
which leads to
\begin{equation}
    R(t) = \frac{1}{(1+\epsilon t)}\label{ana_sol},
\end{equation}
or in terms of the ratio $\dot{R}/R$, which is the relevant quantity in the linear expansion of the radial action (Eq.~\ref{linear_order})
\begin{align}
    \frac{\dot{R}}{R} = -\frac{\epsilon}{\epsilon+t}.\label{firstorder}
\end{align}

This implies that up to first order, the scale factor only depends on time, and there is no residual dependence on the orbital trajectory of individual tracer particles.
In this paper, however, we are interested in both the fully adiabatic regime and the regime in which the evolution of radial actions is non-adiabatic. Thus, adopting Eq.~(\ref{ana_sol}) as the scale factor for all tracer particles might be a bad approximation, in particular when $R/\ddot{R}$ is of the order of the radial period squared. 
%
To avoid this issue, we solve Eq.~(\ref{scalefactor}) numerically along the phase space trajectory of each particle, using a KDK leapfrog algorithm. First, we rewrite Eq.~(\ref{scalefactor}) for the Kepler case as 
\begin{align}
    \ddot{R}R^3r' + R^3\frac{GM(t)}{R^2r'^2} = \frac{GM(0)}{r'^2},\label{linappr}
\end{align}
which using $r  = Rr'$ can be written as 
\begin{align}
    &\ddot{R} = -R\frac{GM(t)}{r^3}+\frac{GM(0)}{r^3} = G(R(t),r(t),t).\label{lfbase}
\end{align}
During the simulation, we record phase space coordinates at each time step, as well as the simulation time. 
We evolve Eq. \ref{lfbase} between times $t$ and $t+\Delta t$, where  $\Delta t$ is the time step in the simulation, as follows: 
\begin{alignat}{2}
    &\dot{R}\left(t+\frac{\Delta t}{2}\right)&&= \dot{R}(t) +\frac{\Delta t}{2}G(R(t),r(t),t)\label{one}\\
    &R(t+\Delta t)&&= R(t)+\Delta t\,\dot{R}\left(t+\frac{\Delta t}{2}\right)\label{two}\\
    &\dot{R}(t+\Delta t)&&= \dot{R}\left(t+\frac{\Delta t}{2}\right)\label{three}\\
\nonumber    &  &&+\frac{\Delta t}{2}G(R(t+\Delta t),r(t+\Delta t),t+\Delta t)).
\end{alignat}
Since Eqs.~(\ref{one}$-$\ref{three}) explicitly depend on the radius of the tracer at a given time, there is now an orbit-dependent part that enters into the solution for $R(t)$ (contrary to Eqs.~\ref{ana_sol} and \ref{firstorder}). We thus 
expect that the true values of $R(t)$ will be distributed around the analytic solution given by Eq.~(\ref{ana_sol}).

\begin{figure}
    \centering
    \includegraphics[width=\linewidth]{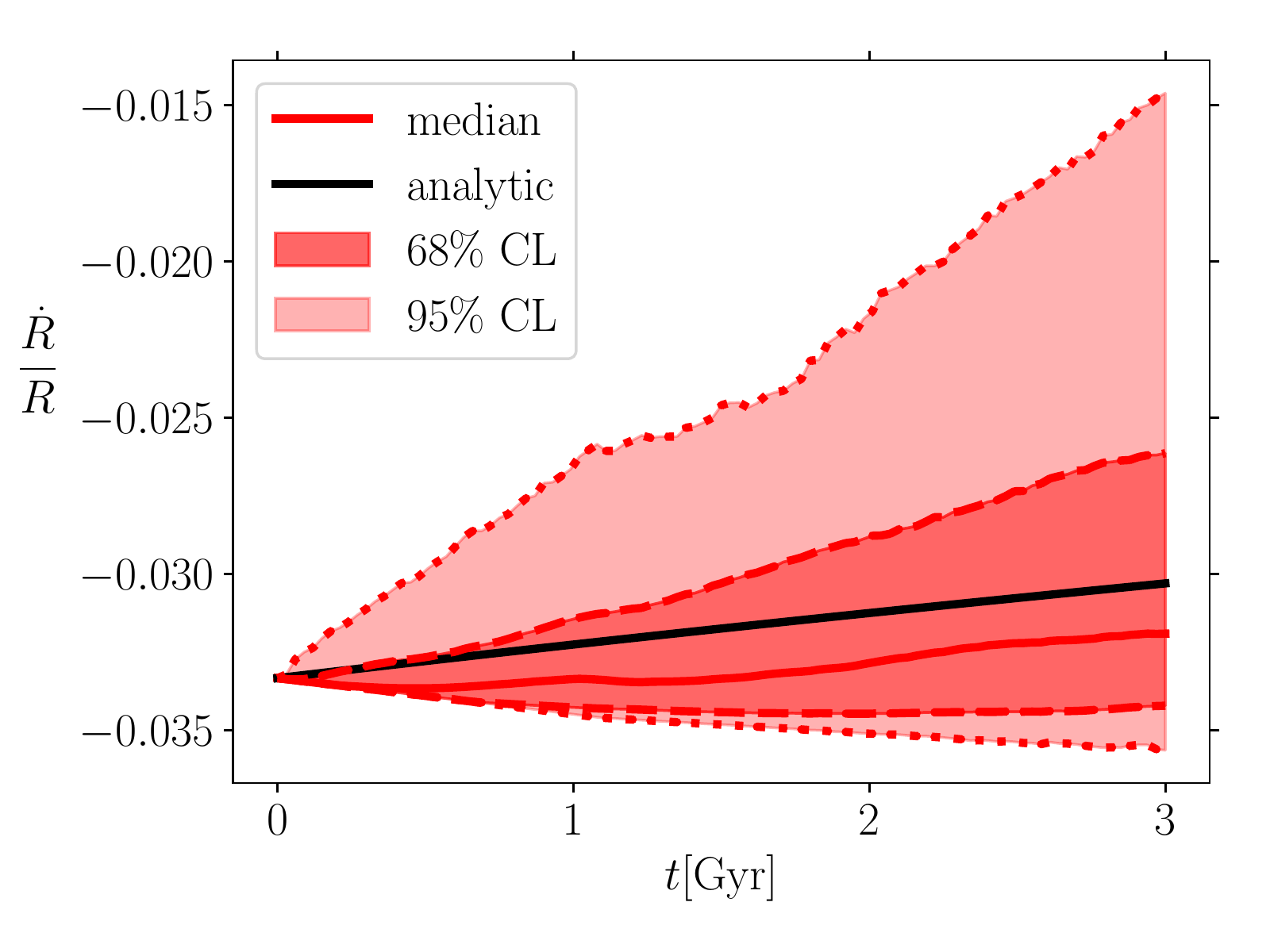}
    \caption{Confidence level regions of the distribution of $\dot{R}/R$ versus time, taken from the ``10-linear" simulation (see Table~\ref{tab_distributions}). The solid red line shows the median value, the dark (light) red shaded region covers 68\% (95\%) CL of the distribution. The analytic solution from Eq.~(\ref{firstorder}) is shown as a solid black line.}
    \label{fig:confidence_levels}
\end{figure}

Fig. \ref{fig:confidence_levels} shows the median of $\dot{R}/R$ versus time as a red solid line, calculated from the "10-linear" case defined in table \ref{tab_distributions}. The 68\% (and 95\%) confidence level regions are shown as dark (and light) red shaded areas. The analytic approximate solution given by Eq.~(\ref{firstorder}) is shown as a black solid line.
Both the median and the analytic solution lie well within the 68\% confidence level. Overall, we find that the analytic solution is a reasonable approximation for most particles. 
%
A small fraction of particles tends towards significantly (50 per cent) larger values of $\dot{R}/R$ as time progresses, yet the median stays close to the analytic solution at all times. This suggests that in the ``10-linear" case, which is the closest to being adiabatic, the analytic solution to Eq.~(\ref{linappr}), i.e. Eq.~(\ref{firstorder}), provides a good approximation of the true scale factor.

\section{Higher order perturbation theory}\label{app2}
The formalism presented in Sections \ref{sec_form} and \ref{sec_results} is based on linear perturbation theory.
Here we calculate the radial action evolution to higher orders in perturbation theory, allowing for a slightly more accurate description of radial actions inhabiting the ``transition" area between the ``linear" and ``fringe" regimes. 


Eq.~(\ref{radial_action_expansion}) is a general expansion of the invariant radial action $J_r'$, which equates the difference between $J_r'$ and $J_r$ to an infinite Taylor expansion. In Eq. (\ref{linear_order}) we truncate this series to first order by assuming that the perturbations to the radial action are sufficiently small on average. However, as we have seen throughout Section \ref{sec_results}, the linear prediction gradually deteriorates in the ``transition" regime and is not valid in the ``fringe" of the potential. 
Hereafter, we calculate the second order expansion and investigate how it changes the results of the diffusion formalism. 

To calculate $J_r$ to second order in perturbation theory, we first need to use the exact (higher order) version for the energy of a particle (Eq.~\ref{linear_energy}), which is (see \citealt{2013MNRAS.433.2576P})
\begin{align}
    \frac{I}{R^2} = E-\frac{\dot{R}}{R}\left(\mathbf{r}\cdot\mathbf{v}\right)+\frac{1}{2}\left(\frac{\ddot{R}}{R}+\left(\frac{\dot{R}}{R}\right)^2\right).\label{invariant_I}
\end{align}
The invariant action in the time-independent reference frame is
\begin{align}
    J_r' = \frac{1}{\pi}\int_{r'_{\rm peri}}^{r'_{\rm apo}}dr' \sqrt{2I-2\Tilde{\Phi}(\mathbf{r'})-\frac{L^2}{r'^2}}\label{basic_integral}
\end{align}
Since $I$ is a true dynamical invariant, changing between coordinate systems and writing $I$ in terms of time-dependent quantities cannot result in $I$ depending on phase space variables. 
The integral in Eq.~(\ref{basic_integral}) is thus always defined solely in the static frame, but we are allowed to write the constants and functions appearing in the integral as functions of time-dependent quantities, as long as the integrand depends on the integration variable in the same way. The defining equation relating the potentials in both frames is 
\begin{align}
    \Tilde{\Phi}(\mathbf{r')} = \frac{1}{2}R\ddot{R}r^2+R^2\Phi(\mathbf{r},t)\label{potential_shift}
\end{align}
which we can rewrite in terms of $r'$ as 
\begin{align}
      \Tilde{\Phi}(\mathbf{r'}) = \frac{1}{2}R^3\ddot{R}r'^2+R^2\Phi(R\mathbf{r'},t)\label{potential_shift2}  
\end{align}
If the potential is scale-free, this enables us to write the integrand of Eq.~(\ref{basic_integral}) in terms of time-dependent quantities while still keeping all dependence on the integration variable explicit and analytic. Note that to do so we have to use $\ddot{R} =  0$  in Eq.~(\ref{potential_shift2}) for consistency, since we originally used this assumption in Eq.~(\ref{scalefactor}) to obtain scale factors which do not depend on the phase-space trajectory of each particle.  

In potentials that are not scale-free, the scale factor cannot, in general, be treated as independent of the phase space trajectory. This implies that writing the potential in the time-independent frame in terms of time-dependent quantities introduces an implicit and thus non-analytic dependence on the integration variable (the radius) through the scale factor. This does not alter the linear correction, provided we take the following (adiabatic) limit. 

The key assumption is that even if the potential is not scale-free, we can approximate the scale factor as independent of the particle's trajectory on time-scales which are of the order of the dynamical time. This means that the scale factor, and hence the potential, evolves slowly compared to the particle's orbital period, which is exactly how the adiabatic regime is usually defined. Mathematically, this means that since we evaluate Eq.~(\ref{invariant_I}) at a fixed time and we treat $R$ in Eq.~(\ref{potential_shift2}) as dependent on time only, we can treat $R$ as a simple scalar when evaluating the integral \ref{basic_integral}. 

\begin{figure*}
    \centering
    \includegraphics[width=0.48\linewidth]{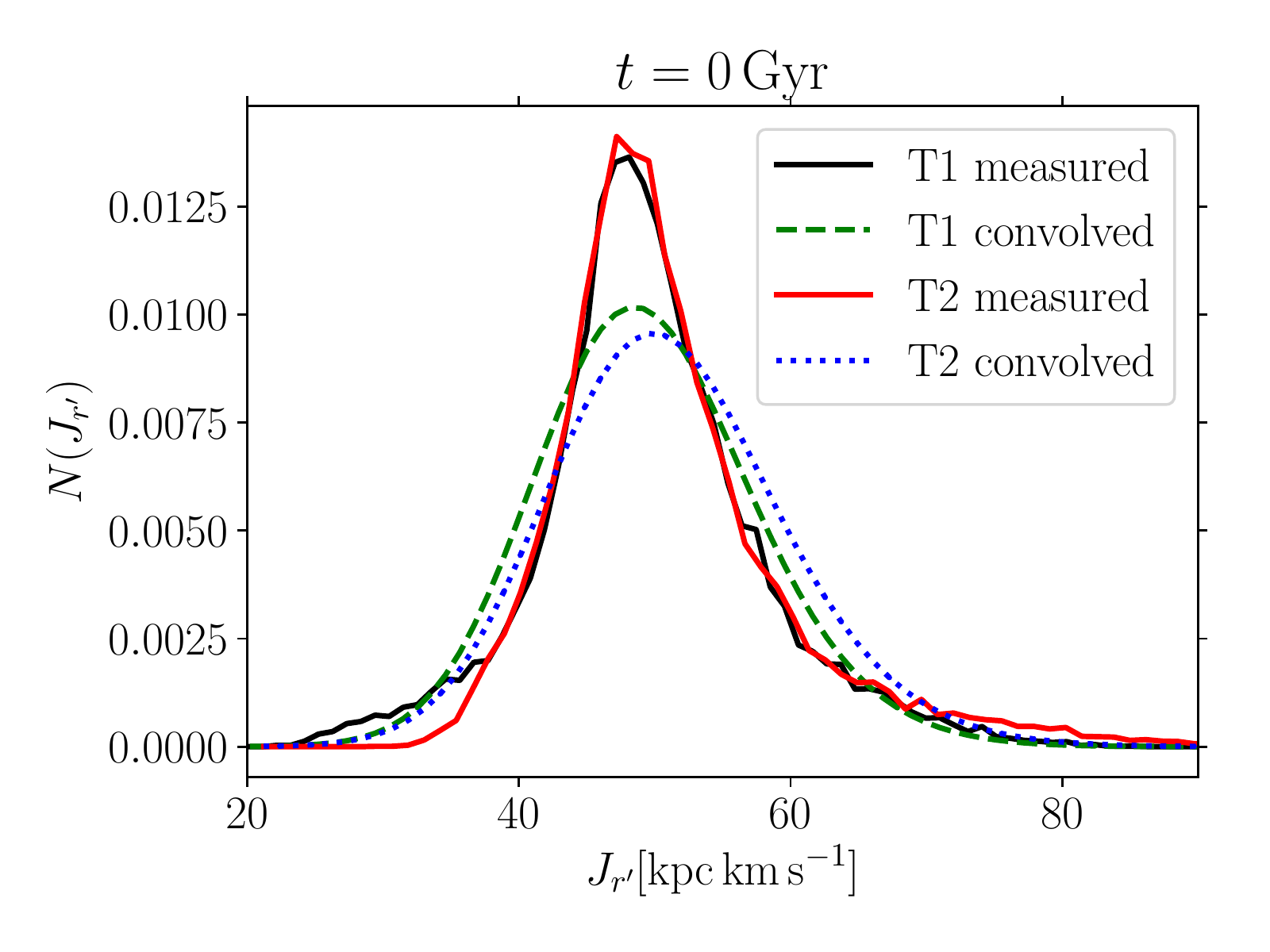}
    \includegraphics[width=0.48\linewidth]{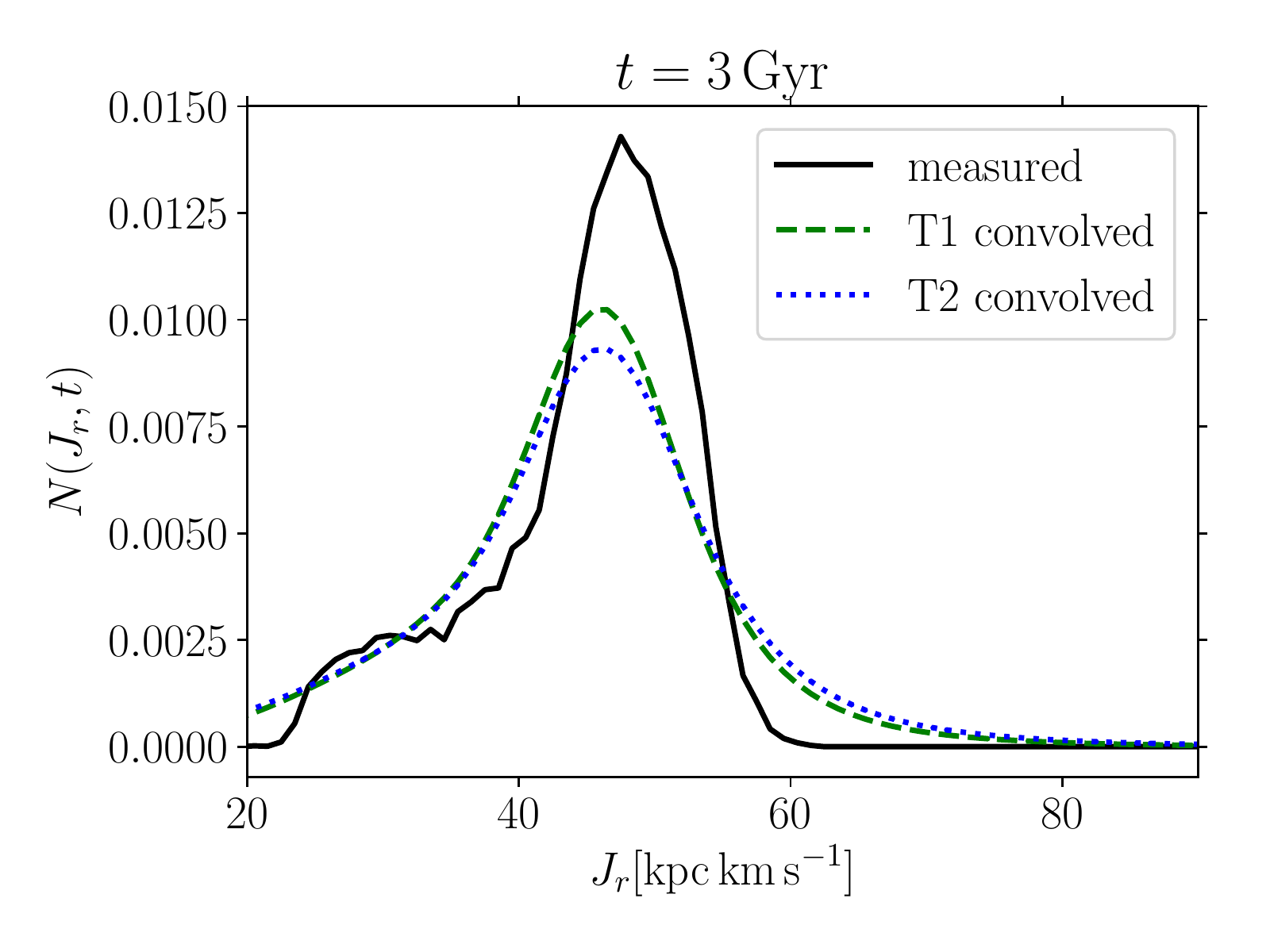}
    \caption{Comparison between the first (T1) and second (T2) order Taylor expansions (T1) in the "50-transition" case (see Table~\ref{tab_distributions} and Section~\ref{sec_results}). The left panel shows the invariant distribution calculated at $t = 0$. The black and red lines are distributions measured directly by calculating $J_r'$ for each particle and binning the results. The red line is calculated applying the second order Taylor expansion to calculate $J_r'$, whereas the black line is calculated using the first order expansion. The green dashed line is the result of a Gaussian convolution, with drift and diffusion coefficients calculated from the first order Taylor expansion. The blue dotted line is the result of a convolution using the second order Taylor expansion. On the right panel, we see the measured radial action distribution in the time-dependent frame in black, calculated at $t=3\,{\rm Gyr}$. The result of the first (second) order diffusion formalism is shown as a green dashed (blue dotted) line.} 
    \label{fig:app_b1}
\end{figure*}

To expand Eq.~(\ref{basic_integral}), we first introduce a few definitions. We define 
\begin{align}
    f'(r') = 2I-2\Tilde{\Phi}(\mathbf{r'})-\frac{L^2}{r'^2}\label{int_sq}
\end{align}
as the square of the integrand in the static frame. Then we define the functions $g'(I,L)$ and $h'(I,L)$ as the peri- and apocentre radii of particles with energy $I$ and angular momentum $L$, respectively. Next, we define
\begin{align}
\nonumber f(\mathbf{r'},y,z) &= 2E - 2 y \left(\mathbf{r}\cdot\mathbf{v}\right)+r^2\left(z+y^2\right)\\
&-R^2 z r'^2-2\Phi(R\mathbf{r'},t)-\frac{L^2}{R^2r'^2}, 
\end{align}
which is $f = R^2f'$ with the energy and the potential written in terms of time-dependent quantities. Here we have further introduced $y = \dot{R}/R$ and $z = \ddot{R}/R$ as the variables upon which the Taylor expansion will be made. Finally, we define $g(E,L,y,z)$ and $h(E,L,y,z)$ as $g=Rg',\,h =Rh'$, which 
reduce, respectively, to the  peri- and apocenter radii in the time-dependent frame when taking the limit $y\to 0,\,z\to 0$.
Under the assumption that $R$ is independent of the orbital trajectory, we can write $R\, dr' = dx$ and thus Eq.~(\ref{basic_integral}) is 
\begin{align}
    J_r' = \frac{1}{\pi}\int_{g(E,L,y,z)}^{h(E,L,y,z)}dx\, k(\mathbf{x},y,z),
\end{align}
where $k(\mathbf{x},y,z) = \sqrt{f(\mathbf{x},y,z)}$ and 
\begin{align}
\nonumber f(\mathbf{x},y,z) &= 2E - 2 y \left(\mathbf{r}\cdot\mathbf{v}\right)+r^2\left(z+y^2\right)\\
&-z x^2-2\Phi(\mathbf{x},t)-\frac{L^2}{x^2}, 
\end{align}
It is thus evident that $J_r'\vert_{y,z\to 0} = J_r$.
For the expansion to first order in $y$, we have to calculate 
\begin{align}
 \nonumber   \frac{\partial J_r'}{\partial y} &= \frac{1}{\pi}\int_{g(E,L,y,z)}^{h(E,L,y,z)}dx \frac{\partial k(\mathbf{x},y,z)}{\partial y}\\
    &+\frac{1}{\pi}k(h(E,L,y,z),y,z)\frac{\partial h}{\partial y} -\frac{1}{\pi} k(g(E,L,y,z),y,z)\frac{\partial g}{\partial y},
\end{align}
which satisfies the following limit
\begin{align}
    \left.\frac{\partial J_r'}{\partial y}\right|_{y,z\to 0} = -\left(\mathbf{r}\cdot\mathbf{v}\right)\frac{P(E,L)}{2\pi}
\end{align}
To continue the expansion to higher orders, we have to calculate 
\begin{align}
 \nonumber   \frac{\partial J_r'}{\partial z} &= \frac{1}{\pi}\int_{g(E,L,y,z)}^{h(E,L,y,z)}dx \frac{\partial k(\mathbf{x},y,z)}{\partial z}\\
    &+ \frac{1}{\pi}k(h(E,L,y,z),y,z)\frac{\partial h}{\partial z} -\frac{1}{\pi} k(g(E,L,y,z),y,z)\frac{\partial g}{\partial z}
\end{align}
and 
\begin{align}
 \nonumber   \frac{\partial^2 J_r'}{\partial y^2} &= \frac{1}{\pi}\int_{g(E,L,y,z)}^{h(E,L,y,z)}dx\, \frac{\partial^2 k(\mathbf{x},y,z)}{\partial y^2}\\
\nonumber &+\frac{2}{\pi}\frac{\partial k(h(E,L,y,z),y,z)}{\partial y}\frac{\partial h}{\partial y}\\
 \nonumber   &-\frac{2}{\pi}\frac{\partial k(g(E,L,y,z),y,z)}{\partial y}\frac{\partial g}{\partial y}\\
 \nonumber &+ \frac{1}{\pi}\frac{\partial k(h(E,L,y,z),y,z)}{\partial h(E,L,y,z)}\left(\frac{\partial h}{\partial y}\right)^2 \\
  \nonumber &- \frac{1}{\pi}\frac{\partial k(g(E,L,y,z),y,z)}{\partial g(E,L,y,z)}\left(\frac{\partial g}{\partial y}\right)^2 \\
\nonumber  &+\frac{1}{\pi}k(h(E,L,y,z),y,z)\frac{\partial^2 h}{\partial y^2}\\
    &- \frac{1}{\pi}k(g(E,L,y,z),y,z)\frac{\partial^2 g}{\partial y^2}.
\end{align}
We can then write to second order
\begin{align}
    J_r' = J_r +A\frac{\dot{R}}{R} + B\frac{\ddot{R}}{R} +\frac{1}{2}C\left(\frac{\dot{R}}{R}\right)^2,\label{hoexp}
\end{align}
with 
\begin{align}
    A = \left.\frac{\partial J_r'}{\partial y}\right|_{y,z\to 0},  \ B = \left.\frac{\partial J_r'}{\partial z}\right|_{y,z\to 0},  \ C = \left.\frac{\partial^2 J_r'}{\partial y^2}\right|_{y,z\to 0}.
\end{align}
%
We note that in Eq.~(\ref{hoexp}), we implicitly assume that $\ddot{R}/R$ is of the same order as $(\dot{R}/R)^2$. In the Kepler example, we can verify this and find from Eq.~(\ref{ana_sol}) that $\ddot{R}/R = 2 \dot{R}/R$ in the adiabatic limit.

In general, the integrals defining the coefficients $A,B,C$ in Eq.~(\ref{hoexp}), have to be evaluated numerically, 
which is a complicated task, 
especially considering that all we hope to achieve is a slightly better understanding of the evolution of radial action distributions in the ``transition" regime of Fig. \ref{fig:keyplot}. Fortunately, in the case of the Kepler potential, we can calculate the invariant action explicitly as 
\begin{align}
    J_r' = \frac{GM_0}{\sqrt{-2I}}-L
\end{align}
If we Taylor expand the invariant action up to the second order, we find
\begin{align}
\nonumber    J_r &= J_r'+\frac{GM}{\sqrt{2\mathcal{E}}^3}\left(\mathbf{r}\cdot\mathbf{v}\right)\frac{\dot{R}}{R}\\
\nonumber&-\frac{1}{2}\frac{GM}{\sqrt{2\mathcal{E}}^3}r^2\left(\left(\frac{\dot{R}}{R}\right)^2+\frac{\ddot{R}}{R}\right)\\
&-\frac{3}{2}\frac{GM}{\sqrt{2\mathcal{E}}^5}\left(\mathbf{r}\cdot\mathbf{v}\right)^2\left(\frac{\dot{R}}{R}\right)^2.\label{2nd_order}
\end{align}
From Eq.~(\ref{2nd_order}) we see that second order corrections become important at larger energies, i.e., closer to the ``fringe". Here we investigate the impact of including second order corrections on the ``$50$-transition" case of Section \ref{sec_results} (see Table~\ref{tab_distributions}).

In Fig. \ref{fig:app_b1} we compare the results of the diffusion formalism for the ``50-transition" distribution when using first order coefficients (T1) and second order coefficients (T2). The left panel compares the invariant distributions at the beginning of the simulation. The ``measured" curves refer to a direct measurement of the distributions, where in order to measure the T1 (T2) distribution we have applied the first order (second order) correction to the action of each individual particle.
The ``convolved" curves are the result of applying the Gaussian convolution defined in Eq.~(\ref{fwd_conv}), using drift and diffusion coefficients obtained from the first or second order Taylor expansions. 
Overall, we find that ``T1 measured" and ``T2 measured" look very similar when compared to each other. The same is true for ``T1 convolved" and ``T2 convolved", implying that including second order corrections does not improve the accuracy of our formalism. 

The effect of including the second order coefficients is mainly to reduce (increase) the tail at the lower (higher) end of the ``measured" invariant distribution 
(left panel of Fig.~\ref{fig:app_b1}).
This is to some extent expected, 
as the second order corrections all increase the value of $J_r'$ (see Eq.~\ref{2nd_order}). 
Comparing the ``T1" curves to the ``T2" curves, we find no significant improvement of the match between the ``measured" and the ``convolved" curves. 
Since the tails of the invariant distribution seem to be quite sensitive to the order of the perturbative expansion, we expect higher orders to have a sizeable impact as  well.  

On the right panel of Fig. \ref{fig:app_b1} we show the final result for the distribution of radial actions in the time-dependent frame obtained from Eq.~(\ref{rev_conv}), using the distributions on the left panel and drift and diffusion coefficients calculated at $t = 3\,{\rm Gyr}$. Once more, T1 (T2) refers to the first (second) order Taylor expansion. 
Evidently, the agreement between the measured distribution and the result of the diffusion formalism does not improve significantly when using second order coefficients. 
The most obvious effect is that the second order result has a slightly more extended tail towards larger radial actions, which is, however, hardly significant. The tail on the lower action side is still resolved quite well in both cases. 
Overall, the diffusion formalism's prediction of the evolution of the `50-transition" distribution does not improve when including second order terms.  


We conclude that the evolution of the  ``50-transition" distribution is
is at least partly non-perturbative and cannot be fully captured by the diffusion formalism derived in Section \ref{sec_form}. 
Nonetheless, the second order correction still represents a more accurate approximation of the invariant action of individual particles inhabiting the transition regime. We can thus use it to gain additional insight into the non-linear evolution of $N(J_r,t)$ in the ``200-transition" simulation.

\section{An approximate scale factor for generic spherical potentials}\label{app3}

\begin{figure*}
	\includegraphics[height=6.5cm,width=8.5cm,trim=1.0cm 0.5cm 0.75cm 0.0cm, clip=true]{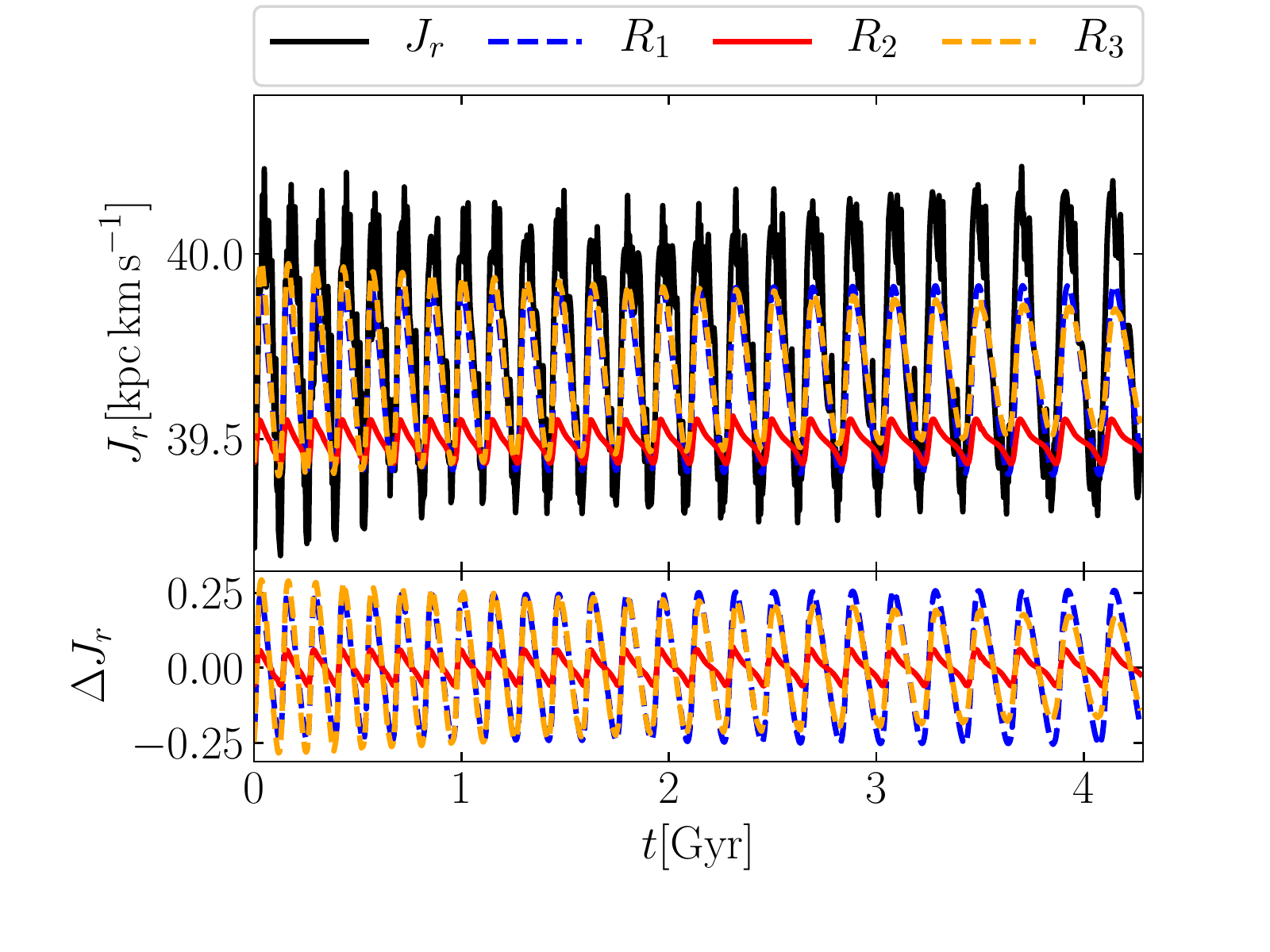}
	\includegraphics[height=6.5cm,width=8.5cm,trim=1.0cm 0.5cm 0.75cm 0.0cm, clip=true]{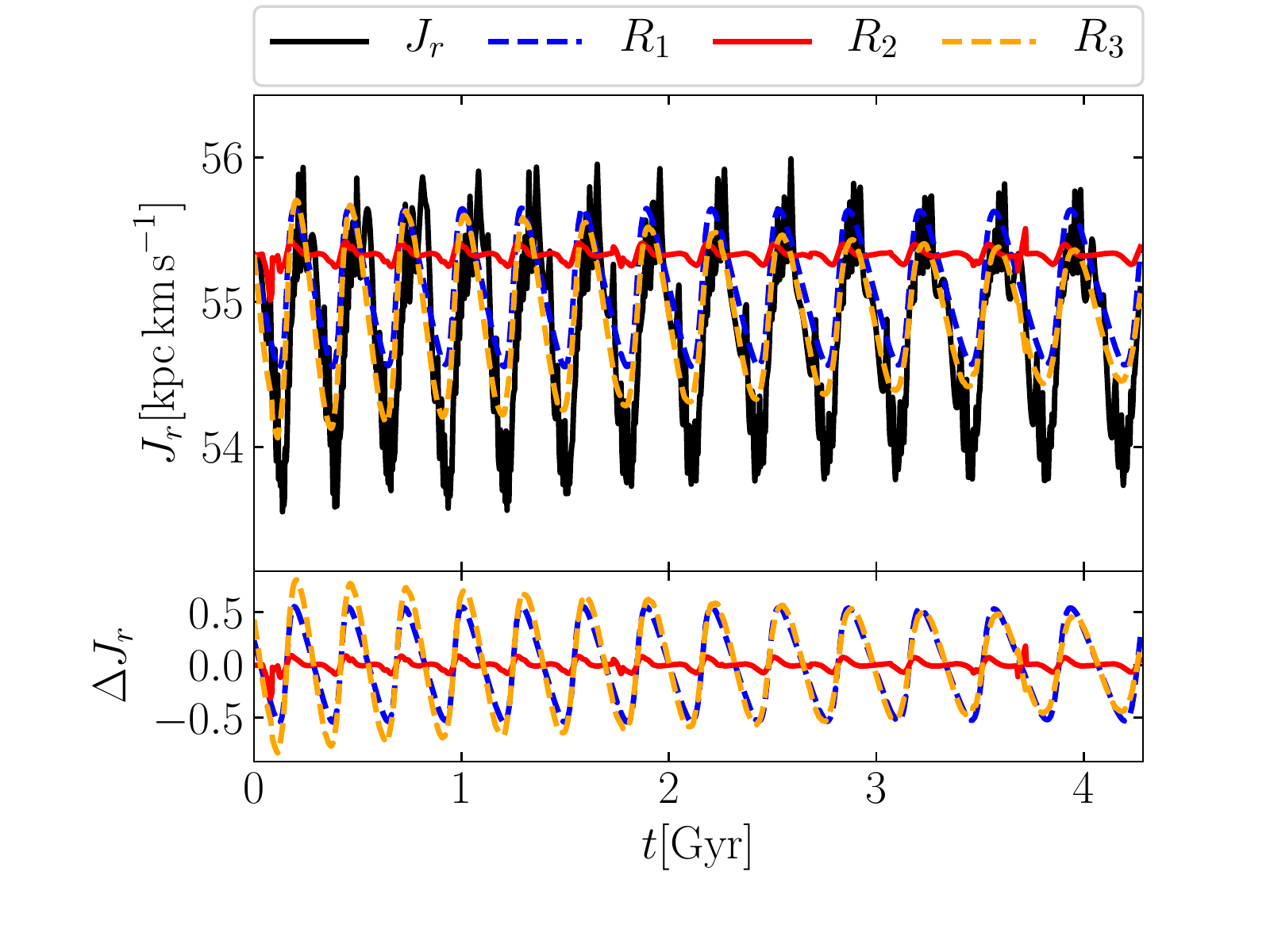}
	\caption{The top panels show the evolution of the radial action of two tracers in a time-dependent Dehnen potential; see text for details. The black line denotes a direct measurement of $J_r$. The dashed blue line, the solid red line, and the dashed orange line indicate the first order Taylor expansion of $J_r$ (Eq.~(\ref{linear_order}), using different approximations for the scale factor $R(t)$ (Eqs.~\ref{R_1}, \ref{R_2} and \ref{R_3}). In the bottom panels, we show the first order corrections $\Delta J_r$ corresponding to each of the three scale factors.  }\label{fig:dehnen_evolution}
\end{figure*}
For the case of scale-free power law potentials, \citet{2013MNRAS.433.2576P} showed that in the adiabatic limit, an approximate solution to Eq.~(\ref{scalefactor}) is given by Eq.~(\ref{decoupled}). In Appendix \ref{app1}, we compare this approximate scale factor to a distribution of numerically calculated scale factors of a set of tracer particles orbiting in a time-dependent Kepler potential; overall, we find a good agreement.
For several applications of physical interest, it is desirable to obtain an approximate scale factor for generic spherical potentials which are not scale-free. Under certain assumptions, we can find an approximate solution to Eq.~(\ref{scalefactor}) as we show in the following. If we assume that the force can be written as 
\begin{align}
    \mathbf{F}(\mathbf{r},t) = F(t)\,r^{\beta(r,t)}\,\frac{\mathbf{r}}{r},
\end{align}
where $\beta(r)$ is a local power law index, we can find an approximate solution to Eq.~(\ref{scalefactor}) in the adiabatic limit, which is 
\begin{align}
    R_1(t) = \left(\frac{F(r,t)}{F(r,0)}\right)^{-\frac{1}{(3+\beta(r,t))}},\label{R_1}
\end{align}
where $F(r,t)$ denotes the magnitude of the local time-dependent force. \citet{2013MNRAS.433.2576P} shows that Eq.~(\ref{scalefactor}) can also be integrated, leading to an equation that relates the time-dependent potential to the constant potential in the time-independent frame. An approximate solution to this integrated equation is 
\begin{align}
    R_2(t) = \frac{1}{2+\alpha(r,t)} \left(\frac{\Phi(r,t)}{\Phi(r,0)}\right)^{-\frac{1}{2+\alpha(r,t)}}, \label{R_2}
\end{align}
where 
\begin{align}
    \alpha(r,t) = \frac{d\log(-\Phi(r,t))}{d\log r}.
\end{align}
It is worth noting, however, that the potential $\Phi(r,t)$ is only defined up to an integration constant. In most $N$-body codes, this constant is used to fix the limit of the potential at large radii to $\Phi(r\to\infty,t) = 0$. For the potentials of most self-gravitating systems, e.g. Plummer, Hernquist or NFW potential, this implies $\Phi(r\to 0,t) = {\rm const}$, and thus a local power law index of $\alpha(0,t) = 0$, irrespective of the behaviour of the force. However, to make Eq.~(\ref{R_2}) consistent with Eq.~(\ref{R_1}), we demand that $\alpha(0,t) = 1+\beta(0,t)$, in order to match the case of scale-free potentials. We can recover this behaviour by setting the integration constant to zero when calculating the potential, or equivalently, by defining
\begin{align}
    \Psi(r,t) = \Phi(r,t)-\Phi(0,t). 
\end{align}
Then, we can calculate the scale factor as 
\begin{align}
    R_3(t) = \frac{1}{2+\alpha(r)} \left(\frac{\Psi(r,t)}{\Psi(r,0)}\right)^{-\frac{1}{2+\alpha(r)}}, \label{R_3}
\end{align}
where 
\begin{align}
    \alpha(r) = \frac{d\log(\Psi(r,t))}{d\log r}.
\end{align}

When calculating the first order correction term to the radial action in Eq.~(\ref{linear_order}), the relevant quantity is not the scale factor itself, but its normalized time derivative $\dot{R}/R$.
For cusp-core transformation in a SIDM halo, which we discuss in Section \ref{sec_sidm}, both the time evolution of the enclosed mass at a given radius and the time-evolution of the potential's logarithmic slope can be significant. This can cause two potential problems. Firstly, the approximate solutions (Eqs.~\ref{R_1}, \ref{R_2} and \ref{R_3}) are derived under the assumption of constant local power law slopes, and it is thus not clear whether they will hold in a potential with a time-dependent shape. Moreover, the time-dependence of the power law slopes differs between Eqs.~(\ref{R_1}), (\ref{R_2}) and (\ref{R_3}), both due to the impact of the integration constant on the power law index of the potential, and because in generic spherical potentials, there is not a single one-to-one correspondence between the power law indices of the force and the potential, i.e., $\alpha = 1+\beta$ is never true for all radii in potentials that are not scale free. 

Here we perform a short numerical test to determine whether we can use the approximate scale factors introduced above to describe the evolution of potentials with a time-dependent shape. We follow the orbits of tracers in the potential generated by a time-dependent Dehnen density profile \citep{1993MNRAS.265..250D}:
\begin{align}
    \rho(r,t) \propto r^{-\gamma(t)}\left(r+r_s\right)^{\gamma(t)-4}
\end{align}
We fix the normalization of the density profile by enforcing mass conservation. At each point in time we normalize the density profile such that 
\begin{align}
    M(<r_{200}) = 1.43\times 10^{10}M_{\odot},
\end{align}
where $r_{200} = 50.1\,{\rm kpc}$. The scale radius $r_s$ is then fixed by setting $c_{200} = 15$. We model the transition from a cuspy to a cored profile with the following time dependence for the power law slope: 
\begin{align}
    \gamma(t) = 1-\frac{t}{t_{\rm max}},
\end{align}
where $t_{\rm max} = 4.3\,{\rm Gyr}$ is the total simulation time. We then closely follow the orbits of kinematic tracers and numerically calculate the radial action at each time-step (a similar exercise as in Fig. \ref{3particles}). Additionally, we also calculate the first order Taylor expansion of $J_r$ defined in Eq.~(\ref{linear_order}) using the scale factors $R_1$, $R_2$ and $R_3$ defined above (Eqs.~\ref{R_1}, \ref{R_2} and \ref{R_3}). To that end, we estimate their time derivatives as 
\begin{align}
 \nonumber   \frac{{\rm d}R_1}{{\rm d}t} &=  -\frac{1}{\left(3+\beta(r,t)\right)}\times
    \left(\frac{F(r,t)}{F(r,0)}\right)^{-\frac{1}{3+\beta(r,t)}}\\
    &\times \left\{\frac{\dot{F}}{F}-\frac{1}{(3+\beta(r,t)}\ln\left(\frac{F(r,t)}{F(r,0)}\right)\times\frac{{\rm d}\beta}{{\rm d}t}\right\}\\
\nonumber    \frac{{\rm d}R_2}{{\rm d}t} &=  -\frac{1}{\left(2+\alpha(r,t)\right)^2}\times \left(\frac{\Phi(r,t)}{\Phi(r,0)}\right)^{-\frac{1}{2+\alpha(r,t)}}\\
    &\times \left\{\left[1-\frac{1}{2+\alpha(r,t)}\ln\left(\frac{\Phi(r,t)}{\Phi(r,0)}\right) \right]\times\frac{{\rm d}\alpha}{{\rm d}t} + \frac{\dot{\Phi}}{\Phi} \right\}\\  
\nonumber    \frac{{\rm d}R_3}{{\rm d}t} &= -\frac{1}{\left(2+\alpha(r,t)\right)^2}\times \left(\frac{\Psi(r,t)}{\Psi(r,0)}\right)^{-\frac{1}{2+\alpha(r,t)}}\\
    &\times \left\{\left[1-\frac{1}{2+\alpha(r,t)}\ln\left(\frac{\Psi(r,t)}{\Psi(r,0)}\right) \right]\times\frac{{\rm d}\alpha}{{\rm d}t} + \frac{\dot{\Psi}}{\Psi} \right\},
\end{align}
where we have taken into account that the local power law slopes are time-dependent.

Fig. \ref{fig:dehnen_evolution} shows the evolution of the radial action of two tracers, along with the prediction of the first order Taylor expansion using each of the three scale factors. Apart from some numerical issues that occur when calculating the radial action close to the apo- and pericentre of the particle's orbits, the evolution of $J_r$ for both tracers is accurately described by the Taylor expansion up to linear order, provided we either use $R_1$ (Eq.~\ref{R_1}) or $R_3$ (Eq.~\ref{R_3}) as the scale factor. In both of these cases, the amplitude and oscillation period of the radial actions are well described by the Taylor expansion. Comparing the linear correction terms directly (lower panels of Fig. \ref{fig:dehnen_evolution}) reveals that the corrections obtained when using $R_1$ and $R_3$ are almost identical. The result of the Taylor approximation using $R_2$, however, provides a rather bad description of the evolution of $J_r$ for both tracers. This confirms that the most important effect is the shape evolution of the potential. Given that in the inner part of the potential, where the impact of the modelled core formation is most substantial, there is a one-to-one correspondence between the logarithmic slopes of $R_1$ and $R_3$, it is no surprise that they yield very similar values for $\Delta J_r$. $R_2$ fails because the logarithmic slope is not related to the radial dependence of the force in the innermost region of the potential.  

Since there is no physical preference of one scale factor over the other, we conclude that for the purpose of modeling core formation in a SIDM halo, as done in Section \ref{sec_sidm}, both $R_1$ and $R_3$ can be used to obtain a valid approximation of the evolution of radial actions of particles orbiting in the time-dependent host potential.
\bsp	
\label{lastpage}
\end{document}